\documentclass[11pt,superscriptaddress,amsmath,amssymb,nofootinbib,longbibliography]{revtex4-2}
\usepackage{graphicx}
\usepackage{dcolumn}
\usepackage{bm}
\usepackage{amssymb}
\usepackage{amsmath}
\usepackage{color}
\usepackage[normalem]{ulem}
\usepackage{dsfont}

\newcommand{\bwt}{\begin{widetext}}
\newcommand{\ewt}{\end{widetext}}
\newcommand{\newc}{\newcommand}
\newc{\hc}{\dagger}
\newc{\pd}{\partial}
\newc{\beq}{\begin{equation}}
\newc{\eeq}{\end{equation}}
\newc{\beqa}{\begin{eqnarray}}
\newc{\eeqa}{\end{eqnarray}}
\newc{\bi}{\begin{itemize}}
\newc{\ei}{\end{itemize}}
\newc{\ra}{\rightarrow}
\newc{\la}{\leftarrow}
\newc{\lra}{\longrightarrow}
\newc{\lla}{\longleftarrow}
\newc{\Lra}{\Longrightarrow}
\newc{\Lla}{\Longleftarrow}
\newc{\half}{\frac{1}{2}}
\newc{\fth}{\frac{1}{4}}
\newc{\hchecked}{\backslash\!\!\!\!\checkmark }
\newc{\del}{\delta}
\newc{\Del}{\Delta}
\newc{\gm}{\gamma}
\newc{\Gm}{\Gamma}
\newc{\lam}{\lambda}
\newc{\kap}{\kappa}
\newc{\tri}{\triangle}
\newc{\eps}{\epsilon}
\newc{\epsp}{\epsilon^\prime}
\newc{\ot}{\frac{1}{3}}
\newc{\tth}{\frac{2}{3}}
\newc{\ft}{\frac{4}{3}}
\newc{\wt}{\widetilde}
\newc{\ovl}{\overline}
\newc{\tchi}{\tilde{\chi}}
\newc{\ds}{\displaystyle}
\newc{\pmt}{\pm\!\pm}
\newc{\PL}{\hat{L}}
\newc{\PR}{\hat{R}}

\newc{\msm}{\mathrm{SM}}
\newc{\msh}{\mathrm{sh}}
\newc{\mtev}{\mathrm{TeV}}
\newc{\mgev}{\mathrm{GeV}}
\newc{\mmev}{\mathrm{MeV}}
\newc{\mkev}{\mathrm{keV}}
\newc{\mev}{\mathrm{eV}}
\newc{\Tr}{\mathrm{Tr}}
\newc{\nonr}{\nonumber}
\newc{\clbl}{\color{blue}}
\newc{\clg}{\color{green}}
\newc{\clr}{\color{red}}
\mathchardef\mhyphen="2D
\newc{\SL}{\not\!\!}
\usepackage{bm}% bold math
\usepackage{amsfonts}
\usepackage{xcolor}
\usepackage{amscd}
\usepackage{amsmath}

\begin{document}

\title{Possibly heteroclite electron Yukawa coupling and small $\triangle a_\mu$ in a hidden  Abelian gauge model for neutrino masses}
\date{\today}

\author{We-Fu Chang}
\email{wfchang@phys.nthu.edu.tw}
\affiliation{Department of Physics, National Tsing Hua University, Hsinchu, Taiwan 30013, R.O.C. }

\author{Shih-Hsien Kuo}
\affiliation{Department of Physics, National Tsing Hua University, Hsinchu, Taiwan 30013, R.O.C. }

\begin{abstract}
We attempt  to simultaneously explain the neutrino oscillation data and the observed $(g-2)_{e,\mu}$ in a hidden gauge $U(1)_X$  model where all the Standard Model(SM) fields are $U(1)_X$ singlets.  The minimal version of this model calls for four exotic scalars and two pairs of vector fermions, and all are charged under $U(1)_X$. We carefully consider the experimental limits on charge lepton flavor violation without assuming any flavor symmetry and explore the viable model parameter space. The model can accommodate the neutrino oscillation data for both the normal and the inverted mass ordering while
explaining the central value of $\triangle a_e$ by adopting the fine structure constant determined by using either Cesium or Rubidium atoms.
However, mainly constrained by the current experimental bound on ${\cal B}(\tau\rightarrow \mu \gamma)$, this model predicts $\triangle a_\mu <5.5(8.0)\times 10^{-10}$ for the normal(inverted) neutrino ordering.
Moreover, while the muon Yukawa coupling is close to the SM one, we find the magnitude of the electron Yukawa coupling could be one order of magnitude larger than the SM prediction. This abnormal electron Yukawa could be probed in the future FCC-ee collider and plays an essential role in testing flavor physics.

\end{abstract}
\maketitle

%%%%%%%%%%%%%%%%%%%%%%%%%%%%%%%%%%%%%%%%%%%%%%%%%%
\section{Introduction}
%%%%%%%%%%%%%%%%%%%%%%%%%%%%%%%%%%%%%%%%%%%%%%%%%%
In the SM of particle physics, the charged fermions acquire their masses after the spontaneous breaking of the $SU(2)_L\times U(1)_Y$ symmetry via the Higgs mechanism.
The fermion masses, $M_F$, are fixed by their Higgs Yukawa couplings, $y_F$, and $M_F =y_F v_0/\sqrt2$, where $v_0\sim 246 \mgev$ is the SM Higgs vacuum expectation value(VEV).
The determination of the Higgs Yukawa couplings of top\cite{ATLAS:2018hxb}, bottom\cite{ATLAS:2018kot, CMS:2018nsn}, tau\cite{ATLAS:2015xst,CMS:2017zyp}, and muon\cite{ATLAS:2020fzp,CMS:2020xwi}  are consistent with the predicted relationship within the errors, typically around a few $\times 10\%$ \cite{ATLAS:2019nkf, CMS:2018uag, Zyla:2020zbs}.
 The consistency  suggests that the origin of masses of the second and the third generation charged fermions can be well accounted by the SM Yukawa interaction.
Although theoretically economic and technically natural, the SM does not explain the origin of the observed puzzling mass hierarchy among the charged fermions.
Meanwhile, the measurements of Yukawa couplings of other light charged fermions are very challenging due to their smallness or(and) colossal experimental backgrounds.
 The current upper limit on electron-Yukawa $|y_e /y_e^{SM}|<260$ \cite{CMS:2014dqm,ATLAS:2019old},  obtained from the branching ratio of ${\cal B}(h\ra e^+e^-)$, is relatively poor, and it leaves considerable room for the signal of new physics, see for example\cite{Botella:2016krk, Ghosh:2015gpa, Altmannshofer:2015esa, Dery:2017axi, Chiang:2021pma}, which might shed light on the origin of flavor.

The flavor puzzle is not limited to the charged fermion sector only: with undoubtful evidence, at least two out of the three active neutrinos are massive\cite{Zyla:2020zbs}.
The observed neutrino masses are about twelve orders of magnitude more diminutive than the electroweak scale. Except for the Dirac CP phase and  whether the neutrino mass spectrum ordering is normal or inverted, the neutrino oscillation parameters, three mixing angles and two mass squared differences, have been determined to the precision at the percent level\cite{Esteban:2020cvm}.
Despite the tremendous success of the minimal SM\footnote{ Namely, there is no extra DOF beyond the three generations of quarks and leptons, gauge bosons of $SU(3)_c\times SU(2)_L\times U(1)_Y$, and the SM Higgs.  }, the origin of neutrino masses calls for new degrees of freedom beyond the SM, and we do not know what they are yet. Regardless of the neutrino mass generation mechanism, the well-measured neutrino oscillation parameters are invaluable guides in exploring the unknown territory of flavor physics.

Moreover, the recently measured $(g-2)_{e,\mu}$ appear to deviate from the SM predictions\footnote{See \cite{Aoyama:2020ynm} for a comprehensive review of the SM prediction of $(g-2)_\mu$.}.
Combining data from BNL E821 and the result of FNAL gives \cite{Abi:2021gix}
 \beq
 \tri a_\mu^{ F\!N\!A\!L} \simeq (25.1\pm 5.9)\times 10^{-10}\,.
 \eeq
A new measurement at the J-PARC\cite{Saito:2012zz}
is expected to improve the experimental uncertainty in the near future.
Two recent lattice estimations\cite{Borsanyi:2020mff, Ce:2022kxy} on the hadronic vacuum polarization contribution to $\tri a_\mu$ suggest a result
consistent with the experimental data. However, these lattice estimations differ significantly from those based on the dispersion relation\cite{Aoyama:2020ynm}.
More theoretical studies, see, for example, \cite{Crivellin:2020zul, Keshavarzi:2020bfy, Colangelo:2020lcg} and references therein, are needed to understand this discrepancy and its consequences.

As for the electron, $\tri a_e$ can be deduced with the input of fine-structure constant $\alpha_{em}$.
By adapting  $\alpha_{em}$ determined by using Cesium atoms\cite{Parker:2018vye}, $\tri a_e$  takes the value
\beq
\tri a_e^{Cs}  \simeq (-8.7\pm 3.6)\times 10^{-13}\,.
\label{eq:deltaAe_minus}
\eeq
However, the measurement of $\alpha_{em}$ by using Rubidium atoms\cite{Morel:2020dww} yields a result
\beq
\tri a_e^{Rb}  \simeq (+4.8\pm 3.0)\times 10^{-13}\,,
\label{eq:deltaAe_plus}
\eeq
which differs from $\tri a_e^{Cs}$ by $\sim 4\sigma$.
It is unclear how to resolve those theoretical and experimental discrepancies associated with $\tri a_{e,\mu}$  mentioned above.
More investigations are needed to settle down the issues. In this work, we try to accommodate $\tri a_e^{Cs}$ and $\tri a_e^{Rb}$ separately and scan our model parameter space to see the prediction of $\tri a_\mu$.

Aiming for a unified explanation for the neutrino mass generation and the anomalous magnetic moments of muon and electron,
we consider a dark gauge $U(1)_X$ model for the one-loop radiative neutrino mass mechanism. For recent attemps to explain both $\tri a_{e,\mu}$ and neutrino mass generation, see for example, Refs.\cite{Abdullah:2019ofw, Chen:2020jvl, Dutta:2020scq, Arbelaez:2020rbq, Jana:2020joi, Cao:2021lmj, Mondal:2021vou, Escribano:2021css, Hernandez:2021iss, Chang:2021axw, Borah:2021khc, Julio:2022ton, Julio:2022bue,  Chowdhury:2022jde}.
Although we focus on the flavor physics in this work, the potential connection to dark matter is another motivation for us to consider the hidden gauge $U(1)_X$. With proper arrangement,  the residual parity after  spontaneous breaking of $U(1)_X$ \cite{Krauss:1988zc}  can be utilized to ensure the stability of the dark matter candidate(s). See \cite{Gu:2007ug, Ma:2013yga, Kanemura:2011vm, Chang:2011kv, Baek:2015fea, Ho:2016aye, Ma:2016nnn, Nomura:2017vzp, Geng:2017foe, Han:2018zcn,  CentellesChulia:2019gic, Ma:2019yfo, Kang:2019sab, Jana:2019mez, Ma:2019iwj, CentellesChulia:2019xky, Jana:2019mgj, delaVega:2020jcp, Wong:2020obo, Chowdhury:2022jde} for the example implementations of the residual gauge parity on dark matter and neutrino mass generation.

As we will show, the new physics responsible for the neutrino mass generation and $(g-2)_{e,\mu}$ might leave a footprint in the electron-Yukawa.
The radiative mechanism for $M_\nu$ and $\tri a_{e,\mu}$ could also lead to sizable charged lepton flavor violation(CLFV) couplings and contradict  the stringent experimental constraints, ${\cal B}(l_i \ra  \gamma l_{j\neq i})$. % ${\cal B}(\mu\ra e\gamma)<10^{-13}$.
To avoid introducing further ad hoc assumptions on the flavor pattern, we take a bottom-up approach to investigate what the data say on the model parameter space. We should show that the resulting electron-Yukawa coupling, even its sign, may differ significantly from the SM prediction. Also, the model predicts that $\tri a_\mu\sim 5.5(8.0)\times 10^{-10}$ at most by saturating the current upper limit of ${\cal B}(\tau\ra \mu \gamma)$( and ${\cal B}(\mu\ra e \gamma)$), for normal ordering ( inverted ordering) neutrino mass. If the experimental limits on CLFV processes get improved, this model predicts a even smaller $\tri a_\mu$.

This paper is organized as follows.  Our model is detailed in Sec.\ref{SEC:model}, and we also consider the $(g-2)_{e,\mu}$, neutrino mass generation, and the effective Higgs-Yukawa couplings.
Sec.\ref{SEC:numerical} is devoted to the numerical study in that we scan the model parameter space to accommodate the neutrino oscillation data while all the experimental constraints are taken care. Four benchmark configurations and our findings are given therein.
In Sec.\ref{SEC:Pheno}, we discuss some phenomenological considerations with brief comments on the possible dark matter candidates and the prospect of detecting the new gauge boson.
Finally, we conclude in Sec.\ref{SEC:conclusion}.

%%%%%%%%%%%%%%%%%%%%%%%%%%%%%%%%%%%%%%%%%%%%%%%%%%
\section{Model}
%%%%%%%%%%%%%%%%%%%%%%%%%%%%%%%%%%%%%%%%%%%%%%%%%%
\label{SEC:model}
On top of the SM, our model employs two pairs of vector fermions $N_{1,2}$ and four scalars, $D,C,S_1,S_2$.
The new degrees of freedom(DOF) are charged under a hidden gauge $U(1)_X$ symmetry.
The detailed quantum numbers of the new DOFs are listed in Tab.\ref{table:newparticle}.
On the other hand, all the SM DOF's are singlet under the $U(1)_X$ such that the $U(1)_X$ is ``dark''.
\begin{table}[th]
\begin{center}
\begin{tabular}{|c||cccc|crrr|}\hline
       &   \multicolumn{4}{c|}{ New Fermion}  &  \multicolumn{4}{c|}{ New Scalar}\\\hline%\hline
 Symmetry$\backslash$ Fields & $N_{1L}$& $N_{1R}$ & $N_{2L}$ & $N_{2R}$&  $D=\begin{pmatrix} D^0 \\ D^-\end{pmatrix}$
  & $C $ & $S_1$ & $S_2$   \\\hline \hline
$SU(2)_L$ &  \multicolumn{4}{c|}{  $1$ }& $2$& $1$ & $1$  & $1$ \\%\hline
$U(1)_Y$  &  \multicolumn{4}{c|}{ $0$}  & $-\half$  & $-1$  & $0$ & $0$\\
$U(1)_X$ &  \multicolumn{4}{c|}{$1$}  & $-1$ &  $-1$ &$-1$ & $-2$ \\
\hline
\end{tabular}
\caption{New field content and quantum number assignment under the SM gauge symmetries $SU(2)_L \otimes U(1)_Y $,  and the gauge $U(1)_X$. }
\label{table:newparticle}
\end{center}
\end{table}

The  $U(1)_X$-invariant renormalizble Yukawa interactions and couplings involving new fermions are
\beqa
{\cal L} \supset  y^L_{ij}\, \overline{L_i}D N_{jR} + y^R_{ij}\, \overline{e_{Ri}} C N_{jL}
-\sum_{i}\half \left[\xi_{iL} \overline{N^c_{iL}}N_{iL} + \xi_{iR} \overline{N^c_{iR}}N_{iR} \right]S_2
-\sum_{ij}  \mathfrak{D}_{ij} \overline{ N_{iR}} N_{jL}
+H.c.
\label{eq:L_Yukawa}
\eeqa
where $\mathfrak{D}_{ij}$ are the new dimensionful Dirac couplings among $N$'s.
Here we adopt a convention that $L_i$ and $e_{Ri}$ are in the charged lepton flavor(mass) basis.
Note that one can choose the diagonal $\xi$'s without losing any generality.
The new Yukawa sector enjoys the conventional global lepton number symmetry $U(1)_L$ if  $\{ N, D, C, S_2 \}$ carry lepton number $\{1, 0,0, -2\}$, respectively\footnote{ See \cite{ Chao:2010mp, Schwaller:2013hqa, Chang:2018vdd,Chang:2018wsw, Chang:2018nid} for the discussion if $U(1)_L$ is gauged.}.

For the scalar sector, the Lagrangian is
\beqa
{\cal L} &\supset& (D_\mu H)^\dag (D^\mu H) + \mu_H^2 H^\dag H -\lambda_H (H^\dag H)^2 -V_{NP}\nonr\\
& +& (D_\mu D)^\dag (D^\mu D)+(D_\mu C)^* (D^\mu C)
 +(D_\mu S_1)^* (D^\mu S_1)+(D_\mu S_2)^* (D^\mu S_2)\,,
\eeqa
where the  new scalar potential is
\beqa
V_{NP} & =& -\mu_2^2 |S_2|^2 + M_D^2 D^\dag D+M_C^2|C|^2 + M_{S_1}^2|S_1|^2 \nonr\\
&+& \lambda_D (D^\dag D)^2 +\lambda_C |C|^4 +\lambda_1|S_1|^4+\lambda_2|S_2|^4 + \bar{\lambda}_{HD} (D^\dag H)(H^\dag D) \nonr\\
&+&  H^\dag H \left(\lambda_{HD} D^\dag D +\lambda_{HC} |C|^2+\lambda_{H1}|S_1|^2+\lambda_{H2}|S_2|^2 \right)\nonr\\
&+&  D^\dag D \left(\lambda_{DC} |C|^2+\lambda_{D1}|S_1|^2+\lambda_{D2}|S_2|^2 \right)\nonr\\
&+&  |C|^2\left(\lambda_{C1}|S_1|^2+\lambda_{C2}|S_2|^2 \right )+\lambda_{12}|S_1|^2|S_2|^2 \nonr\\
&+&   \sqrt2 \mu_{DC} H^\dag D C^* + \sqrt2 \mu_{DS} H^\dag \wt{D} S_1 +2 \lambda_4 H^\dag \wt{D} S_2 S_1^*  + \frac{\mu_{12}}{ \sqrt2} S_1^2 S_2^* +H.c.
\label{eq:L_scalar}
\eeqa
The global $U(1)_L$  is now explicitly broken by $\lambda_4$ and $\mu_{12}$, which are crucial for generating the neutrino Majorana masses.
Moreover, since $U(1)_L$ is explicitly broken, there is no massless Majoraon associated with the SSB of $U(1)_X$
\footnote{In some case, the massless Majoraon can play the role of dark radiation, see \cite{Chang:2014lxa, Chang:2016pya}.  }.
However,  with the presence of $\lambda_4$ or $\bar{\lambda}_{HD}$,  no simple analytic expression is available  for the scalar potential to be bounded from below.
In general, one needs to check the positivity condition numerically. For simplicity while keeping the essential physics, we will set $\lambda_4=0$ and $\bar{\lambda}_{HD}>0$ in our numerical study.

We assume a 2-stage symmetry breaking.
At an energy scale higher than the SM electroweak scale, the $U(1)_X$ is broken spontaneously as $S_2$ acquires a VEV $\left\langle S_2 \right\rangle =v_2/\sqrt{2}$.  The new gauge boson $Z_X$ acquires a mass $M_X = 2 g_X v_2$, where the gauge coupling $g_X$ is a free paramter and we assume $v_2 \gg M_W$.
 The imaginary part of $S_2$ is the would-be-Goldstone eaten by the $Z_X$ boson.
 In between  the $U(1)_X$ breaking scale and the SM electroweak scale, the remaining gauge symmetry is the SM $SU(3)_c\times SU(2)_L\times U(1)_Y$.

The mass term of new fermions becomes
${\cal L} \supset \frac{1}{2} \overline{N^c} M_N  N + H.c.$ where $N=(N_{1L}, N^c_{1R}, N_{2L}, N_{2R}^c)^T$, and
\beq
 M_N= \left(\begin{array}{cccc} \frac{v_2\xi_{1L}}{\sqrt2} &\mathfrak{D}_{11}&0&\mathfrak{D}_{12}\\
\mathfrak{D}_{11}&\frac{v_2\xi_{1R}}{\sqrt2} &\mathfrak{D}_{21}&0\\
0&\mathfrak{D}_{21}&\frac{v_2\xi_{2L}}{\sqrt2} &\mathfrak{D}_{22}\\
\mathfrak{D}_{12}&0&\mathfrak{D}_{22}&\frac{v_2\xi_{2R}}{\sqrt2}  \end{array}
\right)\,.
\eeq
The symmetric mass matrix $M_N$ can be diagonalized by a transformation, $O^T M_N O$ with eigenvalues $M_{\chi_i} (i=1..4)$, and $O O^\dag=O^\dag O=1$. We write the mass eigenstates as
$\chi_L=(\chi_{1L},\chi_{2L},\chi_{3L},\chi_{4L} )$ with $N= O \cdot\chi_L$.
From $\chi_L$, one can construct four Majorana states
\beq
\chi_i =\chi_{iL} +\chi^c_{iL}=\chi_i^c
\eeq
such that ${\cal L} \supset \frac{i}{2} \overline{\chi}_i \SL D \chi_i - \frac{1}{2} M_{\chi_i} \overline{\chi_i} \chi_i$.
Reversely, the interaction states can be expressed in terms of the physical Majorana states
\beq
N_{1L}= O_{1i}\PL \chi_i\,,\;N_{2L}= O_{3i}\PL \chi_i\,,\;N_{1R}= O_{2i}^*\PR \chi_i\,,\;N_{2R}= O_{4i}^*\PR \chi_i\,,
\eeq
where $\PL/\PR= (1\mp \gamma^5)/2$ is the chiral projection operator.
And the Yukawa coupling becomes
\beq
{\cal L} \supset \overline{\chi}_k \left( Y_L^{ik}\, D^+\,  \PL
+ Y_R^{ik}\,  C^+\, \PR \right)l_i +\overline{\chi}_k \left( Y_L^{ik}\, (D^0)^*\, \PL
\right)\nu_i +H.c.\,,
\eeq
where
\beq
 Y_L^{ik} = \sum_{j=1}^2\left(y^L_{ij}\right)^* O_{(2j)k}\,,\; Y_R^{ik} =\sum_{j=1}^2\left(y^R_{ij}\right)^* O^*_{(2j-1)k}\,.
\eeq
In the limit that $\mathfrak{D}_{12},\mathfrak{D}_{21}\ll \mathfrak{D}_{11},\mathfrak{D}_{22}, \xi v_2$, the mass eigenstates $\chi_{1,2}(\chi_{3,4})$ are mainly composed by $N_{1L,1R}(N_{2L,2R})$.
For simplicity, we shall set $\mathfrak{D}_{12,21}=0$ in our numerical study and the physics is more transparent.  For a more general case, the full four by four mixing matrix $O$ can be obtained numerically.

Below the SM electroweak symmetry breaking scale, the neutral component of SM Higgs acquires a VEV, $\left\langle H_0 \right\rangle =v_0/\sqrt{2}$. Working in the unitary gauge,  $H=[ 0, (v_0+h)/\sqrt2 ]^T$,  the mixing between $D^-$ and $C^-$ can be described as a term in Lagrangian
\beq
{\cal L} \supset  - (D^+, C^+){\cal M}_C
\left( \begin{array}{c}   D^- \\ C^- \end{array}  \right)\,,\; {\cal M}_C= \left( \begin{array}{cc} \wt{M}_D^2  &  v_0 \mu_{DC} \\
 v_0 \mu_{DC} &  \wt{M}_C^2       \end{array} \right)
\eeq
where
$\wt{M}_D^2= M_D^2 +\half (\lambda_{HD}+\bar{\lambda}_{HD} )v_0^2  + \half\lambda_{D2}v_2^2$, and
$\wt{M}_C^2= M_C^2 +\half \lambda_{HC}v_0^2 + \half \lambda_{C2}v_2^2$.
The charged scalars can be  diagonalized by a two-by-two rotation
\beq
U_C=\left( \begin{array}{cc}
             \cos \alpha_c &   \sin \alpha_c \\  -\sin \alpha_c & \cos \alpha_c
           \end{array}\right)
\eeq
 with an angle satisfying
\beq
\sin 2\alpha_c=  {  2 v_0 \mu_{DC} \over M_{C2}^2-M_{C1}^2 }\,,
\eeq
where $M_{C1,C2}$ are the physical mass eigenvalues, and $M_{C2}>M_{C1}$.

The $\lambda_{H2}$ term in Eq.(\ref{eq:L_scalar}) induces a mixing between $h$ and the real part of $S_2$.
We assume the observed $125\mgev$ Higgs is the lighter physical scalar $h_{SM}$, and the mass of heavier $h_2$ is undetermined and not important in this study.   The  $h\mhyphen \Re( S_2)$ mixing results in a universal suppressing factor to the SM $125\mgev$ Higgs couplings.
Then the Higgs signal strength becomes $\mu=\cos^2 \theta_0$ with $\theta_0$ the mixing angle between $h$ and $\Re(S_2)$.
 From $\mu=1.02\substack{+0.07 \\ -0.06}$ obtained by CMS\cite{CMS:2020gsy} and
$\mu=1.06\pm 0.06$ by ATLAS\cite{ATLAS:2021vrm}, we obtain $\mu=1.042 \substack{+0.047\\-0.044}$ following the suggestion of \cite{Barlow:2004wg}.
Therefore, one has $\sin^2\theta_0 < 0.046$ at 2$\sigma$ C.L.
This amounts to
$ \left|{\lambda_{H2} v_2 v_0 \over M^2_{h_2}-M^2_{h_{SM}}}\right|< 0.21$ which can be easily satisfied  with the model parameters, for example $M_{h_2}, v_2 \simeq {\cal O}(\mtev)$ and $|\lambda_{H2}|<1$, without much fine-tuning.

Next, the mixings among the neutral components of $D$ and $S_{1}$ can be described by
\beq
{\cal L} \supset - (\Re D^0,\Re S_1)\,{\cal M}_R
\left( \begin{array}{c}   \Re D^0 \\ \Re S_1 \end{array}  \right)
  - (\Im D^0,\Im S_1)\,{\cal M}_I \left( \begin{array}{c} \Im D^0 \\ \Im S_1 \end{array} \right)\,,
\eeq
where
\beq
{\cal M}_R= \left( \begin{array}{cc} \overline{M}_D^2 &  -\mu_{DS} v_0\! -\!\lambda_4 v_0 v_2\\
-\mu_{DS} v_0\!-\!\lambda_4 v_0 v_2 &  \overline{M}_S^2 + \mu_{12}v_2          \end{array} \right)\,,\,
{\cal M}_I = \left( \begin{array}{cc} \overline{M}_D^2 & - \mu_{DS} v_0\!+\!\lambda_4 v_0 v_2\\
-\mu_{DS} v_0\!+\!\lambda_4 v_0 v_2 &  \overline{M}_S^2 -\mu_{12}v_2            \end{array} \right)\,,
\eeq
 $\overline{M}_D^2= \wt{M}_D^2-\half \bar{\lambda}_{HD} v_0^2 $, and
$\overline{M}_S^2=M_{S_1}^2+\half \lambda_{H1} v_0^2 + \half \lambda_{12} v_2^2$. Note the sign differences in some mass matrix elements for scalar and pseudoscalar, also the mass splitting between $\overline{M}_D$ and $\wt{M}_D$ is about the electroweak scale.
 Similarly, the real parts and imaginary parts mixing can be diagonalized by the two-by-two rotations $U_R$ and $U_I$ with angles $\alpha_{R,I}$, and
 \beq
\sin 2\alpha_R=  -{  2 (v_0 \mu_{DS} +\lambda_4 v_0 v_2 ) \over M_{R2}^2-M_{R1}^2 }\,,\;\;
\sin 2\alpha_I=  -{  2 (v_0 \mu_{DS} - \lambda_4 v_0 v_2) \over M_{I2}^2-M_{I1}^2 }\,,
\eeq
where $M_{R2,R1}$ and $M_{I2,I1}$ are the physical masses of scalars and pseudo scalars made from $D$ and $S_1$, respectively.
Again, we adopt the convention that $M_{R2}>M_{R1}$ and  $M_{I2}>M_{I1}$.

\subsection{$\tri a_{e,\mu}$ and $l_i \ra l_j \gamma$}
\begin{figure}%[htb]
\centering
\includegraphics[width=0.3\textwidth]{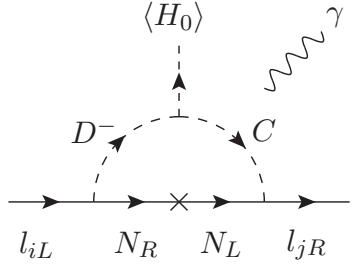}
\caption{ The Feynman diagram, in the interaction basis, for $(g-2)_l$ and the CLFV $l\ra l' \gamma$ process.
The photon attaches to the charged scalar in the loop.}
\label{fig:g-2_mass}
\end{figure}

In this model, the charged lepton $(g-2)$ receives  1-loop contribution, see Fig.\ref{fig:g-2_mass}.
Ignoring the charged lepton mass and summing over all physical states in the loop, the anomalous magnetic dipole moment can be calculated to be\cite{Chang:2021axw}
\beq
\tri a_l= -{m_l \sin(2\alpha_c) \over 16\pi^2}\sum_{k=1}^4 \frac{\Re\left[ Y_L^{lk} (Y_R^{lk})^* \right]}{M_{\chi_k}}
{\cal J}(\beta^k_H,\beta^k_L)\,,
\eeq
where $\beta^k_H=(M_{C_2}/M_{\chi_k})^2$, and $\beta^k_L=(M_{C_1}/M_{\chi_k})^2$.
And the loop function is given as
\beq
{\cal J}(x,y)={\cal J}_0(x)-{\cal J}_0(y)\,,\;
{\cal J}_0(x)={1-x^2+2 x\ln x \over 2(1-x)^3}\,.
\eeq
 The function ${\cal J}_0$ has limits ${\cal J}_0(0)=1/2$ and ${\cal J}_0(1)=1/6$.
 On the other hand, the electric dipole moment will be proportional to $\Im\left[ Y_L^{lk} (Y_R^{lk})^* \right]$ and stringently limited by the experimental bounds. For simplicity, we should assume there is no extra CP violation phases beyond the SM in the work.

Similarly, the $ L(p+q) \ra l(p) +\gamma(q)$ dipole transition amplitude can be read as
\beq
i{\cal M}^\mu_{Ll}=  -i\, e\, \overline{u_l(p) }\left[\left( d_L^{Ll} \PL + d_R^{Ll} \PR \right)
\left( \frac{ i  \sigma^{\mu\alpha}q_\alpha}{2} \right)\right]  u_L(p+q)\,,
\eeq
where
\beqa
 d_L^{Ll}= { \sin(2\alpha_c) \over 16\pi^2 }\left\{\sum_{k=1}^4 \frac{{\cal J}(\beta^k_H,\beta^k_L)}{M_{\chi_k}}\left[  (Y_R^{lk})^* Y_L^{Lk}\right] \right\}\,,\nonr\\
  d_R^{Ll}= { \sin(2\alpha_c) \over 16\pi^2 }\left\{\sum_{k=1}^4 \frac{{\cal J}(\beta^k_H,\beta^k_L)}{M_{\chi_k}}\left[ ( Y_L^{lk} )^* Y_R^{Lk}\right] \right\}\,.
\eeqa
In terms of the two dipole coefficients, the corresponding CLFV transition rate can be calculated to be\cite{Chang:2005ag}
\beq
\Gamma(L \ra l +\gamma) =  {\alpha_{em}\, m_L^3 \over 16} \left(|d_L^{Ll}|^2+ |d_R^{Ll}|^2\right)\,,
\eeq
if ignoring the mass of the lighter charged lepton.
From the current limits: $Br(\mu\ra e \gamma)< 4.2\times 10^{-13}$\cite{TheMEG:2016wtm}, ${\cal B}(\tau\ra e \gamma)< 3.3 \times 10^{-8}$, ${\cal B}(\tau\ra \mu \gamma)< 4.4\times 10^{-8}$\cite{Aubert:2009ag}, $m_\tau=1.77686(12)\mgev$, $\tau_\tau=(290.3\pm 0.5)\times 10^{-15}s$\cite{Zyla:2020zbs}, one has
\beqa
|d_L^{\mu e}|^2 + |d_R^{\mu e}|^2 < 2.35\times 10^{-25} (\mgev)^{-2} \,, \nonr\\
|d_L^{\tau e}|^2 + |d_R^{\tau e}|^2 < 2.93\times 10^{-17} (\mgev)^{-2}\,, \nonr\\
|d_L^{\tau \mu }|^2 + |d_R^{\tau \mu }|^2 < 3.91\times 10^{-17} (\mgev)^{-2}\,.
\eeqa

It is clear that to satisfy the stringent experimental bounds, either some couplings are very small or delicate cancellation among the parameters should be arranged. For simplicity and better understanding the physics, we adopt a simple arrangement ${\cal D}_{12,21}=0$ such that the cross mixing between $N_1$ and $N_2$ vanishes.
In this scheme,  the  charged lepton $g-2$ become
\beqa
\tri a_e  &=& [\cdots] y^L_{11} y^R_{11} + [\cdots] y^L_{12} y^R_{12}\,,\nonr\\
\tri a_\mu  &=& [\cdots] y^L_{21} y^R_{21} + [\cdots] y^L_{22} y^R_{22}\,,\nonr\\
\tri a_\tau  &=& [\cdots] y^L_{31} y^R_{31} + [\cdots] y^L_{32} y^R_{32}\,,
\eeqa
where $[\cdots]$ represent the numerical factors depending on $M_\chi$, $M_{C,R,I}$, and the mixings.
 Moreover, once $y^L$'s and the relevant physical masses and mixings are fixed, the CLFV dipole transition coefficients have simple dependence on the right-handed Yukawa $y^R$:
\beqa
&& d_L^{\mu e} = (\cdots) y^R_{11} + (\cdots) y^R_{12}\,,\;
d_R^{\mu e} = (\cdots) y^R_{21} + (\cdots) y^R_{22}\,,\nonr\\
&& d_L^{\tau e} = (\cdots) y^R_{11} + (\cdots) y^R_{12}\,,\;
d_R^{\tau e} = (\cdots) y^R_{31} + (\cdots) y^R_{32}\,,\nonr\\
&& d_L^{\tau\mu } = (\cdots) y^R_{21} + (\cdots) y^R_{22}\,,\;
d_R^{\tau\mu } = (\cdots) y^R_{31} + (\cdots) y^R_{32}\,,
\eeqa
where $(\cdots)$ represent the numerical factors depending on $y^L$, $M_\chi$, $M_{C,R,I}$, and the mixings.
One can set $y^R_{31}=y^R_{32}=0$ to suppress the CLFV dipole transition, with  vanishing $\tri a_\tau$, $d_R^{\tau\mu }$, and $d_R^{\tau e}$.

\subsection{Radiative contributions to the lepton masses}
The same Feynman diagram displayed in Fig.\ref{fig:g-2_mass} will generate radiative corrections to the charged lepton mass matrix if the external photon line is removed. A simple calculation gives
\beq
\delta M_{ij}^{loop} = - \sum_k {M_k (Y^{ik}_L)^* Y^{jk}_R \over 32 \pi^2} \sin 2\alpha_c \left( {\beta^k_L \log \beta^k_L \over \beta^k_L-1}- {\beta^k_H \log \beta^k_H \over \beta^k_H-1} \right)
\eeq
where the first (second) index stands for the left (right)-handed lepton.

To be consistent with the working assumption that the Yukawa couplings, $y^{L,R}$ in Eq.(\ref{eq:L_Yukawa}), are in the charged leptons' mass basis, it is required that the tree-level Yukawa coupling between the charged leptons and the SM Higgs doublet, ${\cal L}\supset - y^H_{ij} \bar{L}_{i} e_{Rj} H +H.C.$,  must satisfy the following relationship
\beq
\frac{v_0}{\sqrt2}y^H_{ij} + \delta M_{ij}^{loop} = \delta_{ij} m_i\,,\;\;\; i=(e,\mu,\tau)\,.
\label{eq:treeYukawa}
\eeq
Such that the off-diagonal entries of charged lepton mass matrix vanish, and our treatment is self-consistent at the one-loop level.

\begin{figure}%[htb]
\centering
\includegraphics[width=0.64\textwidth]{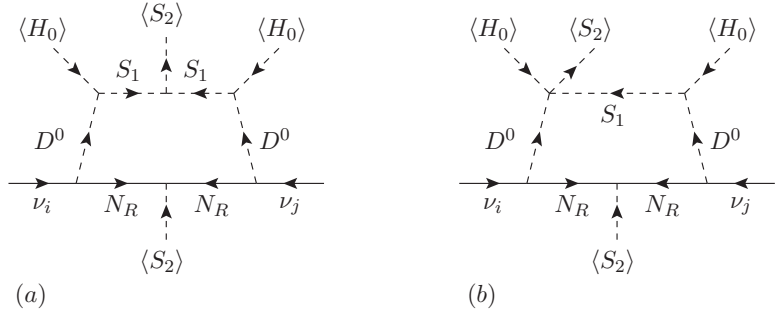}
\caption{ Some leading Feynman diagrams, in the interaction basis, for generating the neutrino masses.
}
\label{fig:nu_mass}
\end{figure}
For neutrino masses, they can be radiatively generated by the 1-loop diagrams shown in Fig.\ref{fig:nu_mass},
in the interaction basis.
Summing over all mass eigenstates in the loop, we obtain the neutrino mass matrix element
\beq
{\cal M}^{\nu}_{ij}= \frac{1}{16\pi^2}\sum_{k=1}^4 M_{\chi_k} \,Y_L^{ik} Y_L^{jk}
% \left[ \left(y^L_{i1}\right)^* O_{2k}+ \left(y^L_{i2}\right)^* O_{4k}\right]
%\, \left[ \left(y^L_{j1}\right)^* O_{2k}+ \left(y^L_{j2}\right)^* O_{4k}\right]\nonr\\
\times \sum_{a=1}^2\left[ \left(U^R_{1a}\right)^2{s_{ak} \ln s_{ak}\over s_{ak}-1}
-\left(U^I_{1a}\right)^2{p_{ak} \ln p_{ak}\over p_{ak}-1} \right]\,,
\label{eq:nu_mass}
\eeq
where $s_{ak} =(M_{Ra}/M_{\chi_k})^2$ and $p_{ak} =(M_{Ia}/M_{\chi_k})^2$ are the mass squared ratios of scalar and pseudoscalar to the Majorana fermion-$k$, respectively.

\subsection{Effective Yukawa coupling of charged lepton }

Note that the one-loop $h_{SM} \mhyphen \bar{l}_i \mhyphen l_j$ couplings can be generated if replacing the external photon in Fig.\ref{fig:g-2_mass} by the $125\mgev$ SM Higgs, and therefore the tree-level SM prediction $y_l= \frac{m_l}{v_0}$ will be modified.
Although the new Yukawa couplings are in the charged lepton mass basis at 1-loop level, the different loop integration involved leads to CLFV Higgs decays $h_{SM}\ra l_i l_j$  in general.

Below the electroweak scale, the dimensionful cubic couplings for ${\cal L}\supset -\wt{\mu}_{ab}\, h_{SM}\, C^+_a C^-_b$ can be read from the scalar potential, Eq.(\ref{eq:L_scalar}), as
\beqa
\wt{\mu}_{11} &=& v_0 (\lambda_{HD} +\bar{\lambda}_{HD})c_\alpha^2 +v_0 \lambda_{HC} s_\alpha^2 - 2 s_\alpha c_\alpha \mu_{DC}\nonr\\
\wt{\mu}_{22} &=& v_0 (\lambda_{HD} +\bar{\lambda}_{HD})s_\alpha^2 +v_0 \lambda_{HC} c_\alpha^2 + 2 s_\alpha c_\alpha \mu_{DC}\nonr\\
\wt{\mu}_{12}&=& \wt{\mu}_{21} = v_0 (\lambda_{HD} +\bar{\lambda}_{HD})s_\alpha c_\alpha -v_0 \lambda_{HC} s_\alpha c_\alpha + ( c_\alpha^2- s_\alpha^2) \mu_{DC}\,,
\label{eq:CCH_mu}
\eeqa
in the mass basis of the charged scalars. If ignoring the charged lepton masses,  the dimensionless $ h_{SM}\mhyphen \bar{l}_i \mhyphen l_j$ Yukawa couplings\footnote{We adopt the convention that ${\cal L}\supset - y^{loop}_{ij}\,\bar{l_i} \PR l_j h_{SM} +h.c.$} can be calculated to be
\beq
y_{ij}^{loop}=
\frac{1}{16\pi^2}\sum_{k=1}^4 (Y_L^{ik})^* Y_R^{jk} \sum_{a,b=1}^2 \frac{\wt{\mu}_{a b}}{M_{\chi_k}}\,  U^C_{2a}\,U^C_{1b}\, {\cal I}_3\left(\beta^a_k\,,\, \beta^b_k\,,\, \frac{q^2}{M_{\chi_k}^2}\right)
% \left[ (y^b_{ik})^* f^a_{jk} \PR + (f^b_{ik})^* y^a_{jk} \PL \right]
\eeq
%where $f^a_{ik}= U^C_{2a} Y_R^{ik}\,,\;y^a_{ik}= U^C_{1a} Y_L^{ik}$, and
where $q$ is the momentum carried by $h_{SM}$, and
\beq
{\cal I}_3(z_1, z_2, z_3) =\int^1_0 dx \int^{1-x}_0 dy ( 1-x-y +x  z_1  +y z_2 -x y z_3)^{-1}\,.
\eeq
The analytic expression for ${\cal I}_3(a_1, a_2, 0)$ can be easily obtained and will not be shown here.

Combining with the tree-level Higgs Yukawa, Eq.(\ref{eq:treeYukawa}), we obtain the effective Yukawa
\beq
y^{eff}_{ij}= \frac{y^H_{ij}}{\sqrt2}+ y_{ij}^{loop}= \frac{1}{v_0}(  \delta_{ij} m_i  - \delta M^{loop}_{ij} )+ y_{ij}^{loop}\,.
\eeq
For $i\neq j$, the CLFV decay width of $h_{SM}\ra l_i l_j$ is given by
\beq
\Gamma(h_{SM}\ra l_i^+ l_j^-) = \Gamma(h_{SM}\ra l_i^- l_j^+) \simeq  \frac{M_h}{16\pi}\left(|y^{eff}_{ij}|^2 +|y^{eff}_{ji}|^2  \right)
\eeq
at tree-level if ignoring the charged lepton masses.
And for the diagonal ones, it is convenient to define the normalized Yukawa
\beq
\zeta_l\equiv \frac{y^{eff}_l}{y^{SM}_l} = 1 - \frac{1}{m_l}\left(\delta M^{loop}_{ll}-v_0\, y_{ll}^{loop}\right)\,.
\eeq

%%%%%%%%%%%%%%%%%%%%%%%%%%%%%%%%%%%%%%%%%%%%%%%%%%
\section{Numerical study}
%%%%%%%%%%%%%%%%%%%%%%%%%%%%%%%%%%%%%%%%%%%%%%%%%%
\label{SEC:numerical}

First, we comment on the number of exotic vector fermions $N$'s.
For the three active neutrinos, one needs six independent parameters to describe the symmetric neutrino mass matrix.
Thus, the minimal setup of our model calls for two pairs of exotic fermions, with six $y^L$'s paramters.
However, it is easy to see from Eq.(\ref{eq:nu_mass}) that the resulting neutrino mass matrix has rank 2.
Namely,  the neutrino mass matrix possesses only five independent parameters.
One of the $y^L$'s is redundant and must be fixed first\footnote{ For example, one can generally require that the value of one of the $y^L$'s must lay within a reasonable range, say, from $-1.0$ to $1.0$. } so the rest five  can be uniquely determined by data. And the remaining six $y^R$'s are used to yield $\tri a_{e,\mu}$ while satisfying all the experimental CLFV constraints.

With three pairs of exotic fermions,  the rank of the neutrino mass matrix is 3.
Hence, three out of the nine $y^L$ free parameters are redundant.
Moreover, with nine $y^R$'s parameters, one can always find a viable solution such that all the CLFV limits are satisfied and both central values of $\tri a_\mu$ and $\tri a_e^{Cs[Rb]}$ are accommodated. Such scenario has more free parameters than experimental constraints and lacks predictability.
Therefore, we focus on the minimal model with two pairs of $N$'s.

As we only consider the CP-conserving scenario  in this work, there is no Dirac nor Majorana CP violation phases in the $U_{PMNS}$ matrix. The case of $\delta_{CP}=0$ is still allowed in the current $3\sigma$ range of global fit of neutrino oscillation data\cite{Esteban:2020cvm}, see Tab.\ref{tab:NuFit3S}.
From the global fit, the pattern of a rank-2 neutrino mass matrix is roughly
\beq
   M_\nu^{NO} \sim \left(
                  \begin{array}{ccc}
                    0.1&0.1&0.1 \\
                    0.1&1&1\\
                     0.1&1&1\\
                  \end{array}
                \right)
    \times (0.03) \mev
\eeq
for Normal Ordering(NO), and
\beq
  M_\nu^{IO}  \sim
    \left(  \begin{array}{ccc}
                    1&-0.1&-0.1 \\
                    -0.1&1&-1\\
                    -0.1&-1&1\\
                  \end{array}
                \right)
    \times (0.03) \mev
\eeq
for Inverted Ordering(IO).

\begin{table}[tbh]
\centering
\begin{tabular}{|c|c|c|c|c|c|}
\hline
 & $\theta_{12}[^\circ]$ & $\theta_{23}[^\circ]$ & $\theta_{13}[^\circ]$ & $\Delta m^2_{2 1}  [10^{-5}  \mev^2]$ & $\Delta m^2_{3 l} [ 10^{-3} \mev^2]$ \\
\hline
Normal Ordering & $31.27 \ra 35.87$ & $39.7 \ra 50.9$ & $8.25 \ra 8.98$ & $6.82 \ra 8.04$ & $+2.430 \ra +2.593$ \\
\hline
Inverted Ordering & $31.27 \ra 35.87$ & $39.8 \ra 51.6$ & $8.24 \ra 9.02$ & $6.82 \ra 8.04$ & $-2.574 \ra -2.410$ \\
\hline
\end{tabular}
\caption{\label{tab:NuFit3S} The $3 \sigma$ ranges of the neutrino oscillation parameters from global fit\cite{Esteban:2020cvm}.
For NO(IO), $\Delta m^2_{3 l} \equiv m_3^2-m_{1(2)}^2$.}
\end{table}

Note that $\tri a_l$, $d^{Ll}_{L,R}$, $\delta M^{loop}$ and $y^{loop}$ are linearly proportional to  $y^R$ in our model.
Moreover,  the physical masses of $\chi$'s and the exotic scalars can be scaled up or down by a common factor $r$ while the mixing angles retain if one performs the following scaling
\beq
M \ra r M\,,\; \mu_{DS,DC}\ra r^2 \mu_{DS,DC}\,,\;\Lambda\ra r^2 \Lambda\,,
\eeq
where $M \supset \{ \overline{M}_{D,S},\wt{M}_{D,C}, {\cal D}, v_2, \mu_{12}\}$, and $\Lambda \supset \{ \lambda_{HD},\bar{\lambda}_{HD},\lambda_{HC},\lambda_{H1}\}$.
Moreover, the resulting ${\cal M}^\nu$, $d^{Ll}_{L,R}$, and $\tri a_l$ will be the same if
\beq
y^L \ra r^{-1/2} y^L\,,\; y^R \ra r^{3/2} y^R\,,
\eeq
such that  both radiative mass corrections and the one-loop effective Yukawa coupling scale like:
\beq
\delta M^{loop}\ra r^2 \delta M^{loop}\,,\;y^{loop}\ra r^2 y^{loop}\,,
\eeq
while $M_\nu$ and $\tri a_{e,\mu}$ remain unchanged.
Therefore, starting from an available solution, one can obtain infinite other possible viable solutions by using this scaling as long as the perturbility and positivity requirements are met.

To avoid  multiple counting of the same solution related by the above mentioned scaling, we adopt the following procedures to find the  numerical solutions:
we first scan the relevant model parameter space spanned by ${\cal M}_{C,R,I}$ and $M_N$. As discussed earlier, we set $\lambda_4=0$ and $\bar{\lambda}_{HD}>0$ to simplify the positivity condition.  We also set ${\cal D}_{12,21}=0$ to speed up the numerical scan\footnote{ We have checked that our main conclusions do not change in the case that ${\cal D}_{12,21} \neq 0$.}.
Explicitly, it is a 13-dimensional parameter space spanned by ${\cal D}_{11,22}$, $(v_2 \xi/\sqrt2)$, $M_{C1,C2}$,$\alpha_C$, $\overline{\lambda}_{HD}$, $\overline{M}_{S}$, $\mu_{DS}$, and $(\mu_{12}v_2)$. We scan the parameters in the ranges: $(v_2 \xi/\sqrt2) \in [10,1000]\mgev$, ${\cal D}_{11,22} \in[0.1, 10]\mtev$,
 $\overline{M}_{S} \in [0.3, 10] \mtev$, $\sqrt{ \mu_{DS} v_0}\in[3.0, 2000]\mgev$, $\sqrt{ \mu_{12} v_2}\in[3.0, 3000]\mgev$, $\overline{\lambda}_{HD} \in[3.3\times 10^{-5}, 0.93]$, $M_{C1}\in[0.5- 10]\mtev$, $M_{C2}/M_{C1}\in[1,10]$, and $\alpha_C\in[0.01,\pi/2]$. Our random samplings are evenly distributed in the logarithmic scale.

After mass diagonalization of $\chi$ and the relevant scalar sector, we obtain the mixing angles and physical masses needed for calculating neutrino masses, Eq.(\ref{eq:nu_mass}).
Then the neutrino mass matrix is fixed by a set of neutrino oscillation parameters randomly picked within the 3$\sigma$ range\cite{Esteban:2020cvm} shown in Tab.\ref{tab:NuFit3S}. Because one of the active neutrino is massless, there are only five independent parameters in the symmetric neutrino mass matrix.
To proceed, we assign $y^L_{11}$ to a value  randomly picked between  $\pm [10^{-5},1.0]$. Then we look for the unique solution  of the other five $y^L$'s in Eq.(\ref{eq:L_Yukawa}) to the neutrino mass matrix.

Only the points which satisfy positivity conditions are kept.
We also require that the resulting physical masses are in the range of $0.5-5\mtev$, and the magnitudes of all dimensionless parameters to be less than $1.0$.

Finally, the six Yukawa couplings $y^R$'s in Eq.(\ref{eq:L_Yukawa}) are determined by finding the minimum of the weighted chi-squared
\beq
\chi^2_{\tri a}= \left( { \tri a_e -\tri a^{Cs[Rb]}_e \over \delta \tri a^{Cs[Rb]}_e}\right)^2 +
\left( { \tri a_\mu -\tri a^{ F\!N\!A\!L}_\mu \over \delta \tri a^{ F\!N\!A\!L}_\mu}\right)^2\,,
\eeq
while complying the experimental bounds on $ {\cal B}(l_i \ra l_j \gamma), (i\neq j)$.
After that, two more model parameters, $\lambda_{HD,HC}$, are required for calculating the effective Higgs Yukawa couplings.
They only associate with the effective Higgs Yukawa couplings and independent of the other observables.
We randomly choose $\lambda_{HD,HC}$ from the conservative range $[-1.0 \ra 1.0 \,]$ which is within the lower(upper) bound imposed by the positivity(perturbility) condition.

\subsection{Benchmark points}
Here we present four viable benchmark points with detailed model parameters.

\subsubsection{Benchmark points--Normal Ordering}
The relevant model parameters are:
\beqa
\left\{ \theta_{12}\,,\,\theta_{23}\,,\,\theta_{13}\,,\,\Delta m^2_{21}\,,\, \Delta m^2_{31} \right\} &=& \left\{ 32.13^\circ\,,\, 48.49^\circ \,,\,8.71^\circ\,,\,
7.362\times10^{-5}\mev^2\,,\, 2.539\times 10^{-3}\mev^2  \right\}\,,\nonr\\
\frac{v_2}{\sqrt2}\times\left \{\xi_{1L}\,,\,\xi_{1R}\,,\,\xi_{2L}\,,\,\xi_{2R} \right\} &=&\left\{ 44.354\,,\, 57.795\,,\,12.301\,,\,27.742\right\}\mgev\,,\nonr\\
\left\{{\cal D}_{11}\,,\,{\cal D}_{22}\,,\,{\cal D}_{12}\,,\,{\cal D}_{21} \right\} &=& \left\{ 2.18072\,,\, 2.41526\,,\,0\,,\,0\right\}\mtev\,,\nonr\\
\left\{ \wt{M}_D\,,\, \wt{M}_C\,,\, \overline{M}_D\,,\, \overline{M}_S \right\}&=&\left\{1.08449\,,\, 1.10508\,,\,1.08446\,,\,1.10210 \right\}\mtev\,,\nonr\\
\left\{ \mu_{DC}\,,\, \mu_{DS}\,,\, \mu_{12} \right\} &=& \left\{2961.99\,,\, 33.722\,,\, 0.67863\times\left( {\mtev \over v_2} \right) \right\}\mgev\,,\nonr\\
\left\{ \overline{\lambda}_{HD}\,,\, \lambda_{HD}\,,\, \lambda_{HC}\right\} &=& \left\{ 2.284\times 10^{-3}\,,\, 0.1\,,\, 0.2 \right\}\,.
\eeqa
From the  parameters listed above, the exotic fermions, $N_{1,2}$, can be diagonalized by the rotation matrix
\beq
O=\begin{pmatrix}
0&0&0.76016&-0.708196 i\\
0&0&0.708196& 0.706016 i\\
0.705976& -0.708236 i&0&0\\
0.708236&0.705976i&0&0\\
 \end{pmatrix}\nonr
\eeq
with mass eigenvalues $ M_{\chi}=\{ 2.43529, 2.39525,2.23181,2.12966 \}\mtev$.
The physical masses and mixings of the exotic scalar relevant to $m_\nu$ and $\tri a_{e,\mu}$ are:
\beqa
M_{R1}=1.08369\mtev\,,\;  M_{R2}=1.10317\mtev\,,\;  \alpha_R=-0.19998\,, \nonr\\
M_{I1}=1.08366\mtev\,,\;  M_{I2}=1.10258\mtev\,,\;  \alpha_I=-0.20637\,, \nonr\\
M_{C1}=0.68533\mtev\,,\;  M_{C2}=1.38841\mtev\,,\;  \alpha_C= 0.76993\,. \nonr
\eeqa
From the above given parameters, the LH Yukawa couplings, $y^L$, can be solved.
On the other hand, $y^R_{Cs[Rb]}$ is determined by the best-fit solution to $(\tri a_e^{Cs[Rb]}, \tri a_\mu)$. The Yukawa couplings are found to be:
\beq
y^L= \begin{pmatrix}
           -0.045143 & 0.252272\\
           -0.300581 & 0.370544\\
           -0.309633 &-0.143563
         \end{pmatrix},\,  y^R_{Cs[Rb]}= \begin{pmatrix}
           -3.124[+1.724] & -2.918[+1.610]\\
           -5.4584 & -1.1237\\
           0 & 0
         \end{pmatrix}\times 10^{-2}\,,
\eeq

The predictions of these two benchmark points are listed in Tab.\ref{tab:BMP1}.
\begin{table}[htb]
\centering
\begin{tabular}{|l|l|l| }
\hline
$\tri a_\mu=+4.617 \times 10^{-10}$\,,\;$\tri a_\tau=0$ & $\frac{y_e^{eff}}{y_e^{SM}}=-1.773[+2.530]$ &$\frac{y_\mu^{eff}}{y_\mu^{SM}}=+1.031$\\
\hline
${\cal B}(\mu\ra e \gamma)=4.2 \times 10^{-13}$ & ${\cal B}(\tau\ra e \gamma)=2.32[0.717]\times 10^{-8}$ & ${\cal B}(\tau\ra \mu \gamma)=4.4\times 10^{-8}$ \\
\hline
${\cal B}(h \ra e \mu)=3.55[1.24] \times 10^{-10}$ & ${\cal B}(h \ra e \tau)=1.25[0.379]\times 10^{-7}$ & ${\cal B}(h \ra \mu \tau)=2.286\times 10^{-7}$ \\
\hline
\end{tabular}
\caption{\label{tab:BMP1} Benchmark point predictions for NO and $\tri a_e= -8.7[4.8]\times 10^{-13}$. }
\end{table}

\subsubsection{Benchmark points--Inverted Ordering}
For IO neutrino masses, the variables take the following values:
\beqa
\left\{ \theta_{12}\,,\,\theta_{23}\,,\,\theta_{13}\,,\,\Delta m^2_{21}\,,\, \Delta m^2_{32} \right\} &=& \left\{ 33.13^\circ\,,\, 43.67^\circ \,,\,8.31^\circ\,,\,
7.318\times10^{-5}\mev^2\,,\, -2.469\times 10^{-3}\mev^2  \right\}\,,\nonr\\
\frac{v_2}{\sqrt2}\times\left \{\xi_{1L}\,,\,\xi_{1R}\,,\,\xi_{2L}\,,\,\xi_{2R} \right\} &=&\left\{ 34.175\,,\, 89.898\,,\,168.230\,,\,253.529\right\}\mgev\,,\nonr\\
\left\{{\cal D}_{11}\,,\,{\cal D}_{22}\,,\,{\cal D}_{12}\,,\,{\cal D}_{21} \right\} &=& \left\{ 2.4416\,,\, 2.3928\,,\,0\,,\,0\right\}\mtev\,,\nonr\\
\left\{ \wt{M}_D\,,\, \wt{M}_C\,,\, \overline{M}_D\,,\, \overline{M}_S \right\}&=&\left\{1.15705\,,\, 1.01073\,,\,1.15702\,,\,1.40638 \right\}\mtev\,,\nonr\\
\left\{ \mu_{DC}\,,\, \mu_{DS}\,,\, \mu_{12} \right\} &=& \left\{3469.71\,,\, 67.306\,,\, 3.3332\times\left( {\mtev \over v_2} \right) \right\}\mgev\,,\nonr\\
\left\{ \overline{\lambda}_{HD}\,,\, \lambda_{HD}\,,\, \lambda_{HC}\right\} &=& \left\{ 2.1959\times 10^{-3}\,,\, 0.1\,,\, 0.2 \right\}\,.
\eeqa
From the  parameters listed above, the $N_{1,2}$ can be diagonalized by the rotation matrix\footnote{ The rotation matrix is not block-diagonal because we reorder the eigenstates such that $M_{\chi_1}>M_{\chi_2}>M_{\chi_3}>M_{\chi_4}$.}
\beq
O=\begin{pmatrix}
0& 0.70306& 0.71113 i& 0\\
0& 0.71113& -0.70306 i& 0\\
0.70078&0 &0&0.71338i\\
0.71338 &0&0&-0.70078i\\
 \end{pmatrix}\nonr
\eeq
with mass eigenvalues  $ M_{\chi}=\{ 2.60403, 2.50388, 2.37981, 2.18227 \}\mtev$.
The physical masses and mixings of the exotic scalar are:
\beqa
M_{R1}=1.15684\mtev\,,\;  M_{R2}=1.40772\mtev\,,\;  \alpha_R=-0.02575\,, \nonr\\
M_{I1}=1.15683\mtev\,,\;  M_{I2}=1.40535\mtev\,,\;  \alpha_I=-0.02602\,, \nonr\\
M_{C1}=0.55858\mtev\,,\;  M_{C2}=1.43199\mtev\,,\;  \alpha_C= 0.87725\,. \nonr
\eeqa
The Yukawa couplings can be found to be:
\beq
y^L \!=\! % \left(
      %   \begin{array}{cc}
        \begin{pmatrix}
           6.5398\times 10^{-4} &  8.2975\times 10^{-2}\\
           0.19400 & -8.099\times 10^{-3}\\
           -0.18533 &-9.022\times 10^{-3}
        \end{pmatrix},\,
        % \end{array}  \right)
        y^R_{Cs[Rb]}\!=\! % \left(         \begin{array}{cc}
         \begin{pmatrix}
           -0.2344[0.1294] & -5.399[+2.979]\\
           9.241 & -0.07156\\
           0 & 0
        \end{pmatrix}\times\! 10^{-2}
%         \end{array}       \right)
\eeq
for $\tri a_e^{Cs[Rb]}$.
And the resulting predictions are listed in Tab.\ref{tab:BMP2}.
\begin{table}[htb]
\centering
\begin{tabular}{|l|l|l| }
\hline
$\tri a_\mu=+6.92 \times 10^{-10}$\,,\;$\tri a_\tau=0$ & $\frac{y_e^{eff}}{y_e^{SM}}=-3.366[+3.409]$ &$\frac{y_\mu^{eff}}{y_\mu^{SM}}=+1.083$\\
\hline
${\cal B}(\mu\ra e \gamma)=4.2 [4.2] \times 10^{-13} $ & ${\cal B}(\tau\ra e \gamma)=1.33[0.405]\times 10^{-10}$ & ${\cal B}(\tau\ra \mu \gamma)=4.4\times 10^{-8}$ \\
\hline
${\cal B}(h \ra e \mu)=2.33[0.710] \times 10^{-13}$ & ${\cal B}(h \ra e \tau)=2.07[0.632]\times 10^{-9}$ & ${\cal B}(h \ra \mu \tau)=7.032\times 10^{-7}$ \\
\hline
\end{tabular}
\caption{\label{tab:BMP2} Benchmark point predictions for IO and $\tri a_e= -8.7[4.8]\times 10^{-13}$. }
\end{table}

\subsection{Numerical results and discussion}

From the benchmark points, we observe the follows:
\bi
\item $\mu_{DC}\sim {\cal O}(\mtev)$, $v_2 \xi \lesssim v_0$, and  $\mu_{DS}, \mu_{12} \ll v_0$.
This is expected because from  Fig.\ref{fig:g-2_mass}, we have a ball-park estimation
\beq
\tri a \sim \frac{y^L y^R}{16\pi^2} {m_l \mu_{DC} v_0 {\cal D} \over M^4 }\, [\cdots]\,,
\eeq
where $M$ represents the relevant mass scale and $[\cdots]$ represent the order one numerical factor of the loop function.
Note that the Dirac mass ${\cal D}$ is called for the necessary chiral flipping on the internal fermion.
So roughly we have
\beq
\tri a_l \sim 10^{-10} \left(\frac{m_l}{m_\mu}\right)
\left(\frac{\mu_{DC}}{\mtev}\right) \left(\frac{y^R y^L }{ 0.001}\right)
\left(\frac{ (\mtev)^3 }{M^4/{\cal D} }\right)\,,\;\; l=(e,\mu)\,,
\eeq
and $\mu_{DC}\sim {\cal O}(\mtev)$ is about right to reproduce the observed $\tri a_{e,\mu}$.

Similarly, from  Fig.\ref{fig:nu_mass}, the back-of-the-envelope estimation for neutrino mass is
\beq
m_\nu \sim \frac{(y^L)^2 }{16\pi^2} {(v_0 v_2)^2\xi \mu_{DS}^2 \mu_{12}\over M^6 }\, [\cdots]\,.
\eeq
Here, both $\mu_{12}$ and  $v_2 \xi$, the Majorana mass insertion on the heavy fermion, are required to break the global lepton number.
Plugging in the reasonable values, we have
\beq
m^\nu \sim  0.4 \mev \times \left(\frac{y^L }{ 0.1}\right)^2
\left(\frac{v_2 \xi}{100 \mgev}\right)
\left(\frac{\mu_{DS}^2 \mu_{12}}{ (10\mgev)^3}\right)
\left(\frac{ (\mtev)^5 }{M^6/v_2 }\right)\,,
\eeq
which agrees with the numerical result that $ \sqrt{ \mu_{DS} \mu_{12} } \ll v_0 $.
The consequences are: (1) the $\chi$'s are pseudo-Dirac fermions,(2) $|\alpha_C|\sim{\cal O}(1)$, and (3) $M_R$ and $M_I$ are nearly degenerate with $|\alpha_{R,I}|\ll 1$.

\item Both $\tri a_e^{Cs}$ and $\tri a_e^{Rb}$ can be easily fitted by only adjusting $y^R_{11}$ and $y^R_{12}$ while all other parameters kept unchanged. In fact, by tuning up $|y^R_{11,12}|$,  $|\tri a_e|$ can be as large as $\sim{\cal O}(10^{-11})$ without affecting $\tri a_{\mu}$ and $M^\nu$ nor upsetting the current experimental limits in the benchmark points.

\item Both ${\cal B}(\tau\ra \mu \gamma)$ and ${\cal B}(\mu \ra e\gamma)$  nearly saturate the current experimental bounds.
\item The resulting $\tri a_\mu\sim 6 \times 10^{-10}$  is roughly  $3 \sigma$ away from $\tri a^{F\!N\!A\!L}_\mu$.
This is because  $\tri a_\mu$ is tightly bounded by ${\cal B}(\tau\ra \mu \gamma)$ (and ${\cal B}(\mu \ra e\gamma)$).
\item Also note that $y^R_{31,32}=0$ which minimizes the CLFV ${\cal B}(l_i \ra  \gamma l_{j\neq i})$ and it predicts $\tri a_\tau=0$ and $y^{eff}_\tau =y^{SM}_\tau$ at one-loop level.

\item Since $y^R_{11,12}$ have little constrain from experiment, the effective electron-Higgs Yukawa coupling can be very different from the SM one. On the contrary, the muon-Higgs Yukawa is close to the SM one because the relevant parameters are stringently constrained by ${\cal B}(\tau\ra \mu \gamma)$ ( and ${\cal B}(\mu \ra e\gamma)$ ).

\item Comparing with the current limits\cite{Zyla:2020zbs},
\beq
{\cal B}(h_{SM}\ra \tau \mu)<2.5\times 10^{-3}\,,\,{\cal B}(h_{SM}\ra \tau e)<4.7\times 10^{-3}\,,\,{\cal B}(h_{SM}\ra e \mu)<6.1\times 10^{-5}\,,
\eeq
the branching ratios of CLFV Higgs decays $h_{SM}\ra l_i l_{j\neq i}$ are small, $\lesssim 10^{-7}$, and experimentally insignificant.
\ei

Besides the displayed four benchmark points, we have generated 3000 viable points for each neutrino mass hierarchy according to the stated scan strategy to better explore the model.
In the followings, we present more results and distribution plots extracted from the data sets found in our numerical study.

\begin{figure}%[htb]
\centering
\includegraphics[width=0.4\textwidth]{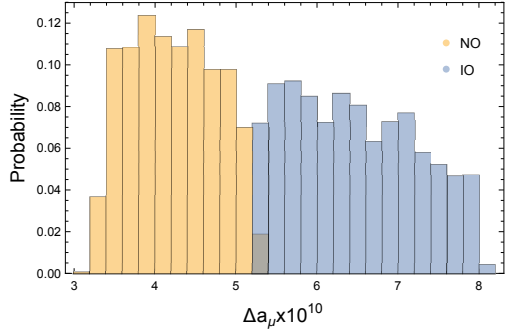}
\caption{ The probability histogram of $\Delta a_\mu$. The yellow(blue) bar is for normal(inverted) hierarchy.
}
\label{fig:deltaAmu}
\end{figure}

Fist of all, by using $y^R_{11,12}$ both central values of $\tri a_e^{Cs}$ and $\tri a_e^{Rb}$ can be well fitted in our model, as have been demonstrated in the benchmark points.
On the other hand, the histogram of predicted $\tri a_\mu$ of our model is shown in Fig.\ref{fig:deltaAmu}. The distribution of $\tri a_\mu$ peaks at around $\sim 4(6)\times 10^{-10}$ for NO(IO) neutrino mass. The largest available $\tri a_\mu$ is about  $\lesssim5(8)\times 10^{-10}$ for NO(IO) neutrino mass.  It is clear that our model cannot reproduce  $\tri a_\mu^{F\!N\!A\!L}$.
This can be understood as following: our radiative neutrino mass generation mechanism implies that $M^\nu_{ij} \sim (\cdots) y^L_i y^L_j$.
Similarly, $\tri a_l \sim (\cdots) y^L_l y^R_l$.
Here we use $(\cdots)$ to collectedly denote the  numerical factors and summing over all the contributions from the relevant physical masses and mixings.
So roughly speaking, it is expected that
\beqa
&&{\cal B}(\mu\ra e \gamma) \sim |(\cdots)y^R_2 y^L_1|^2 +|(\cdots)y^R_1 y^L_2|^2\,,\nonr\\
&&\sim  \frac{M^\nu_{11}}{M^\nu_{22}}\left[ (\cdots) (\tri a_\mu)^2 + (\cdots)\left( \frac{M^\nu_{22}}{M^\nu_{11}}\tri a_e\right)^2\right]
\sim \frac{M^\nu_{11}}{M^\nu_{22}} (\cdots) (\tri a_\mu)^2\,.
\eeqa
Since $|\tri a_\mu| \gg |\tri a_e|$, the second term is dropped in the last approximation.
Likewise,
\beqa
{\cal B}(\tau\ra \mu \gamma) &\sim& \left|(\cdots)y^R_2 y^L_3\right|^2 \sim  (\cdots) \frac{M^\nu_{33}}{M^\nu_{22}} (\tri a_\mu)^2\,,
\label{eq:BTM_HV}\\
{\cal B}(\tau\ra e \gamma) &\sim& \left|(\cdots)y^R_1 y^L_3\right|^2
\sim  (\cdots) \frac{M^\nu_{33}}{M^\nu_{11}} (\tri a_e)^2\,.
\label{eq:BTE_HV}
\eeqa
From the above, we expect that $\tri a_\mu^{NO} \sim \tri a_\mu^{IO}$ due to  the ratio $M^\nu_{33}/ M^\nu_{22}\sim 1$ for both cases.
Because of the $M^\nu_{33}/M^\nu_{11}$ factor, the branching ratio of ${\cal B}(\tau\ra e \gamma)$ in NO is expected to be larger than that in the IO.
For the precise evaluation, we definitely  must consult the full numerical study.

As shown in the benchmark points previously, the CLFV branching ratios ${\cal B}(\mu \ra e \gamma)$ and ${\cal B}(\tau \ra \mu \gamma)$ nearly saturate the current experimental upper bounds.
We define two dimensionless variables,
\beq
L_{\mu e} \equiv { {\cal B}(\mu \ra e \gamma)_{c.l.} \over  {\cal B}(\mu \ra e \gamma)_{f.l.}}\,,\;
L_{\tau \mu } \equiv { {\cal B}(\tau\ra \mu  \gamma)_{c.l.} \over  {\cal B}(\tau \ra \mu  \gamma)_{f.l.}}\,,
\eeq
to characterize the sensitivity improvement of the future experimental bounds, and the subscription  ``c.l.(f.l.)'' stands for the current(future) limit.
\begin{figure}[htbp]
    \centering
    \includegraphics[width=.40\textwidth]{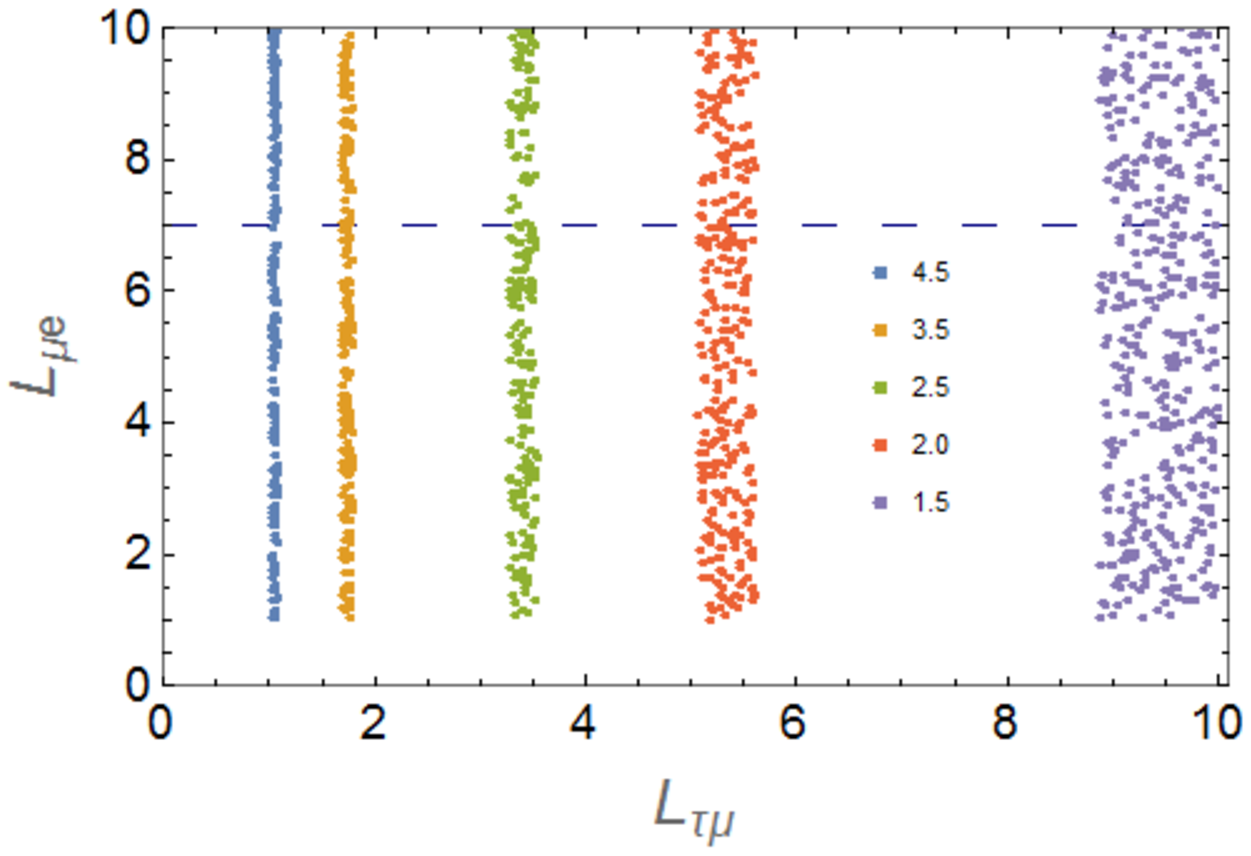}\;
    \includegraphics[width=.40\textwidth]{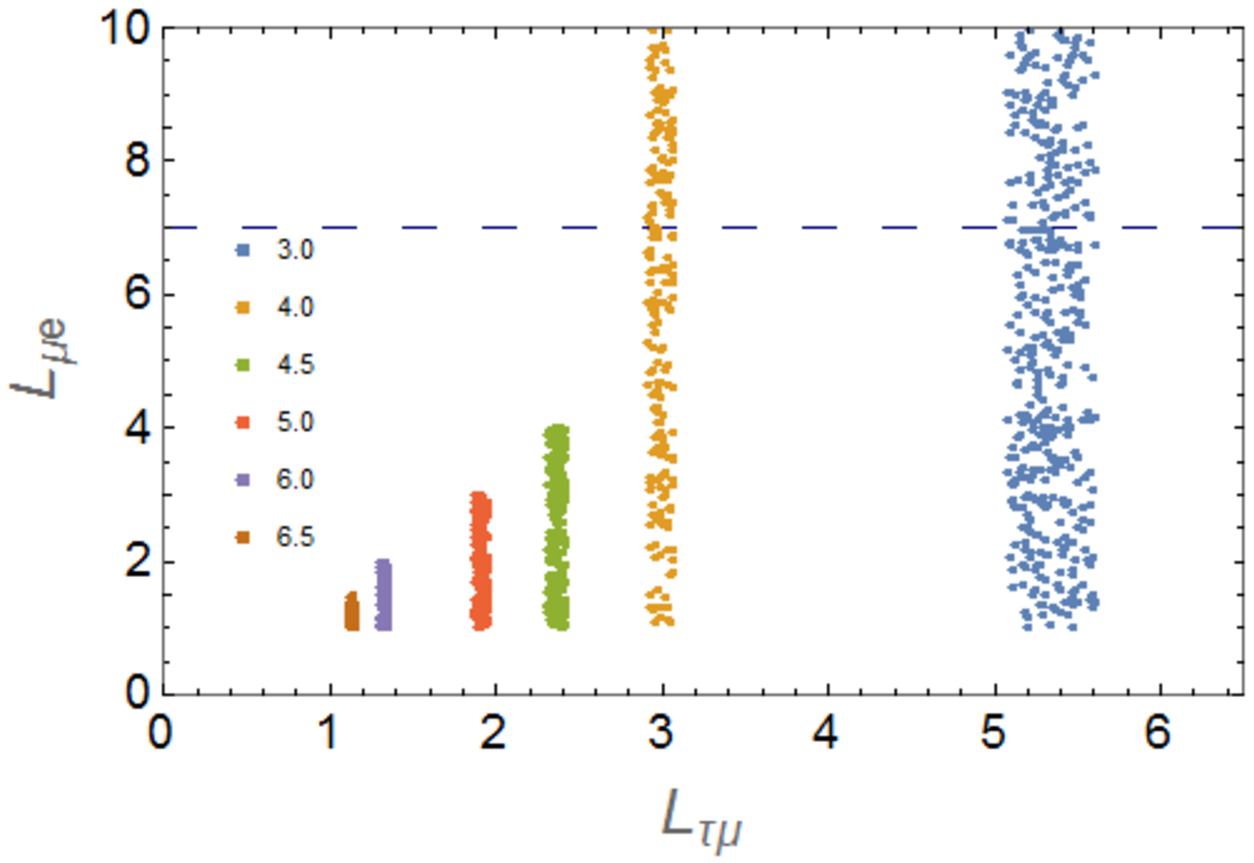}
    \caption{ The dependence of $\tri a_\mu$ (in unit of $10^{-10}$) on $L_{\mu e}$ and $L_{\tau\mu}$. The left(right) panel is for the NO(IO) benchmark point.
    The horizontal dash line represents the projected reach of MEGII\cite{MEGII:2018kmf},
   $ {\cal B}(\mu\ra e \gamma) <6\times 10^{-14}$ or $L_{\mu e}=7.0$.
     }
    \label{fig:BMPFCNCAmu}
\end{figure}
We use the benchmark points to explore the dependence of $\tri a_\mu$ on $L_{\mu e}$ and $L_{\tau\mu}$ by varying the $L$ values.
From Fig.\ref{fig:BMPFCNCAmu}, it is clear that $\tri a_\mu$ is mainly controlled by $L_{\tau\mu}$, and $ \tri a_\mu \propto \sqrt{L_{\tau\mu}}$ as expected in Eq.(\ref{eq:BTM_HV}).
In some parameter space, as in the IO benchmark point, both $L_{\mu e}$ and $L_{\tau\mu}$ place comparable constraint on $\tri a_\mu$.
The projection limit on ${\cal B}( \tau \to \mu \gamma )$ by Belle II is $\sim 10^{-9}$\cite{Belle-II:2018jsg}, or $L_{\tau\mu}\sim 42$.
If no CLFV $\tau\ra \mu \gamma$ transition is observed before reaching that sensitivity, our model predicts $\tri a_\mu\lesssim 1\times 10^{-10}$.

\begin{figure}[htbp]
    \centering
    \includegraphics[width=.50\textwidth]{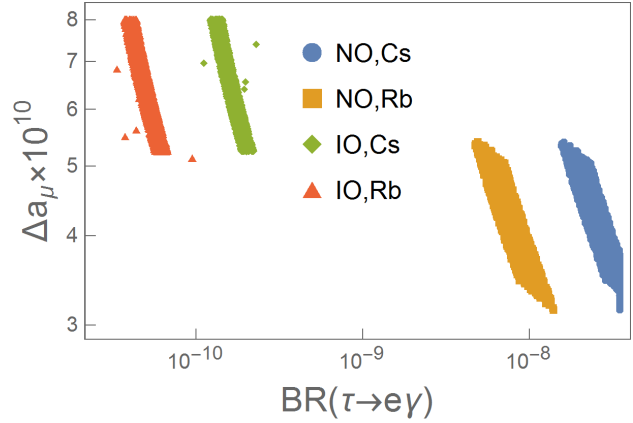}
    \caption{Correlation between ${\cal B}( \tau \to e \gamma )$ and $\Delta a_\mu$. The blue, orange, green, and red points stand for NO $\Delta a_e^{Cs}$, NO $\Delta a_e^{Rb}$, IO $\Delta a_e^{Cs}$, and NO$\Delta a_e^{Rb}$
    respectively.
    }
    \label{fig:AmudLdR31}
\end{figure}
 On the other hand, ${\cal B}( \tau \to e \gamma )$ can be much below the current limit, $< 3.3 \times 10^{-8}$\cite{Zyla:2020zbs}, for IO while it is close to the current experimental constraint for NO. The numerical result, as shown in Fig.\ref{fig:AmudLdR31}, agrees with our naive expectation, Eq.(\ref{eq:BTE_HV}), that ${\cal B}( \tau \to e \gamma )_{NO}>{\cal B}( \tau \to e \gamma )_{IO}$.
 Therefore, ${\cal B}( \tau \to e \gamma )$ could serve as an indirect probe to determine the type of neutrino mass hierarchy in our model.

\begin{figure}[htbp]
    \centering
    \includegraphics[width=.50\textwidth]{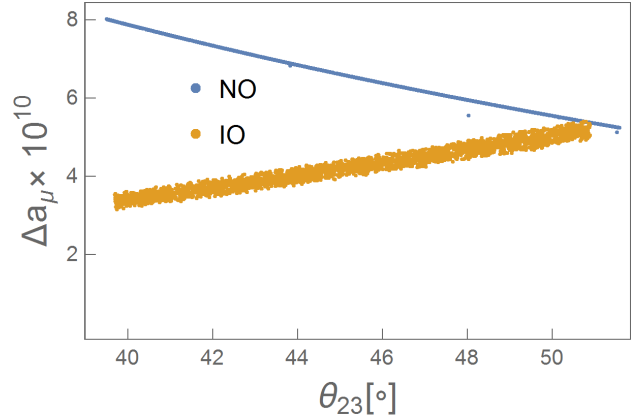}
    \caption{Correlation between $\theta_{23}$ and $\Delta a_\mu$. The blue(orange) dots stand for NO(IO) viable points. }
    \label{fig:AmuTh23}
\end{figure}

In Fig.\ref{fig:AmuTh23}, another observed correlation between $\tri a_\mu$ and $\theta_{23}$ is displayed.
This correlation can be understood as follows. Note that increasing $\theta_{23}$, while keeping all other neutrino oscillation parameters unchanged, results to a larger(smaller) $M^\nu_{22}$ for NO(IO) neutrino masses.
On the other hand, since  $M^\nu_{22} \propto (y^L_2)^2$ and $\tri a_\mu \propto y^L_2 y^R_2$,  the observed correlation follows.
Back to  Fig.\ref{fig:AmudLdR31}, it is clear  that ${\cal B}( \tau \to e \gamma )$ anti-correlates with  $\tri a_\mu$  for both neutrino mass hierarchy.  It is because increasing $\theta_{23}$ leads to smaller(larger) $M^\nu_{33}$(thus $y^L_3$) for NO(IO).
So together with the positive(negative) correlation between $\tri a_\mu^{NO[IO]}$ and $\theta_{23}$, Fig.\ref{fig:AmuTh23}, ${\cal B}( \tau \to e \gamma )\propto |y^L_3 y^R_1|^2$ always anti-correlates with  $\tri a_\mu$.

\begin{figure}[htbp]
    \centering
    \includegraphics[width=0.5\textwidth]{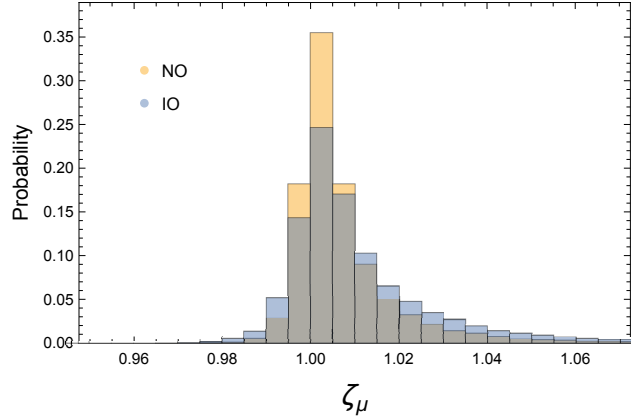}
    \caption{The probability histogram of the predicted normalized Higgs muon-Yukawa. }
    \label{fig:Higgsmumu}
\end{figure}
\begin{figure}
    \centering
    \includegraphics[width=0.45\textwidth]{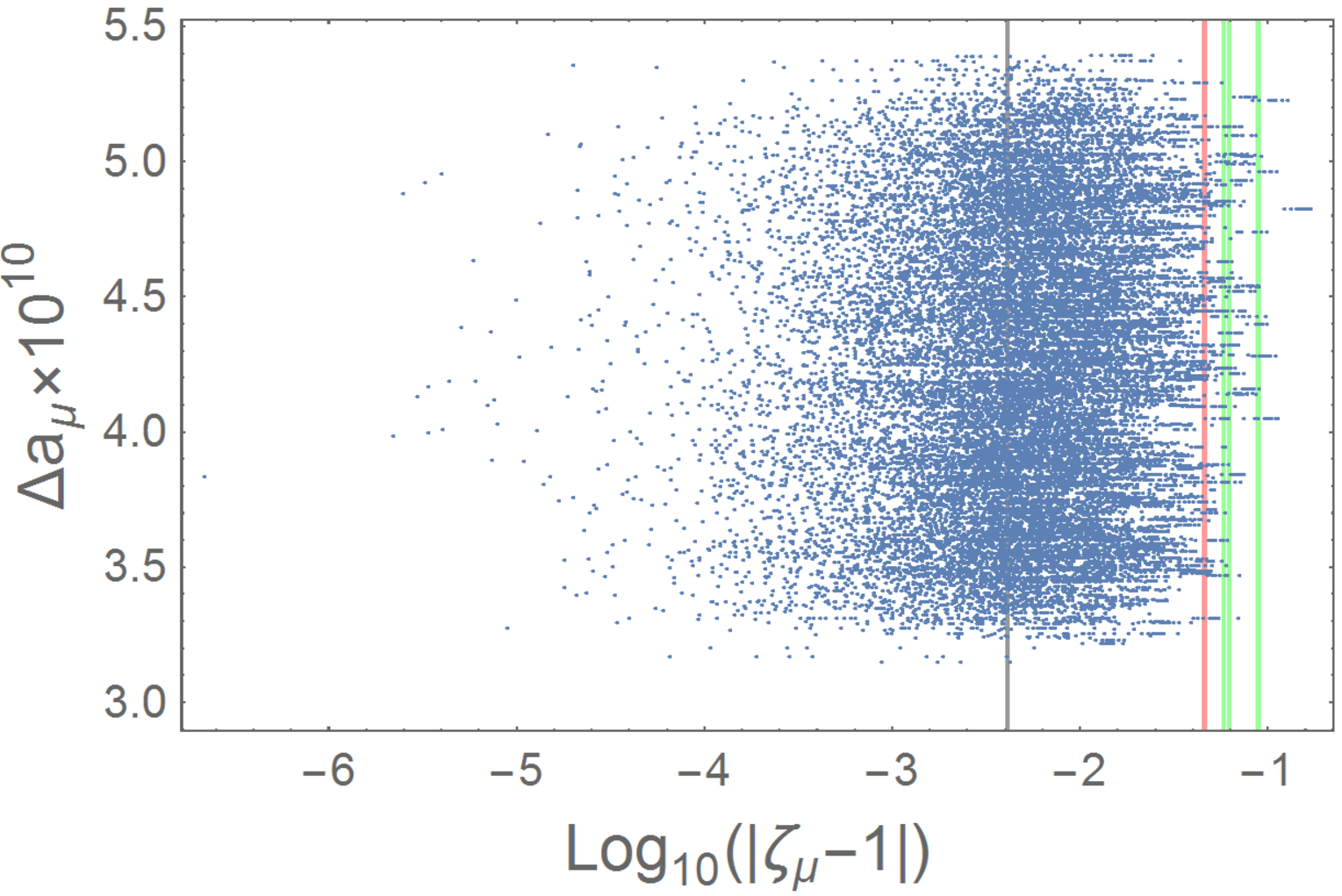}
       \includegraphics[width=0.45\textwidth]{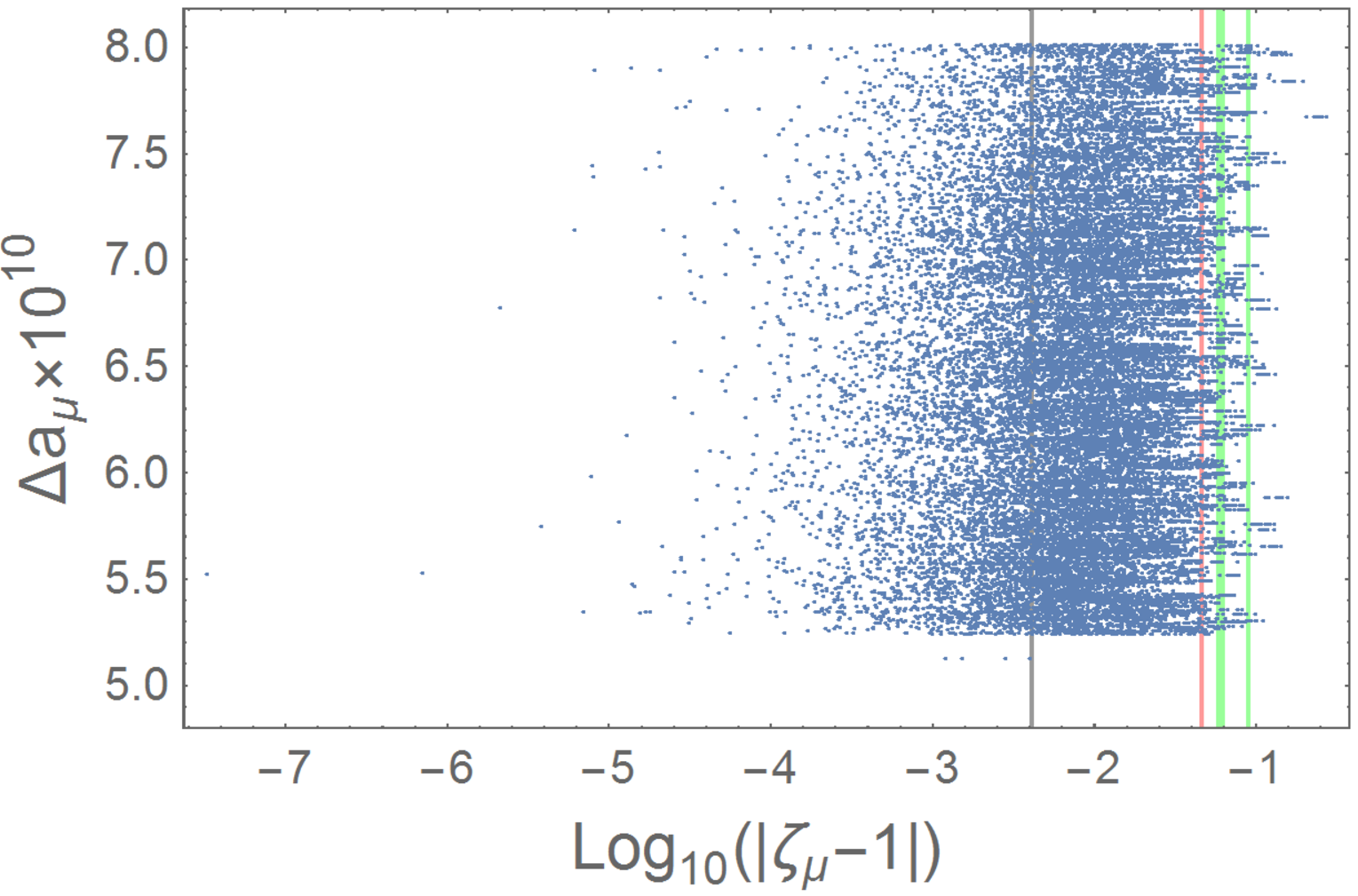}
    \caption{ The left/right panel is the scatted plot of $\tri a_\mu$ v.s. $\log_{10}( |\zeta_\mu-1|)$ for the NO/IO case.  The vertical lines, from right to left, indicate the projected sensitivity\cite{deBlas:2019rxi} at CEPC\cite{CEPCStudyGroup:2018ghi}, ILC\cite{Bambade:2019fyw}/CLIP\cite{CLICdp:2018cto}, HL-HLC(red), and FCC\cite{FCC:2018byv}(black), respectively.
    }
    \label{fig:ymu_amu}
\end{figure}

From all the viable points we found the effective muon-Higgs Yukawa coupling is close to the SM one.
The range of the normalized muon-Yukawa, $\zeta_\mu$, is roughly between $0.98- 1.05$. The distribution variance of IO is slightly larger than NO, but both peak at around $1.005$, as seen in Fig.\ref{fig:Higgsmumu}. It is well within the current $2\sigma$ constraint,  $0.6 \lesssim \zeta_\mu \lesssim 1.5 $\cite{ATLAS:2020fzp,CMS:2020xwi}\footnote{ Note that the $h_{SM}\mhyphen g\mhyphen g$ vertex does not change from the SM prediction at 1-loop level in our model. Therefore, from the signal strength $\mu_\mu$ for $p p \ra h_{SM}\ra \mu \mu$, the $\zeta_\mu$ value is simply estimated as $\zeta_\mu \simeq \sqrt{\mu_\mu} $. }.
For the future updates of $\zeta_\mu$\cite{deBlas:2019rxi}, we also show  the scatter plot of $\tri a_\mu$ vs $\zeta_\mu$ in Fig.\ref{fig:ymu_amu}.
It is clear that $\zeta_\mu$ weakly correlates with $\tri a_\mu$, and roughly half of the parameter space can be probed
at the future FCC\cite{FCC:2018byv} with a projected precision of $\sim 0.4\%$ on muon-Yukawa.

\begin{figure}
    \centering
    \includegraphics[width=0.45\textwidth]{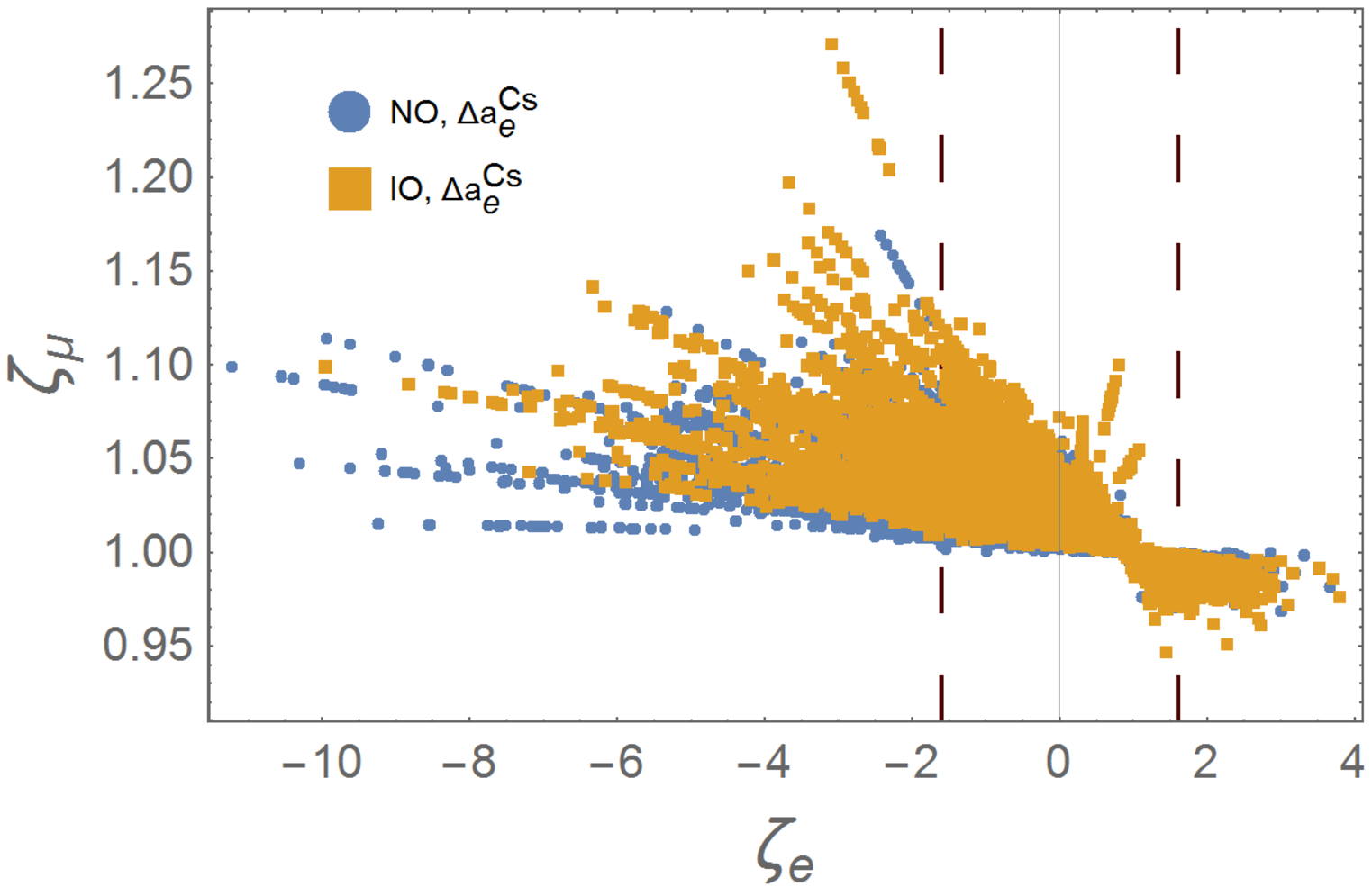}
       \includegraphics[width=0.45\textwidth]{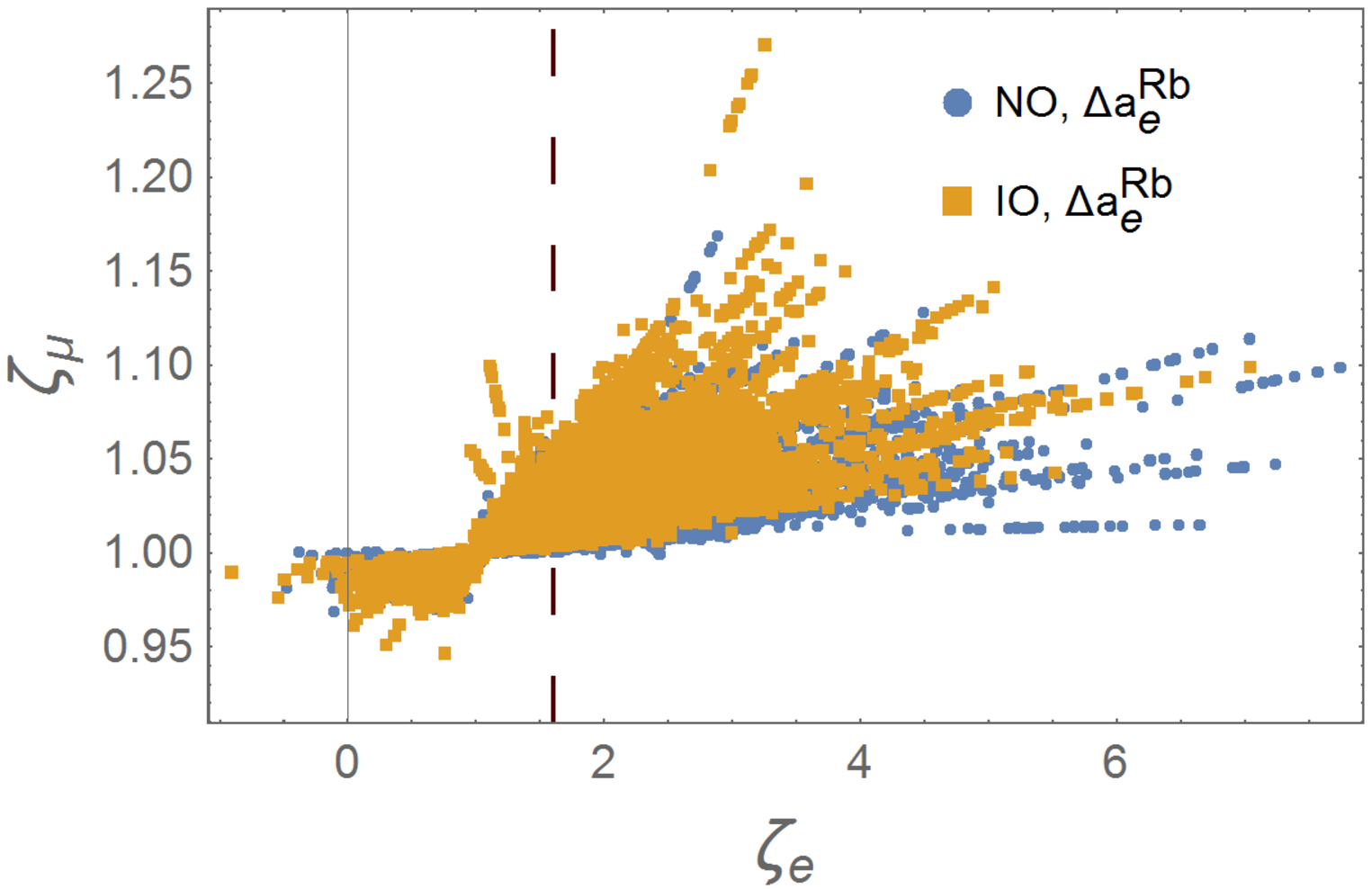}
    \caption{$\zeta_e$ v.s. $\zeta_\mu$. The vertical dashed lines indicate the projected sensitivity at FCC-ee\cite{dEnterria:2021xij}. }
    \label{fig:Higgsmmee}
\end{figure}
Interestingly, this model predicts a possible unconventional electron-Higgs Yukawa coupling.
This is due to $ |\delta M_{ee}^{loop}|\gtrsim m_e$, and that can be traced back to the less constrained $y^R_{11,12}$.
From Fig.\ref{fig:Higgsmmee}, it is clear that the magnitude of electron-Higgs Yukawa could be one order of magnitude larger than the SM prediction.
The projected sensitivity, $|\zeta_e|<1.6$ at $95\%$ CL, at FCC-ee\cite{dEnterria:2021xij} is shown as the vertical dashed line.
For the case of $\tri a_e^{CS}$, the sign of electron-Higgs Yukawa could even be negative. However, the sign determination requires an interference with the tiny electron-Higgs Yukawa coupling, and thus very challenging in the foreseeable future.

In summary, from our numerical study, the viable  model parameter space only requires
(1) Majorana masses $\ll$ Dirac masses $\sim$ TeV,
and (2) order $1(0.1)$ mixings among charged(neutral) scalars.
Since we take a bottom-up approach, this model cannot address the pattern of neutrino mass matrix.
It is taken as the experimental input. However, we find  that the hierarchy among the model parameters $y^L$'s and $y^R$'s is less than 5 orders of magnitude for most of the cases. Thus, this model is technically natural, and no
 extreme fine-tuning is required to accommodate the neutrino oscillation data and $\tri a_e^{Cs[Rb]}$.

%%%%%%%%%%%%%%%%%%%%%%%%%%%%%%%%%%%%%%%%%%%%%%%%%%
\section{Discussion and Phenomenology}
%%%%%%%%%%%%%%%%%%%%%%%%%%%%%%%%%%%%%%%%%%%%%%%%%%
\label{SEC:Pheno}

\subsection*{SM Higgs decay }
Due to the violation of lepton number, we also have $h_{SM}\ra \nu_i \nu_j^c$ decays generated at one-loop level.
By dimensional analysis, the effective $h_{SM}\mhyphen \nu_i\mhyphen \nu_j^c$ Yukawa coupling can be
estimated to be $\sim \frac{1}{4\pi^2} \frac{m_\nu}{M}$, where $M$ is the typical mass of exotic degrees of freedom.
The coupling strength is then $ \sim  {\cal O}(10^{-15})$ if taking  $M \sim{\cal O}( 1 \mtev)$,  and being experimentally  insignificant.
Since we assume all the exotic DOF's are heavier than $0.5\mtev$, there is no modification to the SM invisible Higgs decay width either.
Because all new fields  in our model are color singlet, the $ g g h_{SM}$ vertex does not receive any correction at 1-loop level.

However, the two additional charged scalars contribute to di-photon Higgs decay width.
The decay width is given by\cite{Shifman:1979eb, Chang:2012ta}
\beq
\Gamma_{\gamma\gamma}= {G_F \alpha_{em}^2 M^3_{h_{SM}} \over 128\sqrt{2}\pi^3} \left| \frac{4}{3} F_{1/2}\left( \tau_t \right)+
 F_{1}(\tau_W ) + \sum_{a=1}^2\frac{v_0 \widetilde{\mu}_{aa}}{2 M_{C_a}^2} F_0(\tau_{C_a})
   \right|^2\,,
\eeq
where $\wt{\mu}_{aa}$ is the charged scalar-Higgs cubic coupling given in Eq.(\ref{eq:CCH_mu}).
The $F_{1/2}$ and $F_0$ terms are the dominate SM 1-loop contributions from top quark and $W^\pm$ boson, respectively.
We define $\tau_i= ( M_{h_{SM}}/2 M_i)^2$, and the one-loop functions are given by
\beq
F_0(x) = {f(x)-x \over x^2}\,,\;
F_{1/2}(x)={ 2[x+(x-1)f(x)] \over x^2}\,,\;
F_1(x)=-{ 2x^2+3x+3(2x-1)f(x) \over x^2}\,,
\eeq
with $f(x)=[\arcsin(\sqrt{x})]^2$ for $x<1$. For $x \ll 1$, $F_0(x) \sim \frac{1}{3}+ \frac{8}{45}x $.
Plugging in the masses of top, $W^\pm$, and $h_{SM}$, we have $ F_{1/2}\left( \tau_t \right)= 1.38$ and $F_{1}(\tau_W )=-8.32$.
Assuming that $\mu_{DC}\gg v_0$, we have approximately  $\widetilde{\mu}_{11}\sim -2 s_\alpha c_\alpha \mu_{DC}$ and  $\widetilde{\mu}_{22}\sim +2 s_\alpha c_\alpha \mu_{DC}$. So the width becomes
\beq
\Gamma_{\gamma\gamma} %&\simeq& {G_F \alpha_{em}^2 M^3_{h_{SM}} \over 128\sqrt{2}\pi^3} \left| (-6.49)_{SM}+  \frac{v_0 \sin(2\alpha_C)\mu_{DC} }{6 M_{C_1}^2}\left(\frac{M_{C_1}^2}{M_{C_2}^2}-1\right)  \right|^2\nonr\\
\simeq {G_F \alpha_{em}^2 M^3_{h_{SM}} \over 128\sqrt{2}\pi^3} \left| (-6.49)_{SM} - \frac{1}{3} \left( {v_0 \mu_{DC} \over M_{C_1} M_{C_2} }\right)^2  \right|^2\,.
\eeq
It is clear that  $\Gamma_{\gamma\gamma}/\Gamma^{SM}_{\gamma\gamma}>1$.
Assuming $M_{C_1}=0.5\mtev$ and  $\mu_{DC}\simeq M_{C_2}$, then
the NP will only modify the di-photon decay width by a magnitude $\sim 2\%$.
Comparing to the latest coupling scaling factor $\kappa_\gamma =1.06\pm0.05$ obtained by \cite{ATLAS:2020qdt}, we expect a weak  constraint on this model from the Higgs diphoton decay.

\subsection*{CLFV $\mu \ra e$ conversion, and $\mu \ra 3e$ }

Concerning on the other CLFV experimental bounds on the model parameter space,
we comment on the $\mu\mhyphen e$ conversion on nuclei and $\mu\ra 3 e$.

For  $\mu\mhyphen e$ conversion on nuclei, the current upper bound of the ratio of $\mu \ra e$ conversion rate normalized to the muon capture rate\cite{Kuno:1999jp},
\beq
R_{\mu e}={\Gamma(\mu+(A,Z)\ra e+(A,Z)) \over \Gamma(\mu+(A,Z)\ra \nu_\mu+(A,Z-1))} < 7\times 10^{-13}\,,
\eeq
is given by SINDRUM II with gold as target\cite{SINDRUMII:2006dvw}.
Since the hidden sector does not couple to the SM quark sector, the $\mu \ra e$ conversion is dominated by the $\mu e \gamma$ dipole.
For a given target nuclei $N$, the $\mu \ra e$ conversion rate can be expressed as
\beq
\Gamma_{\mu\ra e} \sim \pi D_N^2 \Gamma_\mu {\cal B}(\mu\ra e \gamma)\,,
\label{eq:Gmeg}
\eeq
where $\Gamma_\mu=0.45517\times 10^6 s^{-1}$\cite{Zyla:2020zbs} is the muon decay rate, and $D_N$ is the lepton-nucleus overlap integral coefficient.
For gold and aluminum, $D_{Au} =0.189$ and $D_{Al}=0.0362$
\cite{Kitano:2002mt}, while
the muon capture rates are $13.07\times 10^6 s^{-1}$ and $0.71\times 10^6 s^{-1}$, respectively\cite{Kitano:2002mt,Suzuki:1987jf}.
Plugging in the values of $\Gamma_\mu$ and the current upper limit of ${\cal B}(\mu\ra e \gamma)$, one finds
$R_{\mu e} < 10^{-15}$ with Au or Al target, roughly three orders of magnitude below the current experimental limit.
In the near future, the model prediction, Eq.(\ref{eq:Gmeg}), could be checked with improved sensitivity\cite{Teshima:2018ise, COMET:2018auw, Mu2e:2014fns, Kuno:2005mm}.

\begin{figure}
    \centering
    \includegraphics[width=0.35\textwidth]{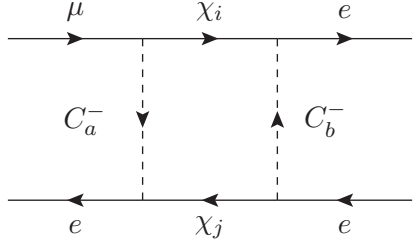}
    \caption{ The box diagram for $\mu\ra 3e$ transition, where $a,b \in\{ 1,2\}$ and $i,j\in\{1,2,3,4\}$.
    }
    \label{fig:mu3e}
\end{figure}

In this model, $\mu \ra 3e$ can be induced by the CLFV $\mu e\gamma$ dipole coupling  and the box diagram shown in Fig.\ref{fig:mu3e}.
The contribution form the $\mu e\gamma$ dipole can be calculated to be\cite{Petcov:1977ab, Kuno:1999jp, Chang:2005ag}
\beq
\Gamma(\mu\ra 3e)_{dip} \simeq \frac{2\alpha_{em}}{3 \pi}\left[\ln\frac{m_\mu}{m_e}-\frac{11}{8}\right]\Gamma(\mu\ra e \gamma)\,.
\eeq
This part is about three orders of magnitude below the current bound of ${\cal B}(\mu\ra 3 e)<10^{-12}$\cite{SINDRUM:1987nra}.
In addition to the dipole contribution, the box-diagram gives rise to FCNC 4-fermi interactions\cite{Chang:2005ag}:
\beq
{\cal L} \supset -\frac{4G_F}{\sqrt{2}} \left[ g_1 \left(\bar{e}\PL \mu\right)\left(\bar{e}\PL e\right)
+g_3 \left(\bar{e}\gamma^\alpha \PR \mu\right)\left(\bar{e}\gamma_\alpha\PR e\right)
+g_5 \left(\bar{e}\gamma^\alpha \PR \mu\right)\left(\bar{e}\gamma_\alpha\PL e\right)
+ ( \PL \Leftrightarrow \PR) \right] +H.c.\,,
\eeq
and leads to
\beq
{\cal B}(\mu\ra 3e)_{4 fermi} \simeq \frac{|g_1|^2}{8} +2 |g_3|^2+ |g_5|^2 + ( \PL \Leftrightarrow \PR)\,,
\eeq
if ignoring the interference between the dipole and 4-fermi interactions.
By dimension analysis, the dimensionless Wilson coefficients can be estimated as
\beq
|g|\sim \frac{1}{16 \pi^2} \frac{\sqrt{2}}{4 G_F} { Y_e^3 Y_\mu \over  M^2}\times(0.5)\,,
\eeq
where $M$ is the relevant highest mass in the loop, $0.5$ is the typical value from the box-diagram loop integration, and $Y_e^3 Y_\mu$ denotes the product of four relevant LH or RH Yukawa couplings.
Thus, ${\cal B}(\mu\ra 3e)_{4 fermi} \simeq 10^{-15} (2\mtev/ M)^4(Y_e^3 Y_\mu/ 0.001)^2 $ is also expected to be safely below the current experimental limit. The $\mu\ra 3e$ decay could be a relevant constraint and probed by the planned Mu3e experiment with a sensitivity of $10^{-16}$\cite{Blondel:2013ia} in the near future.
For CLFV $\tau\ra l_i l_j l_k\,(l_{i,j,k}=e,\mu)$ decay branching ratios, the current bounds, $ (\mbox{a few})\times 10^{-8} $\cite{Hayasaka:2010np}, and the future $\sim 10^{-10}$ sensitivity \cite{Belle-II:2018jsg, LHCb:2018roe, Beacham:2019nyx} do not post further constraint on this model.

\subsection*{$Z_X$: The smoking gun of the gauge $U(1)_X$}
One robust prediction of gauge $U(1)_X$ is the existence of the gauge boson, $Z_X$.
The direct detection of $Z_X$ will be the smoking gun of the gauge $U(1)_X$.
 Here we give a brief remark on the prospects on its discovery.
As discussed earlier, after $S_2$ acquires a VEV $v_2$, $Z_X$ also acquires a mass $M_{Z_X} =2 g_X v_2$.
But $Z_X$ does not play any role in our flavor physics discussion.
 So far, both the gauge coupling $g_X$ and $M_{Z_X}$ are unknown free parameters.
Although $Z_X$ does not couple to any of the SM fields at tree-level, its couplings to $\gamma Z, ZZ, W^+W^-, h_{SM} h_{SM}, \gamma h_{SM}, Z h_{SM}$, and lepton pairs
can be generated at one-loop level. Also, the one-loop vacuum polarization diagrams can generate the $Z_X\mhyphen Z$ and $Z_X\mhyphen\gamma$ mixings.
The $Z_X\mhyphen Z$ and $Z_X\mhyphen\gamma$ mixings can also be induced through the tree-level kinematic mixing between the field strengthes of $U(1)_X$ and $U(1)_Y$, ${\cal L} \supset -\frac{\epsilon}{2} B^{\mu\nu}X_{\mu\nu}$, see \cite{Holdom:1985ag, Chang:2006fp, Chang:2007ki}.
Since the kinematic mixing term is gauge invariant and renormalizable, $\epsilon$ is not required to be small.  The kinematic mixing term can be rotated away by a $GL(2)$ transformation, and the two massive eigenstates couple to SM fields.
From electroweak precision measurements, $h_{SM}\ra Z Z_X$, and the Drell-Yan process, the mixing is constrained to be $\epsilon \lesssim 10^{-2}$ for $ 10\, \mgev \lesssim M_{Z_X}\lesssim 1\, \mtev$\cite{Curtin:2014cca}.
In the future,  the  HL-LHC and HE-LHC can probe the effective mixing to the level about $\sim 10^{-3}$ for
$ 10\, \mgev \lesssim M_{Z_X}\lesssim 90\, \mgev$ and $ 0.2\, \mtev \lesssim M_{Z_X}\lesssim 2\, \mtev$\cite{Curtin:2014cca}.
In this model, the $\epsilon$ parameter is the combination of tree-level and 1-loop contributions. Moreover, the $Z_X$ to lepton pair couplings receive flavor dependent quantum corrections.
At the LHC, the Drell-Yan processes $pp\ra Z^* \ra Z Z_X$, or  $pp\ra W^* \ra W Z_X$ will be the dominate $Z_X$ production mechanism for $M_{Z_X}\gtrsim 180\, \mgev$.
And the decays $Z_X\ra l_i^+ l_j^-$ with di-lepton invariant mass peaked at around $m_{l_il_j} \simeq M_{Z_X}$ will be the clear signal.
Incidentally, in this model the CLFV di-lepton decay connects to the flavor physics that we have discussed.
A comprehensive study on this topic is beyond the scope of this paper, and we will leave it to the future works.

\subsection*{Dark matter}
Finally, we give a sketchy discussion on DM in this model.
After the SSB of $U(1)_X$ by $S_2$, the remaining gauge discrete parity\cite{Krauss:1988zc} stabilizes the
the lightest neutral DOF carrying one unit of $U(1)_X$ charge.
Apparently, this DOF is a DM candidate.
There are two possible candidates in our model: (1) $S_D$, the lighter of $R_1$ and $I_1$, and (2) $\chi_4$, the lightest mass eigenstate of the exotic fermions. From our numerical scan, about $\sim 70(30)\%$ of the solutions yield scalar(fermionic) DM candidate.
And about $\sim 90\%$ of the potential DM mass is in the range of $[0.5,1.0]\mtev$.
For a DM in that mass range, the current observational upper limits on the spin-independent DM-nucleon cross section is  $\sigma_{SI}\lesssim 10^{-46} (cm)^2$ \cite{LUX:2016ggv, XENON:2018voc,PandaX-II:2020oim}.
In our model, both $S_D$ and $\chi_4$ do not couple to the SM quark sector  at tree-level and all the new DOFs are color singlets.
Hence, for both DM candidates the $\sigma_{SI}$ can be easily arranged to stay below the direct detection bounds.

If $\chi_4$ is the DM candidate, the relic density is determined mainly by $t$- and $u$-channel annihilation $\chi_4 \chi_4\ra l \bar{l} (l=l^-, \nu)$, and the $s-$channel $ \chi_4 \chi_4\ra h_2\, \mbox{or}\, h_{SM}\ra (SM)(SM)$ processes, where $(SM)$ stands for any SM field coupled to the SM Higgs.
On the other hand, for the bosonic DM case, the relic density is mainly controlled by the $S_D S_D \ra h_{SM} h_{SM}$,  $S_D S_D \ra h_2\, \mbox{or}\, h_{SM} \ra (SM)(SM)$, and
$S_D S_D \ra W^+W^-, Z Z$. One needs to take into account the co-annihilation  $R_1 I_1 \ra (SM)(SM)$ when $R_1$ and $I_1$ are nearly degenerate.

It should be emphasised that although the DM candidate is stable, phenomenology only demands that its relic density must not exceed the observed DM relic density, $\Omega_{DM}h^2 =0.120\pm 0.001$\cite{Planck:2018vyg}.
However, the full evaluation of the relic density requires more model parameters which are independent of the flavor physics we are focusing on, and the comprehensive analysis of the extended parameter space is beyond the scope of this paper.

%%%%%%%%%%%%%%%%%%%%%%%%%%%%%%%%%%%%%%%%%%%%%%%%%%
\section{Conclusion}
%%%%%%%%%%%%%%%%%%%%%%%%%%%%%%%%%%%%%%%%%%%%%%%%%%
\label{SEC:conclusion}

Motivated by the recently measured anomalous magnetic moments of muon  and electron, we studied a model with gauge hidden $U(1)_X$ symmetry
as the unified framework to accommodate both the radiative neutrino mass generation mechanism and the measured $\tri a_{\mu, e}$.
This UV-complete model employs four exotic scalars and a minimum of two pairs of exotic vector fermions, all charged under $U(1)_X$, as seen in Table-\ref{table:newparticle}.   The $U(1)_X$ is assumed to be spontaneously broken at an energy scale higher then the SM electroweak scale when one of the exotic singlet bosons gets a non-zero VEV.  After the SSB of $U(1)_X$, the new fermions acquire Majorana masses, which are crucial to the radiative neutrino mass generation. Moreover, the new vector fermions admit tree-level Dirac masses which are essential for chirality flipping in explaining $\tri a_l$.

Any mechanism that gives rise to charged lepton anomalous magnetic moments could potentially lead to CLFV
which are stringently constrained by experiments.
Usually, additional flavor symmetries or assumptions are summoned to suppress the unwanted CLFV.
Contrarily, in this work we took a bottom-up approach and asked what restrictions the experimental constraints would impose upon the model parameter space. We have carefully taken into account the neutrino oscillation data and the experimental CLFV limits in our numerical study.
We found that the model can explain the observed neutrino oscillation data, either normal or inverted ordering, and the central value of $\tri a_e^{Cs[Rb]}$ without much fine-tuning on the model parameters. However, in the minimal model, the current experimental  CLFV limit on ${\cal B}(\tau\ra \mu \gamma)<4.4\times 10^{-8}$\cite{Aubert:2009ag}  results in $\tri a_\mu \sim (4-8) \times 10^{-10}$, which differs from $\tri a_\mu^{F\!N\!A\!L}\simeq (25.1\pm 5.9)\times 10^{-10}$\cite{Abi:2021gix} by $\sim 3 \sigma$'s, but agrees with the SM prediction using the recent lattice QCD evaluation on hardronic contribution to $a_\mu$\cite{Borsanyi:2020mff,Ce:2022kxy}. Of course, more theoretical and experimental investigations are needed to settle down the $\tri a_\mu$ issue. We pointed out that the future improvement on the CLFV limit will further suppress our predicted $\tri a_\mu$ because of $ (\tri a_\mu)^2 \propto {\cal B}(\tau\ra \mu \gamma)$ in our model.

The unconventional electron-Higgs Yukawa is another intriguing feature of this model.
We found that the magnitude of electron-Higgs Yukawa can be one order of magnitude larger than the SM prediction.
This unusual electron Yukawa can be probed at the future FCC-ee\cite{dEnterria:2021xij}.
More interestingly, if $\tri a_e^{Cs}$ is confirmed in the future, a negative electron-Higgs Yukawa could be allowed in this model.
Due to the smallness of electron-Higgs Yukawa, the determination of its sign is extremely challenging in the foreseeable future.
On the other hand, the muon-Higgs Yukawa is found to be close to the SM prediction with the deviation fraction $\lesssim 10^{-2}$ for most of the viable model parameter space.  Although  the deviation of muon-Higgs Yukawa is small, it could be probed at the FCC with a  projected precision of $\sim 0.4\%$\cite{deBlas:2019rxi}.

Since our approach is bottom-up, the prediction $\tri a_\mu <\tri a_\mu^{F\!N\!A\!L}$ is robust and applies to
the minimal model with arbitrary add-on flavor symmetry.
One has to go beyond the minimal model if the deviation between the experimentally and theoretically improved $\tri a_\mu$ and the model prediction persists, or a few$\times 10\%$ muon Yukawa deviation is confirmed in the future.
If that is the case, a trivial extension of this model by utilizing the third pair of vector fermion will do the job, and an even more enormous electron-Higgs Yukawa coupling strength is possible.

From the exercise, we have demonstrated that due to the smallness of the SM electron Yukawa, the effective electron-Higgs coupling strength is sensitive to new physics and plays a vital role in testing our understanding of flavor physics.

\section*{Acknowledgments}
This research is supported by MOST 109-2112-M-007-012 and
110-2112-M-007-028 of Taiwan.

\bibliography{U1G_Ref}

\begin{thebibliography}{108}%
\makeatletter
\providecommand \@ifxundefined [1]{%
 \@ifx{#1\undefined}
}%
\providecommand \@ifnum [1]{%
 \ifnum #1\expandafter \@firstoftwo
 \else \expandafter \@secondoftwo
 \fi
}%
\providecommand \@ifx [1]{%
 \ifx #1\expandafter \@firstoftwo
 \else \expandafter \@secondoftwo
 \fi
}%
\providecommand \natexlab [1]{#1}%
\providecommand \enquote  [1]{``#1''}%
\providecommand \bibnamefont  [1]{#1}%
\providecommand \bibfnamefont [1]{#1}%
\providecommand \citenamefont [1]{#1}%
\providecommand \href@noop [0]{\@secondoftwo}%
\providecommand \href [0]{\begingroup \@sanitize@url \@href}%
\providecommand \@href[1]{\@@startlink{#1}\@@href}%
\providecommand \@@href[1]{\endgroup#1\@@endlink}%
\providecommand \@sanitize@url [0]{\catcode `\\12\catcode `\$12\catcode
  `\&12\catcode `\#12\catcode `\^12\catcode `\_12\catcode `\%12\relax}%
\providecommand \@@startlink[1]{}%
\providecommand \@@endlink[0]{}%
\providecommand \url  [0]{\begingroup\@sanitize@url \@url }%
\providecommand \@url [1]{\endgroup\@href {#1}{\urlprefix }}%
\providecommand \urlprefix  [0]{URL }%
\providecommand \Eprint [0]{\href }%
\providecommand \doibase [0]{https://doi.org/}%
\providecommand \selectlanguage [0]{\@gobble}%
\providecommand \bibinfo  [0]{\@secondoftwo}%
\providecommand \bibfield  [0]{\@secondoftwo}%
\providecommand \translation [1]{[#1]}%
\providecommand \BibitemOpen [0]{}%
\providecommand \bibitemStop [0]{}%
\providecommand \bibitemNoStop [0]{.\EOS\space}%
\providecommand \EOS [0]{\spacefactor3000\relax}%
\providecommand \BibitemShut  [1]{\csname bibitem#1\endcsname}%
\let\auto@bib@innerbib\@empty
%</preamble>
\bibitem [{\citenamefont {Aaboud}\ \emph
  {et~al.}(2018{\natexlab{a}})\citenamefont {Aaboud} \emph
  {et~al.}}]{ATLAS:2018hxb}%
  \BibitemOpen
  \bibfield  {author} {\bibinfo {author} {\bibfnamefont {M.}~\bibnamefont
  {Aaboud}} \emph {et~al.} (\bibinfo {collaboration} {ATLAS}),\ }\bibfield
  {title} {\bibinfo {title} {{Measurements of Higgs boson properties in the
  diphoton decay channel with 36 fb$^{-1}$ of $pp$ collision data at $\sqrt{s}
  = 13$ TeV with the ATLAS detector}},\ }\href
  {https://doi.org/10.1103/PhysRevD.98.052005} {\bibfield  {journal} {\bibinfo
  {journal} {Phys. Rev. D}\ }\textbf {\bibinfo {volume} {98}},\ \bibinfo
  {pages} {052005} (\bibinfo {year} {2018}{\natexlab{a}})},\ \Eprint
  {https://arxiv.org/abs/1802.04146} {arXiv:1802.04146 [hep-ex]} \BibitemShut
  {NoStop}%
\bibitem [{\citenamefont {Aaboud}\ \emph
  {et~al.}(2018{\natexlab{b}})\citenamefont {Aaboud} \emph
  {et~al.}}]{ATLAS:2018kot}%
  \BibitemOpen
  \bibfield  {author} {\bibinfo {author} {\bibfnamefont {M.}~\bibnamefont
  {Aaboud}} \emph {et~al.} (\bibinfo {collaboration} {ATLAS}),\ }\bibfield
  {title} {\bibinfo {title} {{Observation of $H \rightarrow b\bar{b}$ decays
  and $VH$ production with the ATLAS detector}},\ }\href
  {https://doi.org/10.1016/j.physletb.2018.09.013} {\bibfield  {journal}
  {\bibinfo  {journal} {Phys. Lett. B}\ }\textbf {\bibinfo {volume} {786}},\
  \bibinfo {pages} {59} (\bibinfo {year} {2018}{\natexlab{b}})},\ \Eprint
  {https://arxiv.org/abs/1808.08238} {arXiv:1808.08238 [hep-ex]} \BibitemShut
  {NoStop}%
\bibitem [{\citenamefont {Sirunyan}\ \emph
  {et~al.}(2018{\natexlab{a}})\citenamefont {Sirunyan} \emph
  {et~al.}}]{CMS:2018nsn}%
  \BibitemOpen
  \bibfield  {author} {\bibinfo {author} {\bibfnamefont {A.~M.}\ \bibnamefont
  {Sirunyan}} \emph {et~al.} (\bibinfo {collaboration} {CMS}),\ }\bibfield
  {title} {\bibinfo {title} {{Observation of Higgs boson decay to bottom
  quarks}},\ }\href {https://doi.org/10.1103/PhysRevLett.121.121801} {\bibfield
   {journal} {\bibinfo  {journal} {Phys. Rev. Lett.}\ }\textbf {\bibinfo
  {volume} {121}},\ \bibinfo {pages} {121801} (\bibinfo {year}
  {2018}{\natexlab{a}})},\ \Eprint {https://arxiv.org/abs/1808.08242}
  {arXiv:1808.08242 [hep-ex]} \BibitemShut {NoStop}%
\bibitem [{\citenamefont {Aad}\ \emph {et~al.}(2015)\citenamefont {Aad} \emph
  {et~al.}}]{ATLAS:2015xst}%
  \BibitemOpen
  \bibfield  {author} {\bibinfo {author} {\bibfnamefont {G.}~\bibnamefont
  {Aad}} \emph {et~al.} (\bibinfo {collaboration} {ATLAS}),\ }\bibfield
  {title} {\bibinfo {title} {{Evidence for the Higgs-boson Yukawa coupling to
  tau leptons with the ATLAS detector}},\ }\href
  {https://doi.org/10.1007/JHEP04(2015)117} {\bibfield  {journal} {\bibinfo
  {journal} {JHEP}\ }\textbf {\bibinfo {volume} {04}},\ \bibinfo {pages}
  {117}},\ \Eprint {https://arxiv.org/abs/1501.04943} {arXiv:1501.04943
  [hep-ex]} \BibitemShut {NoStop}%
\bibitem [{\citenamefont {Sirunyan}\ \emph
  {et~al.}(2018{\natexlab{b}})\citenamefont {Sirunyan} \emph
  {et~al.}}]{CMS:2017zyp}%
  \BibitemOpen
  \bibfield  {author} {\bibinfo {author} {\bibfnamefont {A.~M.}\ \bibnamefont
  {Sirunyan}} \emph {et~al.} (\bibinfo {collaboration} {CMS}),\ }\bibfield
  {title} {\bibinfo {title} {{Observation of the Higgs boson decay to a pair of
  $\tau$ leptons with the CMS detector}},\ }\href
  {https://doi.org/10.1016/j.physletb.2018.02.004} {\bibfield  {journal}
  {\bibinfo  {journal} {Phys. Lett. B}\ }\textbf {\bibinfo {volume} {779}},\
  \bibinfo {pages} {283} (\bibinfo {year} {2018}{\natexlab{b}})},\ \Eprint
  {https://arxiv.org/abs/1708.00373} {arXiv:1708.00373 [hep-ex]} \BibitemShut
  {NoStop}%
\bibitem [{\citenamefont {Aad}\ \emph {et~al.}(2021)\citenamefont {Aad} \emph
  {et~al.}}]{ATLAS:2020fzp}%
  \BibitemOpen
  \bibfield  {author} {\bibinfo {author} {\bibfnamefont {G.}~\bibnamefont
  {Aad}} \emph {et~al.} (\bibinfo {collaboration} {ATLAS}),\ }\bibfield
  {title} {\bibinfo {title} {{A search for the dimuon decay of the Standard
  Model Higgs boson with the ATLAS detector}},\ }\href
  {https://doi.org/10.1016/j.physletb.2020.135980} {\bibfield  {journal}
  {\bibinfo  {journal} {Phys. Lett. B}\ }\textbf {\bibinfo {volume} {812}},\
  \bibinfo {pages} {135980} (\bibinfo {year} {2021})},\ \Eprint
  {https://arxiv.org/abs/2007.07830} {arXiv:2007.07830 [hep-ex]} \BibitemShut
  {NoStop}%
\bibitem [{\citenamefont {Sirunyan}\ \emph {et~al.}(2021)\citenamefont
  {Sirunyan} \emph {et~al.}}]{CMS:2020xwi}%
  \BibitemOpen
  \bibfield  {author} {\bibinfo {author} {\bibfnamefont {A.~M.}\ \bibnamefont
  {Sirunyan}} \emph {et~al.} (\bibinfo {collaboration} {CMS}),\ }\bibfield
  {title} {\bibinfo {title} {{Evidence for Higgs boson decay to a pair of
  muons}},\ }\href {https://doi.org/10.1007/JHEP01(2021)148} {\bibfield
  {journal} {\bibinfo  {journal} {JHEP}\ }\textbf {\bibinfo {volume} {01}},\
  \bibinfo {pages} {148}},\ \Eprint {https://arxiv.org/abs/2009.04363}
  {arXiv:2009.04363 [hep-ex]} \BibitemShut {NoStop}%
\bibitem [{\citenamefont {Aad}\ \emph {et~al.}(2020{\natexlab{a}})\citenamefont
  {Aad} \emph {et~al.}}]{ATLAS:2019nkf}%
  \BibitemOpen
  \bibfield  {author} {\bibinfo {author} {\bibfnamefont {G.}~\bibnamefont
  {Aad}} \emph {et~al.} (\bibinfo {collaboration} {ATLAS}),\ }\bibfield
  {title} {\bibinfo {title} {{Combined measurements of Higgs boson production
  and decay using up to $80$ fb$^{-1}$ of proton-proton collision data at
  $\sqrt{s}=$ 13 TeV collected with the ATLAS experiment}},\ }\href
  {https://doi.org/10.1103/PhysRevD.101.012002} {\bibfield  {journal} {\bibinfo
   {journal} {Phys. Rev. D}\ }\textbf {\bibinfo {volume} {101}},\ \bibinfo
  {pages} {012002} (\bibinfo {year} {2020}{\natexlab{a}})},\ \Eprint
  {https://arxiv.org/abs/1909.02845} {arXiv:1909.02845 [hep-ex]} \BibitemShut
  {NoStop}%
\bibitem [{\citenamefont {Sirunyan}\ \emph {et~al.}(2019)\citenamefont
  {Sirunyan} \emph {et~al.}}]{CMS:2018uag}%
  \BibitemOpen
  \bibfield  {author} {\bibinfo {author} {\bibfnamefont {A.~M.}\ \bibnamefont
  {Sirunyan}} \emph {et~al.} (\bibinfo {collaboration} {CMS}),\ }\bibfield
  {title} {\bibinfo {title} {{Combined measurements of Higgs boson couplings in
  proton\textendash{}proton collisions at $\sqrt{s}=13\,\text {Te}\text {V}
  $}},\ }\href {https://doi.org/10.1140/epjc/s10052-019-6909-y} {\bibfield
  {journal} {\bibinfo  {journal} {Eur. Phys. J. C}\ }\textbf {\bibinfo {volume}
  {79}},\ \bibinfo {pages} {421} (\bibinfo {year} {2019})},\ \Eprint
  {https://arxiv.org/abs/1809.10733} {arXiv:1809.10733 [hep-ex]} \BibitemShut
  {NoStop}%
\bibitem [{\citenamefont {Zyla}\ \emph {et~al.}(2020)\citenamefont {Zyla} \emph
  {et~al.}}]{Zyla:2020zbs}%
  \BibitemOpen
  \bibfield  {author} {\bibinfo {author} {\bibfnamefont {P.~A.}\ \bibnamefont
  {Zyla}} \emph {et~al.} (\bibinfo {collaboration} {Particle Data Group}),\
  }\bibfield  {title} {\bibinfo {title} {{Review of Particle Physics}},\ }\href
  {https://doi.org/10.1093/ptep/ptaa104} {\bibfield  {journal} {\bibinfo
  {journal} {PTEP}\ }\textbf {\bibinfo {volume} {2020}},\ \bibinfo {pages}
  {083C01} (\bibinfo {year} {2020})}\BibitemShut {NoStop}%
\bibitem [{\citenamefont {Khachatryan}\ \emph {et~al.}(2015)\citenamefont
  {Khachatryan} \emph {et~al.}}]{CMS:2014dqm}%
  \BibitemOpen
  \bibfield  {author} {\bibinfo {author} {\bibfnamefont {V.}~\bibnamefont
  {Khachatryan}} \emph {et~al.} (\bibinfo {collaboration} {CMS}),\ }\bibfield
  {title} {\bibinfo {title} {{Search for a standard model-like Higgs boson in
  the $\mu^+ \mu^-$ and $e^+ e^-$ decay channels at the LHC}},\ }\href
  {https://doi.org/10.1016/j.physletb.2015.03.048} {\bibfield  {journal}
  {\bibinfo  {journal} {Phys. Lett. B}\ }\textbf {\bibinfo {volume} {744}},\
  \bibinfo {pages} {184} (\bibinfo {year} {2015})},\ \Eprint
  {https://arxiv.org/abs/1410.6679} {arXiv:1410.6679 [hep-ex]} \BibitemShut
  {NoStop}%
\bibitem [{\citenamefont {Aad}\ \emph {et~al.}(2020{\natexlab{b}})\citenamefont
  {Aad} \emph {et~al.}}]{ATLAS:2019old}%
  \BibitemOpen
  \bibfield  {author} {\bibinfo {author} {\bibfnamefont {G.}~\bibnamefont
  {Aad}} \emph {et~al.} (\bibinfo {collaboration} {ATLAS}),\ }\bibfield
  {title} {\bibinfo {title} {{Search for the Higgs boson decays $H \rightarrow
  ee$ and $H \rightarrow e\mu$ in $pp$ collisions at $\sqrt{s} = 13$ TeV with
  the ATLAS detector}},\ }\href
  {https://doi.org/10.1016/j.physletb.2019.135148} {\bibfield  {journal}
  {\bibinfo  {journal} {Phys. Lett. B}\ }\textbf {\bibinfo {volume} {801}},\
  \bibinfo {pages} {135148} (\bibinfo {year} {2020}{\natexlab{b}})},\ \Eprint
  {https://arxiv.org/abs/1909.10235} {arXiv:1909.10235 [hep-ex]} \BibitemShut
  {NoStop}%
\bibitem [{\citenamefont {Botella}\ \emph {et~al.}(2016)\citenamefont
  {Botella}, \citenamefont {Branco}, \citenamefont {Rebelo},\ and\
  \citenamefont {Silva-Marcos}}]{Botella:2016krk}%
  \BibitemOpen
  \bibfield  {author} {\bibinfo {author} {\bibfnamefont {F.~J.}\ \bibnamefont
  {Botella}}, \bibinfo {author} {\bibfnamefont {G.~C.}\ \bibnamefont {Branco}},
  \bibinfo {author} {\bibfnamefont {M.~N.}\ \bibnamefont {Rebelo}},\ and\
  \bibinfo {author} {\bibfnamefont {J.~I.}\ \bibnamefont {Silva-Marcos}},\
  }\bibfield  {title} {\bibinfo {title} {{What if the masses of the first two
  quark families are not generated by the standard model Higgs boson?}},\
  }\href {https://doi.org/10.1103/PhysRevD.94.115031} {\bibfield  {journal}
  {\bibinfo  {journal} {Phys. Rev. D}\ }\textbf {\bibinfo {volume} {94}},\
  \bibinfo {pages} {115031} (\bibinfo {year} {2016})},\ \Eprint
  {https://arxiv.org/abs/1602.08011} {arXiv:1602.08011 [hep-ph]} \BibitemShut
  {NoStop}%
\bibitem [{\citenamefont {Ghosh}\ \emph {et~al.}(2016)\citenamefont {Ghosh},
  \citenamefont {Gupta},\ and\ \citenamefont {Perez}}]{Ghosh:2015gpa}%
  \BibitemOpen
  \bibfield  {author} {\bibinfo {author} {\bibfnamefont {D.}~\bibnamefont
  {Ghosh}}, \bibinfo {author} {\bibfnamefont {R.~S.}\ \bibnamefont {Gupta}},\
  and\ \bibinfo {author} {\bibfnamefont {G.}~\bibnamefont {Perez}},\ }\bibfield
   {title} {\bibinfo {title} {{Is the Higgs Mechanism of Fermion Mass
  Generation a Fact? A Yukawa-less First-Two-Generation Model}},\ }\href
  {https://doi.org/10.1016/j.physletb.2016.02.059} {\bibfield  {journal}
  {\bibinfo  {journal} {Phys. Lett. B}\ }\textbf {\bibinfo {volume} {755}},\
  \bibinfo {pages} {504} (\bibinfo {year} {2016})},\ \Eprint
  {https://arxiv.org/abs/1508.01501} {arXiv:1508.01501 [hep-ph]} \BibitemShut
  {NoStop}%
\bibitem [{\citenamefont {Altmannshofer}\ \emph {et~al.}(2016)\citenamefont
  {Altmannshofer}, \citenamefont {Gori}, \citenamefont {Kagan}, \citenamefont
  {Silvestrini},\ and\ \citenamefont {Zupan}}]{Altmannshofer:2015esa}%
  \BibitemOpen
  \bibfield  {author} {\bibinfo {author} {\bibfnamefont {W.}~\bibnamefont
  {Altmannshofer}}, \bibinfo {author} {\bibfnamefont {S.}~\bibnamefont {Gori}},
  \bibinfo {author} {\bibfnamefont {A.~L.}\ \bibnamefont {Kagan}}, \bibinfo
  {author} {\bibfnamefont {L.}~\bibnamefont {Silvestrini}},\ and\ \bibinfo
  {author} {\bibfnamefont {J.}~\bibnamefont {Zupan}},\ }\bibfield  {title}
  {\bibinfo {title} {{Uncovering Mass Generation Through Higgs Flavor
  Violation}},\ }\href {https://doi.org/10.1103/PhysRevD.93.031301} {\bibfield
  {journal} {\bibinfo  {journal} {Phys. Rev. D}\ }\textbf {\bibinfo {volume}
  {93}},\ \bibinfo {pages} {031301} (\bibinfo {year} {2016})},\ \Eprint
  {https://arxiv.org/abs/1507.07927} {arXiv:1507.07927 [hep-ph]} \BibitemShut
  {NoStop}%
\bibitem [{\citenamefont {Dery}\ \emph {et~al.}(2018)\citenamefont {Dery},
  \citenamefont {Frugiuele},\ and\ \citenamefont {Nir}}]{Dery:2017axi}%
  \BibitemOpen
  \bibfield  {author} {\bibinfo {author} {\bibfnamefont {A.}~\bibnamefont
  {Dery}}, \bibinfo {author} {\bibfnamefont {C.}~\bibnamefont {Frugiuele}},\
  and\ \bibinfo {author} {\bibfnamefont {Y.}~\bibnamefont {Nir}},\ }\bibfield
  {title} {\bibinfo {title} {{Large Higgs-electron Yukawa coupling in 2HDM}},\
  }\href {https://doi.org/10.1007/JHEP04(2018)044} {\bibfield  {journal}
  {\bibinfo  {journal} {JHEP}\ }\textbf {\bibinfo {volume} {04}},\ \bibinfo
  {pages} {044}},\ \Eprint {https://arxiv.org/abs/1712.04514} {arXiv:1712.04514
  [hep-ph]} \BibitemShut {NoStop}%
\bibitem [{\citenamefont {Chiang}\ and\ \citenamefont
  {Yagyu}(2021)}]{Chiang:2021pma}%
  \BibitemOpen
  \bibfield  {author} {\bibinfo {author} {\bibfnamefont {C.-W.}\ \bibnamefont
  {Chiang}}\ and\ \bibinfo {author} {\bibfnamefont {K.}~\bibnamefont {Yagyu}},\
  }\bibfield  {title} {\bibinfo {title} {{Radiative Seesaw Mechanism for
  Charged Leptons}},\ }\href {https://doi.org/10.1103/PhysRevD.103.L111302}
  {\bibfield  {journal} {\bibinfo  {journal} {Phys. Rev. D}\ }\textbf {\bibinfo
  {volume} {103}},\ \bibinfo {pages} {L111302} (\bibinfo {year} {2021})},\
  \Eprint {https://arxiv.org/abs/2104.00890} {arXiv:2104.00890 [hep-ph]}
  \BibitemShut {NoStop}%
\bibitem [{\citenamefont {Esteban}\ \emph {et~al.}(2020)\citenamefont
  {Esteban}, \citenamefont {Gonzalez-Garcia}, \citenamefont {Maltoni},
  \citenamefont {Schwetz},\ and\ \citenamefont {Zhou}}]{Esteban:2020cvm}%
  \BibitemOpen
  \bibfield  {author} {\bibinfo {author} {\bibfnamefont {I.}~\bibnamefont
  {Esteban}}, \bibinfo {author} {\bibfnamefont {M.~C.}\ \bibnamefont
  {Gonzalez-Garcia}}, \bibinfo {author} {\bibfnamefont {M.}~\bibnamefont
  {Maltoni}}, \bibinfo {author} {\bibfnamefont {T.}~\bibnamefont {Schwetz}},\
  and\ \bibinfo {author} {\bibfnamefont {A.}~\bibnamefont {Zhou}},\ }\bibfield
  {title} {\bibinfo {title} {{The fate of hints: updated global analysis of
  three-flavor neutrino oscillations}},\ }\href
  {https://doi.org/10.1007/JHEP09(2020)178} {\bibfield  {journal} {\bibinfo
  {journal} {JHEP}\ }\textbf {\bibinfo {volume} {09}},\ \bibinfo {pages}
  {178}},\ \Eprint {https://arxiv.org/abs/2007.14792} {arXiv:2007.14792
  [hep-ph]} \BibitemShut {NoStop}%
\bibitem [{\citenamefont {Aoyama}\ \emph {et~al.}(2020)\citenamefont {Aoyama}
  \emph {et~al.}}]{Aoyama:2020ynm}%
  \BibitemOpen
  \bibfield  {author} {\bibinfo {author} {\bibfnamefont {T.}~\bibnamefont
  {Aoyama}} \emph {et~al.},\ }\bibfield  {title} {\bibinfo {title} {{The
  anomalous magnetic moment of the muon in the Standard Model}},\ }\href
  {https://doi.org/10.1016/j.physrep.2020.07.006} {\bibfield  {journal}
  {\bibinfo  {journal} {Phys. Rept.}\ }\textbf {\bibinfo {volume} {887}},\
  \bibinfo {pages} {1} (\bibinfo {year} {2020})},\ \Eprint
  {https://arxiv.org/abs/2006.04822} {arXiv:2006.04822 [hep-ph]} \BibitemShut
  {NoStop}%
\bibitem [{\citenamefont {Abi}\ \emph {et~al.}(2021)\citenamefont {Abi} \emph
  {et~al.}}]{Abi:2021gix}%
  \BibitemOpen
  \bibfield  {author} {\bibinfo {author} {\bibfnamefont {B.}~\bibnamefont
  {Abi}} \emph {et~al.} (\bibinfo {collaboration} {Muon g-2}),\ }\bibfield
  {title} {\bibinfo {title} {{Measurement of the Positive Muon Anomalous
  Magnetic Moment to 0.46 ppm}},\ }\href
  {https://doi.org/10.1103/PhysRevLett.126.141801} {\bibfield  {journal}
  {\bibinfo  {journal} {Phys. Rev. Lett.}\ }\textbf {\bibinfo {volume} {126}},\
  \bibinfo {pages} {141801} (\bibinfo {year} {2021})},\ \Eprint
  {https://arxiv.org/abs/2104.03281} {arXiv:2104.03281 [hep-ex]} \BibitemShut
  {NoStop}%
\bibitem [{\citenamefont {Saito}(2012)}]{Saito:2012zz}%
  \BibitemOpen
  \bibfield  {author} {\bibinfo {author} {\bibfnamefont {N.}~\bibnamefont
  {Saito}} (\bibinfo {collaboration} {J-PARC g-'2/EDM}),\ }\bibfield  {title}
  {\bibinfo {title} {{A novel precision measurement of muon g-2 and EDM at
  J-PARC}},\ }\href {https://doi.org/10.1063/1.4742078} {\bibfield  {journal}
  {\bibinfo  {journal} {AIP Conf. Proc.}\ }\textbf {\bibinfo {volume} {1467}},\
  \bibinfo {pages} {45} (\bibinfo {year} {2012})}\BibitemShut {NoStop}%
\bibitem [{\citenamefont {Borsanyi}\ \emph {et~al.}(2021)\citenamefont
  {Borsanyi} \emph {et~al.}}]{Borsanyi:2020mff}%
  \BibitemOpen
  \bibfield  {author} {\bibinfo {author} {\bibfnamefont {S.}~\bibnamefont
  {Borsanyi}} \emph {et~al.},\ }\bibfield  {title} {\bibinfo {title} {{Leading
  hadronic contribution to the muon magnetic moment from lattice QCD}},\ }\href
  {https://doi.org/10.1038/s41586-021-03418-1} {\bibfield  {journal} {\bibinfo
  {journal} {Nature}\ }\textbf {\bibinfo {volume} {593}},\ \bibinfo {pages}
  {51} (\bibinfo {year} {2021})},\ \Eprint {https://arxiv.org/abs/2002.12347}
  {arXiv:2002.12347 [hep-lat]} \BibitemShut {NoStop}%
\bibitem [{\citenamefont {C\`e}\ \emph {et~al.}(2022)\citenamefont {C\`e} \emph
  {et~al.}}]{Ce:2022kxy}%
  \BibitemOpen
  \bibfield  {author} {\bibinfo {author} {\bibfnamefont {M.}~\bibnamefont
  {C\`e}} \emph {et~al.},\ }\bibfield  {title} {\bibinfo {title} {{Window
  observable for the hadronic vacuum polarization contribution to the muon
  $g-2$ from lattice QCD}},\ }\href@noop {} {\  (\bibinfo {year} {2022})},\
  \Eprint {https://arxiv.org/abs/2206.06582} {arXiv:2206.06582 [hep-lat]}
  \BibitemShut {NoStop}%
\bibitem [{\citenamefont {Crivellin}\ \emph {et~al.}(2020)\citenamefont
  {Crivellin}, \citenamefont {Hoferichter}, \citenamefont {Manzari},\ and\
  \citenamefont {Montull}}]{Crivellin:2020zul}%
  \BibitemOpen
  \bibfield  {author} {\bibinfo {author} {\bibfnamefont {A.}~\bibnamefont
  {Crivellin}}, \bibinfo {author} {\bibfnamefont {M.}~\bibnamefont
  {Hoferichter}}, \bibinfo {author} {\bibfnamefont {C.~A.}\ \bibnamefont
  {Manzari}},\ and\ \bibinfo {author} {\bibfnamefont {M.}~\bibnamefont
  {Montull}},\ }\bibfield  {title} {\bibinfo {title} {{Hadronic Vacuum
  Polarization: $(g-2)_\mu$ versus Global Electroweak Fits}},\ }\href
  {https://doi.org/10.1103/PhysRevLett.125.091801} {\bibfield  {journal}
  {\bibinfo  {journal} {Phys. Rev. Lett.}\ }\textbf {\bibinfo {volume} {125}},\
  \bibinfo {pages} {091801} (\bibinfo {year} {2020})},\ \Eprint
  {https://arxiv.org/abs/2003.04886} {arXiv:2003.04886 [hep-ph]} \BibitemShut
  {NoStop}%
\bibitem [{\citenamefont {Keshavarzi}\ \emph {et~al.}(2020)\citenamefont
  {Keshavarzi}, \citenamefont {Marciano}, \citenamefont {Passera},\ and\
  \citenamefont {Sirlin}}]{Keshavarzi:2020bfy}%
  \BibitemOpen
  \bibfield  {author} {\bibinfo {author} {\bibfnamefont {A.}~\bibnamefont
  {Keshavarzi}}, \bibinfo {author} {\bibfnamefont {W.~J.}\ \bibnamefont
  {Marciano}}, \bibinfo {author} {\bibfnamefont {M.}~\bibnamefont {Passera}},\
  and\ \bibinfo {author} {\bibfnamefont {A.}~\bibnamefont {Sirlin}},\
  }\bibfield  {title} {\bibinfo {title} {{Muon $g-2$ and $\Delta \alpha$
  connection}},\ }\href {https://doi.org/10.1103/PhysRevD.102.033002}
  {\bibfield  {journal} {\bibinfo  {journal} {Phys. Rev. D}\ }\textbf {\bibinfo
  {volume} {102}},\ \bibinfo {pages} {033002} (\bibinfo {year} {2020})},\
  \Eprint {https://arxiv.org/abs/2006.12666} {arXiv:2006.12666 [hep-ph]}
  \BibitemShut {NoStop}%
\bibitem [{\citenamefont {Colangelo}\ \emph {et~al.}(2021)\citenamefont
  {Colangelo}, \citenamefont {Hoferichter},\ and\ \citenamefont
  {Stoffer}}]{Colangelo:2020lcg}%
  \BibitemOpen
  \bibfield  {author} {\bibinfo {author} {\bibfnamefont {G.}~\bibnamefont
  {Colangelo}}, \bibinfo {author} {\bibfnamefont {M.}~\bibnamefont
  {Hoferichter}},\ and\ \bibinfo {author} {\bibfnamefont {P.}~\bibnamefont
  {Stoffer}},\ }\bibfield  {title} {\bibinfo {title} {{Constraints on the
  two-pion contribution to hadronic vacuum polarization}},\ }\href
  {https://doi.org/10.1016/j.physletb.2021.136073} {\bibfield  {journal}
  {\bibinfo  {journal} {Phys. Lett. B}\ }\textbf {\bibinfo {volume} {814}},\
  \bibinfo {pages} {136073} (\bibinfo {year} {2021})},\ \Eprint
  {https://arxiv.org/abs/2010.07943} {arXiv:2010.07943 [hep-ph]} \BibitemShut
  {NoStop}%
\bibitem [{\citenamefont {Parker}\ \emph {et~al.}(2018)\citenamefont {Parker},
  \citenamefont {Yu}, \citenamefont {Zhong}, \citenamefont {Estey},\ and\
  \citenamefont {M\"uller}}]{Parker:2018vye}%
  \BibitemOpen
  \bibfield  {author} {\bibinfo {author} {\bibfnamefont {R.~H.}\ \bibnamefont
  {Parker}}, \bibinfo {author} {\bibfnamefont {C.}~\bibnamefont {Yu}}, \bibinfo
  {author} {\bibfnamefont {W.}~\bibnamefont {Zhong}}, \bibinfo {author}
  {\bibfnamefont {B.}~\bibnamefont {Estey}},\ and\ \bibinfo {author}
  {\bibfnamefont {H.}~\bibnamefont {M\"uller}},\ }\bibfield  {title} {\bibinfo
  {title} {{Measurement of the fine-structure constant as a test of the
  Standard Model}},\ }\href {https://doi.org/10.1126/science.aap7706}
  {\bibfield  {journal} {\bibinfo  {journal} {Science}\ }\textbf {\bibinfo
  {volume} {360}},\ \bibinfo {pages} {191} (\bibinfo {year} {2018})},\ \Eprint
  {https://arxiv.org/abs/1812.04130} {arXiv:1812.04130 [physics.atom-ph]}
  \BibitemShut {NoStop}%
\bibitem [{\citenamefont {Morel}\ \emph {et~al.}(2020)\citenamefont {Morel},
  \citenamefont {Yao}, \citenamefont {Clad\'e},\ and\ \citenamefont
  {Guellati-Kh\'elifa}}]{Morel:2020dww}%
  \BibitemOpen
  \bibfield  {author} {\bibinfo {author} {\bibfnamefont {L.}~\bibnamefont
  {Morel}}, \bibinfo {author} {\bibfnamefont {Z.}~\bibnamefont {Yao}}, \bibinfo
  {author} {\bibfnamefont {P.}~\bibnamefont {Clad\'e}},\ and\ \bibinfo {author}
  {\bibfnamefont {S.}~\bibnamefont {Guellati-Kh\'elifa}},\ }\bibfield  {title}
  {\bibinfo {title} {{Determination of the fine-structure constant with an
  accuracy of 81 parts per trillion}},\ }\href
  {https://doi.org/10.1038/s41586-020-2964-7} {\bibfield  {journal} {\bibinfo
  {journal} {Nature}\ }\textbf {\bibinfo {volume} {588}},\ \bibinfo {pages}
  {61} (\bibinfo {year} {2020})}\BibitemShut {NoStop}%
\bibitem [{\citenamefont {Abdullah}\ \emph {et~al.}(2019)\citenamefont
  {Abdullah}, \citenamefont {Dutta}, \citenamefont {Ghosh},\ and\ \citenamefont
  {Li}}]{Abdullah:2019ofw}%
  \BibitemOpen
  \bibfield  {author} {\bibinfo {author} {\bibfnamefont {M.}~\bibnamefont
  {Abdullah}}, \bibinfo {author} {\bibfnamefont {B.}~\bibnamefont {Dutta}},
  \bibinfo {author} {\bibfnamefont {S.}~\bibnamefont {Ghosh}},\ and\ \bibinfo
  {author} {\bibfnamefont {T.}~\bibnamefont {Li}},\ }\bibfield  {title}
  {\bibinfo {title} {{$(g-2)_{\mu,e}$ and the ANITA anomalous events in a
  three-loop neutrino mass model}},\ }\href
  {https://doi.org/10.1103/PhysRevD.100.115006} {\bibfield  {journal} {\bibinfo
   {journal} {Phys. Rev. D}\ }\textbf {\bibinfo {volume} {100}},\ \bibinfo
  {pages} {115006} (\bibinfo {year} {2019})},\ \Eprint
  {https://arxiv.org/abs/1907.08109} {arXiv:1907.08109 [hep-ph]} \BibitemShut
  {NoStop}%
\bibitem [{\citenamefont {Chen}\ and\ \citenamefont
  {Nomura}(2021)}]{Chen:2020jvl}%
  \BibitemOpen
  \bibfield  {author} {\bibinfo {author} {\bibfnamefont {C.-H.}\ \bibnamefont
  {Chen}}\ and\ \bibinfo {author} {\bibfnamefont {T.}~\bibnamefont {Nomura}},\
  }\bibfield  {title} {\bibinfo {title} {{Electron and muon $g-2$, radiative
  neutrino mass, and $\ell' \to \ell \gamma$ in a $U(1)_{e-\mu}$ model}},\
  }\href {https://doi.org/10.1016/j.nuclphysb.2021.115314} {\bibfield
  {journal} {\bibinfo  {journal} {Nucl. Phys. B}\ }\textbf {\bibinfo {volume}
  {964}},\ \bibinfo {pages} {115314} (\bibinfo {year} {2021})},\ \Eprint
  {https://arxiv.org/abs/2003.07638} {arXiv:2003.07638 [hep-ph]} \BibitemShut
  {NoStop}%
\bibitem [{\citenamefont {Dutta}\ \emph {et~al.}(2020)\citenamefont {Dutta},
  \citenamefont {Ghosh},\ and\ \citenamefont {Li}}]{Dutta:2020scq}%
  \BibitemOpen
  \bibfield  {author} {\bibinfo {author} {\bibfnamefont {B.}~\bibnamefont
  {Dutta}}, \bibinfo {author} {\bibfnamefont {S.}~\bibnamefont {Ghosh}},\ and\
  \bibinfo {author} {\bibfnamefont {T.}~\bibnamefont {Li}},\ }\bibfield
  {title} {\bibinfo {title} {{Explaining $(g-2)_{\mu,e}$, the KOTO anomaly and
  the MiniBooNE excess in an extended Higgs model with sterile neutrinos}},\
  }\href {https://doi.org/10.1103/PhysRevD.102.055017} {\bibfield  {journal}
  {\bibinfo  {journal} {Phys. Rev. D}\ }\textbf {\bibinfo {volume} {102}},\
  \bibinfo {pages} {055017} (\bibinfo {year} {2020})},\ \Eprint
  {https://arxiv.org/abs/2006.01319} {arXiv:2006.01319 [hep-ph]} \BibitemShut
  {NoStop}%
\bibitem [{\citenamefont {Arbel\'aez}\ \emph {et~al.}(2020)\citenamefont
  {Arbel\'aez}, \citenamefont {Cepedello}, \citenamefont {Fonseca},\ and\
  \citenamefont {Hirsch}}]{Arbelaez:2020rbq}%
  \BibitemOpen
  \bibfield  {author} {\bibinfo {author} {\bibfnamefont {C.}~\bibnamefont
  {Arbel\'aez}}, \bibinfo {author} {\bibfnamefont {R.}~\bibnamefont
  {Cepedello}}, \bibinfo {author} {\bibfnamefont {R.~M.}\ \bibnamefont
  {Fonseca}},\ and\ \bibinfo {author} {\bibfnamefont {M.}~\bibnamefont
  {Hirsch}},\ }\bibfield  {title} {\bibinfo {title} {{$(g-2)$ anomalies and
  neutrino mass}},\ }\href {https://doi.org/10.1103/PhysRevD.102.075005}
  {\bibfield  {journal} {\bibinfo  {journal} {Phys. Rev. D}\ }\textbf {\bibinfo
  {volume} {102}},\ \bibinfo {pages} {075005} (\bibinfo {year} {2020})},\
  \Eprint {https://arxiv.org/abs/2007.11007} {arXiv:2007.11007 [hep-ph]}
  \BibitemShut {NoStop}%
\bibitem [{\citenamefont {Jana}\ \emph
  {et~al.}(2020{\natexlab{a}})\citenamefont {Jana}, \citenamefont {Vishnu},
  \citenamefont {Rodejohann},\ and\ \citenamefont {Saad}}]{Jana:2020joi}%
  \BibitemOpen
  \bibfield  {author} {\bibinfo {author} {\bibfnamefont {S.}~\bibnamefont
  {Jana}}, \bibinfo {author} {\bibfnamefont {P.~K.}\ \bibnamefont {Vishnu}},
  \bibinfo {author} {\bibfnamefont {W.}~\bibnamefont {Rodejohann}},\ and\
  \bibinfo {author} {\bibfnamefont {S.}~\bibnamefont {Saad}},\ }\bibfield
  {title} {\bibinfo {title} {{Dark matter assisted lepton anomalous magnetic
  moments and neutrino masses}},\ }\href
  {https://doi.org/10.1103/PhysRevD.102.075003} {\bibfield  {journal} {\bibinfo
   {journal} {Phys. Rev. D}\ }\textbf {\bibinfo {volume} {102}},\ \bibinfo
  {pages} {075003} (\bibinfo {year} {2020}{\natexlab{a}})},\ \Eprint
  {https://arxiv.org/abs/2008.02377} {arXiv:2008.02377 [hep-ph]} \BibitemShut
  {NoStop}%
\bibitem [{\citenamefont {Cao}\ \emph {et~al.}(2021)\citenamefont {Cao},
  \citenamefont {He}, \citenamefont {Lian}, \citenamefont {Zhang},\ and\
  \citenamefont {Zhu}}]{Cao:2021lmj}%
  \BibitemOpen
  \bibfield  {author} {\bibinfo {author} {\bibfnamefont {J.}~\bibnamefont
  {Cao}}, \bibinfo {author} {\bibfnamefont {Y.}~\bibnamefont {He}}, \bibinfo
  {author} {\bibfnamefont {J.}~\bibnamefont {Lian}}, \bibinfo {author}
  {\bibfnamefont {D.}~\bibnamefont {Zhang}},\ and\ \bibinfo {author}
  {\bibfnamefont {P.}~\bibnamefont {Zhu}},\ }\bibfield  {title} {\bibinfo
  {title} {{Electron and muon anomalous magnetic moments in the inverse seesaw
  extended NMSSM}},\ }\href {https://doi.org/10.1103/PhysRevD.104.055009}
  {\bibfield  {journal} {\bibinfo  {journal} {Phys. Rev. D}\ }\textbf {\bibinfo
  {volume} {104}},\ \bibinfo {pages} {055009} (\bibinfo {year} {2021})},\
  \Eprint {https://arxiv.org/abs/2102.11355} {arXiv:2102.11355 [hep-ph]}
  \BibitemShut {NoStop}%
\bibitem [{\citenamefont {Mondal}\ and\ \citenamefont
  {Okada}(2022)}]{Mondal:2021vou}%
  \BibitemOpen
  \bibfield  {author} {\bibinfo {author} {\bibfnamefont {T.}~\bibnamefont
  {Mondal}}\ and\ \bibinfo {author} {\bibfnamefont {H.}~\bibnamefont {Okada}},\
  }\bibfield  {title} {\bibinfo {title} {{Inverse seesaw and (g-2) anomalies in
  B-L extended two Higgs doublet model}},\ }\href
  {https://doi.org/10.1016/j.nuclphysb.2022.115716} {\bibfield  {journal}
  {\bibinfo  {journal} {Nucl. Phys. B}\ }\textbf {\bibinfo {volume} {976}},\
  \bibinfo {pages} {115716} (\bibinfo {year} {2022})},\ \Eprint
  {https://arxiv.org/abs/2103.13149} {arXiv:2103.13149 [hep-ph]} \BibitemShut
  {NoStop}%
\bibitem [{\citenamefont {Escribano}\ \emph {et~al.}(2021)\citenamefont
  {Escribano}, \citenamefont {Terol-Calvo},\ and\ \citenamefont
  {Vicente}}]{Escribano:2021css}%
  \BibitemOpen
  \bibfield  {author} {\bibinfo {author} {\bibfnamefont {P.}~\bibnamefont
  {Escribano}}, \bibinfo {author} {\bibfnamefont {J.}~\bibnamefont
  {Terol-Calvo}},\ and\ \bibinfo {author} {\bibfnamefont {A.}~\bibnamefont
  {Vicente}},\ }\bibfield  {title} {\bibinfo {title}
  {{$\boldsymbol{(g-2)_{e,\mu}}$ in an extended inverse type-III seesaw
  model}},\ }\href {https://doi.org/10.1103/PhysRevD.103.115018} {\bibfield
  {journal} {\bibinfo  {journal} {Phys. Rev. D}\ }\textbf {\bibinfo {volume}
  {103}},\ \bibinfo {pages} {115018} (\bibinfo {year} {2021})},\ \Eprint
  {https://arxiv.org/abs/2104.03705} {arXiv:2104.03705 [hep-ph]} \BibitemShut
  {NoStop}%
\bibitem [{\citenamefont {Hern\'andez}\ \emph {et~al.}(2021)\citenamefont
  {Hern\'andez}, \citenamefont {Kovalenko}, \citenamefont {Maniatis},\ and\
  \citenamefont {Schmidt}}]{Hernandez:2021iss}%
  \BibitemOpen
  \bibfield  {author} {\bibinfo {author} {\bibfnamefont {A.~E.~C.}\
  \bibnamefont {Hern\'andez}}, \bibinfo {author} {\bibfnamefont
  {S.}~\bibnamefont {Kovalenko}}, \bibinfo {author} {\bibfnamefont
  {M.}~\bibnamefont {Maniatis}},\ and\ \bibinfo {author} {\bibfnamefont
  {I.}~\bibnamefont {Schmidt}},\ }\bibfield  {title} {\bibinfo {title}
  {{Fermion mass hierarchy and g \ensuremath{-} 2 anomalies in an extended 3HDM
  Model}},\ }\href {https://doi.org/10.1007/JHEP10(2021)036} {\bibfield
  {journal} {\bibinfo  {journal} {JHEP}\ }\textbf {\bibinfo {volume} {10}},\
  \bibinfo {pages} {036}},\ \Eprint {https://arxiv.org/abs/2104.07047}
  {arXiv:2104.07047 [hep-ph]} \BibitemShut {NoStop}%
\bibitem [{\citenamefont {Chang}(2021)}]{Chang:2021axw}%
  \BibitemOpen
  \bibfield  {author} {\bibinfo {author} {\bibfnamefont {W.-F.}\ \bibnamefont
  {Chang}},\ }\bibfield  {title} {\bibinfo {title} {{One colorful resolution to
  the neutrino mass generation, three lepton flavor universality anomalies, and
  the Cabibbo angle anomaly}},\ }\href@noop {} {\  (\bibinfo {year} {2021})},\
  \Eprint {https://arxiv.org/abs/2105.06917} {arXiv:2105.06917 [hep-ph]}
  \BibitemShut {NoStop}%
\bibitem [{\citenamefont {Borah}\ \emph {et~al.}(2022)\citenamefont {Borah},
  \citenamefont {Dutta}, \citenamefont {Mahapatra},\ and\ \citenamefont
  {Sahu}}]{Borah:2021khc}%
  \BibitemOpen
  \bibfield  {author} {\bibinfo {author} {\bibfnamefont {D.}~\bibnamefont
  {Borah}}, \bibinfo {author} {\bibfnamefont {M.}~\bibnamefont {Dutta}},
  \bibinfo {author} {\bibfnamefont {S.}~\bibnamefont {Mahapatra}},\ and\
  \bibinfo {author} {\bibfnamefont {N.}~\bibnamefont {Sahu}},\ }\bibfield
  {title} {\bibinfo {title} {{Lepton anomalous magnetic moment with
  singlet-doublet fermion dark matter in a scotogenic
  U(1)L\ensuremath{\mu}-L\ensuremath{\tau} model}},\ }\href
  {https://doi.org/10.1103/PhysRevD.105.015029} {\bibfield  {journal} {\bibinfo
   {journal} {Phys. Rev. D}\ }\textbf {\bibinfo {volume} {105}},\ \bibinfo
  {pages} {015029} (\bibinfo {year} {2022})},\ \Eprint
  {https://arxiv.org/abs/2109.02699} {arXiv:2109.02699 [hep-ph]} \BibitemShut
  {NoStop}%
\bibitem [{\citenamefont {Julio}\ \emph
  {et~al.}(2022{\natexlab{a}})\citenamefont {Julio}, \citenamefont {Saad},\
  and\ \citenamefont {Thapa}}]{Julio:2022ton}%
  \BibitemOpen
  \bibfield  {author} {\bibinfo {author} {\bibfnamefont {J.}~\bibnamefont
  {Julio}}, \bibinfo {author} {\bibfnamefont {S.}~\bibnamefont {Saad}},\ and\
  \bibinfo {author} {\bibfnamefont {A.}~\bibnamefont {Thapa}},\ }\bibfield
  {title} {\bibinfo {title} {{A Tale of Flavor Anomalies and the Origin of
  Neutrino Mass}},\ }\href@noop {} {\  (\bibinfo {year}
  {2022}{\natexlab{a}})},\ \Eprint {https://arxiv.org/abs/2202.10479}
  {arXiv:2202.10479 [hep-ph]} \BibitemShut {NoStop}%
\bibitem [{\citenamefont {Julio}\ \emph
  {et~al.}(2022{\natexlab{b}})\citenamefont {Julio}, \citenamefont {Saad},\
  and\ \citenamefont {Thapa}}]{Julio:2022bue}%
  \BibitemOpen
  \bibfield  {author} {\bibinfo {author} {\bibfnamefont {J.}~\bibnamefont
  {Julio}}, \bibinfo {author} {\bibfnamefont {S.}~\bibnamefont {Saad}},\ and\
  \bibinfo {author} {\bibfnamefont {A.}~\bibnamefont {Thapa}},\ }\bibfield
  {title} {\bibinfo {title} {{Marriage between neutrino mass and flavor
  anomalies}},\ }\href@noop {} {\  (\bibinfo {year} {2022}{\natexlab{b}})},\
  \Eprint {https://arxiv.org/abs/2203.15499} {arXiv:2203.15499 [hep-ph]}
  \BibitemShut {NoStop}%
\bibitem [{\citenamefont {Chowdhury}\ \emph {et~al.}(2022)\citenamefont
  {Chowdhury}, \citenamefont {Ehsanuzzaman},\ and\ \citenamefont
  {Saad}}]{Chowdhury:2022jde}%
  \BibitemOpen
  \bibfield  {author} {\bibinfo {author} {\bibfnamefont {T.~A.}\ \bibnamefont
  {Chowdhury}}, \bibinfo {author} {\bibfnamefont {M.}~\bibnamefont
  {Ehsanuzzaman}},\ and\ \bibinfo {author} {\bibfnamefont {S.}~\bibnamefont
  {Saad}},\ }\bibfield  {title} {\bibinfo {title} {{Dark Matter and
  $(g-2)_{\mu,e}$ in radiative Dirac neutrino mass models}},\ }\href@noop {} {\
   (\bibinfo {year} {2022})},\ \Eprint {https://arxiv.org/abs/2203.14983}
  {arXiv:2203.14983 [hep-ph]} \BibitemShut {NoStop}%
\bibitem [{\citenamefont {Krauss}\ and\ \citenamefont
  {Wilczek}(1989)}]{Krauss:1988zc}%
  \BibitemOpen
  \bibfield  {author} {\bibinfo {author} {\bibfnamefont {L.~M.}\ \bibnamefont
  {Krauss}}\ and\ \bibinfo {author} {\bibfnamefont {F.}~\bibnamefont
  {Wilczek}},\ }\bibfield  {title} {\bibinfo {title} {{Discrete Gauge Symmetry
  in Continuum Theories}},\ }\href
  {https://doi.org/10.1103/PhysRevLett.62.1221} {\bibfield  {journal} {\bibinfo
   {journal} {Phys. Rev. Lett.}\ }\textbf {\bibinfo {volume} {62}},\ \bibinfo
  {pages} {1221} (\bibinfo {year} {1989})}\BibitemShut {NoStop}%
\bibitem [{\citenamefont {Gu}\ and\ \citenamefont {Sarkar}(2008)}]{Gu:2007ug}%
  \BibitemOpen
  \bibfield  {author} {\bibinfo {author} {\bibfnamefont {P.-H.}\ \bibnamefont
  {Gu}}\ and\ \bibinfo {author} {\bibfnamefont {U.}~\bibnamefont {Sarkar}},\
  }\bibfield  {title} {\bibinfo {title} {{Radiative Neutrino Mass, Dark Matter
  and Leptogenesis}},\ }\href {https://doi.org/10.1103/PhysRevD.77.105031}
  {\bibfield  {journal} {\bibinfo  {journal} {Phys. Rev. D}\ }\textbf {\bibinfo
  {volume} {77}},\ \bibinfo {pages} {105031} (\bibinfo {year} {2008})},\
  \Eprint {https://arxiv.org/abs/0712.2933} {arXiv:0712.2933 [hep-ph]}
  \BibitemShut {NoStop}%
\bibitem [{\citenamefont {Ma}\ \emph {et~al.}(2013)\citenamefont {Ma},
  \citenamefont {Picek},\ and\ \citenamefont {Radov\v{c}i\'c}}]{Ma:2013yga}%
  \BibitemOpen
  \bibfield  {author} {\bibinfo {author} {\bibfnamefont {E.}~\bibnamefont
  {Ma}}, \bibinfo {author} {\bibfnamefont {I.}~\bibnamefont {Picek}},\ and\
  \bibinfo {author} {\bibfnamefont {B.}~\bibnamefont {Radov\v{c}i\'c}},\
  }\bibfield  {title} {\bibinfo {title} {{New Scotogenic Model of Neutrino Mass
  with $U(1)_D$ Gauge Interaction}},\ }\href
  {https://doi.org/10.1016/j.physletb.2013.09.049} {\bibfield  {journal}
  {\bibinfo  {journal} {Phys. Lett. B}\ }\textbf {\bibinfo {volume} {726}},\
  \bibinfo {pages} {744} (\bibinfo {year} {2013})},\ \Eprint
  {https://arxiv.org/abs/1308.5313} {arXiv:1308.5313 [hep-ph]} \BibitemShut
  {NoStop}%
\bibitem [{\citenamefont {Kanemura}\ \emph {et~al.}(2011)\citenamefont
  {Kanemura}, \citenamefont {Seto},\ and\ \citenamefont
  {Shimomura}}]{Kanemura:2011vm}%
  \BibitemOpen
  \bibfield  {author} {\bibinfo {author} {\bibfnamefont {S.}~\bibnamefont
  {Kanemura}}, \bibinfo {author} {\bibfnamefont {O.}~\bibnamefont {Seto}},\
  and\ \bibinfo {author} {\bibfnamefont {T.}~\bibnamefont {Shimomura}},\
  }\bibfield  {title} {\bibinfo {title} {{Masses of dark matter and neutrino
  from TeV scale spontaneous $U(1)_{B-L}$ breaking}},\ }\href
  {https://doi.org/10.1103/PhysRevD.84.016004} {\bibfield  {journal} {\bibinfo
  {journal} {Phys. Rev. D}\ }\textbf {\bibinfo {volume} {84}},\ \bibinfo
  {pages} {016004} (\bibinfo {year} {2011})},\ \Eprint
  {https://arxiv.org/abs/1101.5713} {arXiv:1101.5713 [hep-ph]} \BibitemShut
  {NoStop}%
\bibitem [{\citenamefont {Chang}\ and\ \citenamefont
  {Wong}(2012)}]{Chang:2011kv}%
  \BibitemOpen
  \bibfield  {author} {\bibinfo {author} {\bibfnamefont {W.-F.}\ \bibnamefont
  {Chang}}\ and\ \bibinfo {author} {\bibfnamefont {C.-F.}\ \bibnamefont
  {Wong}},\ }\bibfield  {title} {\bibinfo {title} {{A Model for Neutrino Masses
  and Dark Matter with the Discrete Gauge Symmetry}},\ }\href
  {https://doi.org/10.1103/PhysRevD.85.013018} {\bibfield  {journal} {\bibinfo
  {journal} {Phys. Rev. D}\ }\textbf {\bibinfo {volume} {85}},\ \bibinfo
  {pages} {013018} (\bibinfo {year} {2012})},\ \Eprint
  {https://arxiv.org/abs/1104.3934} {arXiv:1104.3934 [hep-ph]} \BibitemShut
  {NoStop}%
\bibitem [{\citenamefont {Baek}(2016)}]{Baek:2015fea}%
  \BibitemOpen
  \bibfield  {author} {\bibinfo {author} {\bibfnamefont {S.}~\bibnamefont
  {Baek}},\ }\bibfield  {title} {\bibinfo {title} {{Dark matter and muon
  $(g-2)$ in local $U(1)_{L_\mu-L_\tau}$-extended Ma Model}},\ }\href
  {https://doi.org/10.1016/j.physletb.2016.02.062} {\bibfield  {journal}
  {\bibinfo  {journal} {Phys. Lett. B}\ }\textbf {\bibinfo {volume} {756}},\
  \bibinfo {pages} {1} (\bibinfo {year} {2016})},\ \Eprint
  {https://arxiv.org/abs/1510.02168} {arXiv:1510.02168 [hep-ph]} \BibitemShut
  {NoStop}%
\bibitem [{\citenamefont {Ho}\ \emph {et~al.}(2016)\citenamefont {Ho},
  \citenamefont {Toma},\ and\ \citenamefont {Tsumura}}]{Ho:2016aye}%
  \BibitemOpen
  \bibfield  {author} {\bibinfo {author} {\bibfnamefont {S.-Y.}\ \bibnamefont
  {Ho}}, \bibinfo {author} {\bibfnamefont {T.}~\bibnamefont {Toma}},\ and\
  \bibinfo {author} {\bibfnamefont {K.}~\bibnamefont {Tsumura}},\ }\bibfield
  {title} {\bibinfo {title} {{Systematic $U(1)_{B–L}$ extensions of
  loop-induced neutrino mass models with dark matter}},\ }\href
  {https://doi.org/10.1103/PhysRevD.94.033007} {\bibfield  {journal} {\bibinfo
  {journal} {Phys. Rev. D}\ }\textbf {\bibinfo {volume} {94}},\ \bibinfo
  {pages} {033007} (\bibinfo {year} {2016})},\ \Eprint
  {https://arxiv.org/abs/1604.07894} {arXiv:1604.07894 [hep-ph]} \BibitemShut
  {NoStop}%
\bibitem [{\citenamefont {Ma}\ \emph {et~al.}(2016)\citenamefont {Ma},
  \citenamefont {Pollard}, \citenamefont {Popov},\ and\ \citenamefont
  {Zakeri}}]{Ma:2016nnn}%
  \BibitemOpen
  \bibfield  {author} {\bibinfo {author} {\bibfnamefont {E.}~\bibnamefont
  {Ma}}, \bibinfo {author} {\bibfnamefont {N.}~\bibnamefont {Pollard}},
  \bibinfo {author} {\bibfnamefont {O.}~\bibnamefont {Popov}},\ and\ \bibinfo
  {author} {\bibfnamefont {M.}~\bibnamefont {Zakeri}},\ }\bibfield  {title}
  {\bibinfo {title} {{Gauge $B–L$ model of radiative neutrino mass with
  multipartite dark matter}},\ }\href
  {https://doi.org/10.1142/S0217732316501637} {\bibfield  {journal} {\bibinfo
  {journal} {Mod. Phys. Lett. A}\ }\textbf {\bibinfo {volume} {31}},\ \bibinfo
  {pages} {1650163} (\bibinfo {year} {2016})},\ \Eprint
  {https://arxiv.org/abs/1605.00991} {arXiv:1605.00991 [hep-ph]} \BibitemShut
  {NoStop}%
\bibitem [{\citenamefont {Nomura}\ and\ \citenamefont
  {Okada}(2019)}]{Nomura:2017vzp}%
  \BibitemOpen
  \bibfield  {author} {\bibinfo {author} {\bibfnamefont {T.}~\bibnamefont
  {Nomura}}\ and\ \bibinfo {author} {\bibfnamefont {H.}~\bibnamefont {Okada}},\
  }\bibfield  {title} {\bibinfo {title} {{Radiative neutrino mass in an
  alternative $U(1)_{B-L}$ gauge symmetry}},\ }\href
  {https://doi.org/10.1016/j.nuclphysb.2019.02.025} {\bibfield  {journal}
  {\bibinfo  {journal} {Nucl. Phys. B}\ }\textbf {\bibinfo {volume} {941}},\
  \bibinfo {pages} {586} (\bibinfo {year} {2019})},\ \Eprint
  {https://arxiv.org/abs/1705.08309} {arXiv:1705.08309 [hep-ph]} \BibitemShut
  {NoStop}%
\bibitem [{\citenamefont {Geng}\ and\ \citenamefont
  {Okada}(2018)}]{Geng:2017foe}%
  \BibitemOpen
  \bibfield  {author} {\bibinfo {author} {\bibfnamefont {C.-Q.}\ \bibnamefont
  {Geng}}\ and\ \bibinfo {author} {\bibfnamefont {H.}~\bibnamefont {Okada}},\
  }\bibfield  {title} {\bibinfo {title} {{Neutrino masses, dark matter and
  leptogenesis with $U(1)_{B-L}$ gauge symmetry}},\ }\href
  {https://doi.org/10.1016/j.dark.2018.02.005} {\bibfield  {journal} {\bibinfo
  {journal} {Phys. Dark Univ.}\ }\textbf {\bibinfo {volume} {20}},\ \bibinfo
  {pages} {13} (\bibinfo {year} {2018})},\ \Eprint
  {https://arxiv.org/abs/1710.09536} {arXiv:1710.09536 [hep-ph]} \BibitemShut
  {NoStop}%
\bibitem [{\citenamefont {Han}\ and\ \citenamefont {Wang}(2018)}]{Han:2018zcn}%
  \BibitemOpen
  \bibfield  {author} {\bibinfo {author} {\bibfnamefont {Z.-L.}\ \bibnamefont
  {Han}}\ and\ \bibinfo {author} {\bibfnamefont {W.}~\bibnamefont {Wang}},\
  }\bibfield  {title} {\bibinfo {title} {{$Z'$ Portal Dark Matter in $B-L$
  Scotogenic Dirac Model}},\ }\href
  {https://doi.org/10.1140/epjc/s10052-018-6308-9} {\bibfield  {journal}
  {\bibinfo  {journal} {Eur. Phys. J. C}\ }\textbf {\bibinfo {volume} {78}},\
  \bibinfo {pages} {839} (\bibinfo {year} {2018})},\ \Eprint
  {https://arxiv.org/abs/1805.02025} {arXiv:1805.02025 [hep-ph]} \BibitemShut
  {NoStop}%
\bibitem [{\citenamefont {Centelles~Chuli\'a}\ \emph
  {et~al.}(2020)\citenamefont {Centelles~Chuli\'a}, \citenamefont {Cepedello},
  \citenamefont {Peinado},\ and\ \citenamefont
  {Srivastava}}]{CentellesChulia:2019gic}%
  \BibitemOpen
  \bibfield  {author} {\bibinfo {author} {\bibfnamefont {S.}~\bibnamefont
  {Centelles~Chuli\'a}}, \bibinfo {author} {\bibfnamefont {R.}~\bibnamefont
  {Cepedello}}, \bibinfo {author} {\bibfnamefont {E.}~\bibnamefont {Peinado}},\
  and\ \bibinfo {author} {\bibfnamefont {R.}~\bibnamefont {Srivastava}},\
  }\bibfield  {title} {\bibinfo {title} {{Scotogenic dark symmetry as a
  residual subgroup of Standard Model symmetries}},\ }\href
  {https://doi.org/10.1088/1674-1137/44/8/083110} {\bibfield  {journal}
  {\bibinfo  {journal} {Chin. Phys. C}\ }\textbf {\bibinfo {volume} {44}},\
  \bibinfo {pages} {083110} (\bibinfo {year} {2020})},\ \Eprint
  {https://arxiv.org/abs/1901.06402} {arXiv:1901.06402 [hep-ph]} \BibitemShut
  {NoStop}%
\bibitem [{\citenamefont {Ma}(2019{\natexlab{a}})}]{Ma:2019yfo}%
  \BibitemOpen
  \bibfield  {author} {\bibinfo {author} {\bibfnamefont {E.}~\bibnamefont
  {Ma}},\ }\bibfield  {title} {\bibinfo {title} {{Scotogenic $U(1)_\chi$ Dirac
  neutrinos}},\ }\href {https://doi.org/10.1016/j.physletb.2019.05.006}
  {\bibfield  {journal} {\bibinfo  {journal} {Phys. Lett. B}\ }\textbf
  {\bibinfo {volume} {793}},\ \bibinfo {pages} {411} (\bibinfo {year}
  {2019}{\natexlab{a}})},\ \Eprint {https://arxiv.org/abs/1901.09091}
  {arXiv:1901.09091 [hep-ph]} \BibitemShut {NoStop}%
\bibitem [{\citenamefont {Kang}\ \emph {et~al.}(2019)\citenamefont {Kang},
  \citenamefont {Popov}, \citenamefont {Srivastava}, \citenamefont {Valle},\
  and\ \citenamefont {Vaquera-Araujo}}]{Kang:2019sab}%
  \BibitemOpen
  \bibfield  {author} {\bibinfo {author} {\bibfnamefont {S.~K.}\ \bibnamefont
  {Kang}}, \bibinfo {author} {\bibfnamefont {O.}~\bibnamefont {Popov}},
  \bibinfo {author} {\bibfnamefont {R.}~\bibnamefont {Srivastava}}, \bibinfo
  {author} {\bibfnamefont {J.~W.~F.}\ \bibnamefont {Valle}},\ and\ \bibinfo
  {author} {\bibfnamefont {C.~A.}\ \bibnamefont {Vaquera-Araujo}},\ }\bibfield
  {title} {\bibinfo {title} {{Scotogenic dark matter stability from gauged
  matter parity}},\ }\href {https://doi.org/10.1016/j.physletb.2019.135013}
  {\bibfield  {journal} {\bibinfo  {journal} {Phys. Lett. B}\ }\textbf
  {\bibinfo {volume} {798}},\ \bibinfo {pages} {135013} (\bibinfo {year}
  {2019})},\ \Eprint {https://arxiv.org/abs/1902.05966} {arXiv:1902.05966
  [hep-ph]} \BibitemShut {NoStop}%
\bibitem [{\citenamefont {Jana}\ \emph {et~al.}(2019)\citenamefont {Jana},
  \citenamefont {Vishnu},\ and\ \citenamefont {Saad}}]{Jana:2019mez}%
  \BibitemOpen
  \bibfield  {author} {\bibinfo {author} {\bibfnamefont {S.}~\bibnamefont
  {Jana}}, \bibinfo {author} {\bibfnamefont {P.~K.}\ \bibnamefont {Vishnu}},\
  and\ \bibinfo {author} {\bibfnamefont {S.}~\bibnamefont {Saad}},\ }\bibfield
  {title} {\bibinfo {title} {{Minimal dirac neutrino mass models from $\hbox
  {U}(1)_{\mathrm{R}}$ gauge symmetry and left\textendash{}right asymmetry at
  colliders}},\ }\href {https://doi.org/10.1140/epjc/s10052-019-7441-9}
  {\bibfield  {journal} {\bibinfo  {journal} {Eur. Phys. J. C}\ }\textbf
  {\bibinfo {volume} {79}},\ \bibinfo {pages} {916} (\bibinfo {year} {2019})},\
  \Eprint {https://arxiv.org/abs/1904.07407} {arXiv:1904.07407 [hep-ph]}
  \BibitemShut {NoStop}%
\bibitem [{\citenamefont {Ma}(2019{\natexlab{b}})}]{Ma:2019iwj}%
  \BibitemOpen
  \bibfield  {author} {\bibinfo {author} {\bibfnamefont {E.}~\bibnamefont
  {Ma}},\ }\bibfield  {title} {\bibinfo {title} {{Scotogenic cobimaximal Dirac
  neutrino mixing from $\Delta (27)$ and $U(1)_\chi $}},\ }\href
  {https://doi.org/10.1140/epjc/s10052-019-7440-x} {\bibfield  {journal}
  {\bibinfo  {journal} {Eur. Phys. J. C}\ }\textbf {\bibinfo {volume} {79}},\
  \bibinfo {pages} {903} (\bibinfo {year} {2019}{\natexlab{b}})},\ \Eprint
  {https://arxiv.org/abs/1905.01535} {arXiv:1905.01535 [hep-ph]} \BibitemShut
  {NoStop}%
\bibitem [{\citenamefont {Centelles~Chuli\'a}\ \emph
  {et~al.}(2019)\citenamefont {Centelles~Chuli\'a}, \citenamefont {Cepedello},
  \citenamefont {Peinado},\ and\ \citenamefont
  {Srivastava}}]{CentellesChulia:2019xky}%
  \BibitemOpen
  \bibfield  {author} {\bibinfo {author} {\bibfnamefont {S.}~\bibnamefont
  {Centelles~Chuli\'a}}, \bibinfo {author} {\bibfnamefont {R.}~\bibnamefont
  {Cepedello}}, \bibinfo {author} {\bibfnamefont {E.}~\bibnamefont {Peinado}},\
  and\ \bibinfo {author} {\bibfnamefont {R.}~\bibnamefont {Srivastava}},\
  }\bibfield  {title} {\bibinfo {title} {{Systematic classification of two loop
  $d$ = 4 Dirac neutrino mass models and the Diracness-dark matter stability
  connection}},\ }\href {https://doi.org/10.1007/JHEP10(2019)093} {\bibfield
  {journal} {\bibinfo  {journal} {JHEP}\ }\textbf {\bibinfo {volume} {10}},\
  \bibinfo {pages} {093}},\ \Eprint {https://arxiv.org/abs/1907.08630}
  {arXiv:1907.08630 [hep-ph]} \BibitemShut {NoStop}%
\bibitem [{\citenamefont {Jana}\ \emph
  {et~al.}(2020{\natexlab{b}})\citenamefont {Jana}, \citenamefont {Vishnu},\
  and\ \citenamefont {Saad}}]{Jana:2019mgj}%
  \BibitemOpen
  \bibfield  {author} {\bibinfo {author} {\bibfnamefont {S.}~\bibnamefont
  {Jana}}, \bibinfo {author} {\bibfnamefont {P.~K.}\ \bibnamefont {Vishnu}},\
  and\ \bibinfo {author} {\bibfnamefont {S.}~\bibnamefont {Saad}},\ }\bibfield
  {title} {\bibinfo {title} {{Minimal realizations of Dirac neutrino mass from
  generic one-loop and two-loop topologies at $d = 5$}},\ }\href
  {https://doi.org/10.1088/1475-7516/2020/04/018} {\bibfield  {journal}
  {\bibinfo  {journal} {JCAP}\ }\textbf {\bibinfo {volume} {04}},\ \bibinfo
  {pages} {018}},\ \Eprint {https://arxiv.org/abs/1910.09537} {arXiv:1910.09537
  [hep-ph]} \BibitemShut {NoStop}%
\bibitem [{\citenamefont {de~la Vega}\ \emph {et~al.}(2020)\citenamefont {de~la
  Vega}, \citenamefont {Nath},\ and\ \citenamefont
  {Peinado}}]{delaVega:2020jcp}%
  \BibitemOpen
  \bibfield  {author} {\bibinfo {author} {\bibfnamefont {L.~M.~G.}\
  \bibnamefont {de~la Vega}}, \bibinfo {author} {\bibfnamefont
  {N.}~\bibnamefont {Nath}},\ and\ \bibinfo {author} {\bibfnamefont
  {E.}~\bibnamefont {Peinado}},\ }\bibfield  {title} {\bibinfo {title} {{Dirac
  neutrinos from Peccei-Quinn symmetry: two examples}},\ }\href
  {https://doi.org/10.1016/j.nuclphysb.2020.115099} {\bibfield  {journal}
  {\bibinfo  {journal} {Nucl. Phys. B}\ }\textbf {\bibinfo {volume} {957}},\
  \bibinfo {pages} {115099} (\bibinfo {year} {2020})},\ \Eprint
  {https://arxiv.org/abs/2001.01846} {arXiv:2001.01846 [hep-ph]} \BibitemShut
  {NoStop}%
\bibitem [{\citenamefont {Wong}(2021)}]{Wong:2020obo}%
  \BibitemOpen
  \bibfield  {author} {\bibinfo {author} {\bibfnamefont {C.-F.}\ \bibnamefont
  {Wong}},\ }\bibfield  {title} {\bibinfo {title} {{Anomaly-free chiral
  $U(1)_D$ and its scotogenic implication}},\ }\href
  {https://doi.org/10.1016/j.dark.2021.100818} {\bibfield  {journal} {\bibinfo
  {journal} {Phys. Dark Univ.}\ }\textbf {\bibinfo {volume} {32}},\ \bibinfo
  {pages} {100818} (\bibinfo {year} {2021})},\ \Eprint
  {https://arxiv.org/abs/2008.08573} {arXiv:2008.08573 [hep-ph]} \BibitemShut
  {NoStop}%
\bibitem [{\citenamefont {Chao}(2011)}]{Chao:2010mp}%
  \BibitemOpen
  \bibfield  {author} {\bibinfo {author} {\bibfnamefont {W.}~\bibnamefont
  {Chao}},\ }\bibfield  {title} {\bibinfo {title} {{Pure Leptonic Gauge
  Symmetry, Neutrino Masses and Dark Matter}},\ }\href
  {https://doi.org/10.1016/j.physletb.2010.10.056} {\bibfield  {journal}
  {\bibinfo  {journal} {Phys. Lett. B}\ }\textbf {\bibinfo {volume} {695}},\
  \bibinfo {pages} {157} (\bibinfo {year} {2011})},\ \Eprint
  {https://arxiv.org/abs/1005.1024} {arXiv:1005.1024 [hep-ph]} \BibitemShut
  {NoStop}%
\bibitem [{\citenamefont {Schwaller}\ \emph {et~al.}(2013)\citenamefont
  {Schwaller}, \citenamefont {Tait},\ and\ \citenamefont
  {Vega-Morales}}]{Schwaller:2013hqa}%
  \BibitemOpen
  \bibfield  {author} {\bibinfo {author} {\bibfnamefont {P.}~\bibnamefont
  {Schwaller}}, \bibinfo {author} {\bibfnamefont {T.~M.~P.}\ \bibnamefont
  {Tait}},\ and\ \bibinfo {author} {\bibfnamefont {R.}~\bibnamefont
  {Vega-Morales}},\ }\bibfield  {title} {\bibinfo {title} {{Dark Matter and
  Vectorlike Leptons from Gauged Lepton Number}},\ }\href
  {https://doi.org/10.1103/PhysRevD.88.035001} {\bibfield  {journal} {\bibinfo
  {journal} {Phys. Rev. D}\ }\textbf {\bibinfo {volume} {88}},\ \bibinfo
  {pages} {035001} (\bibinfo {year} {2013})},\ \Eprint
  {https://arxiv.org/abs/1305.1108} {arXiv:1305.1108 [hep-ph]} \BibitemShut
  {NoStop}%
\bibitem [{\citenamefont {Chang}\ and\ \citenamefont
  {Ng}(2018{\natexlab{a}})}]{Chang:2018vdd}%
  \BibitemOpen
  \bibfield  {author} {\bibinfo {author} {\bibfnamefont {W.-F.}\ \bibnamefont
  {Chang}}\ and\ \bibinfo {author} {\bibfnamefont {J.~N.}\ \bibnamefont {Ng}},\
  }\bibfield  {title} {\bibinfo {title} {{Study of Gauged Lepton Symmetry
  Signatures at Colliders}},\ }\href
  {https://doi.org/10.1103/PhysRevD.98.035015} {\bibfield  {journal} {\bibinfo
  {journal} {Phys. Rev. D}\ }\textbf {\bibinfo {volume} {98}},\ \bibinfo
  {pages} {035015} (\bibinfo {year} {2018}{\natexlab{a}})},\ \Eprint
  {https://arxiv.org/abs/1805.10382} {arXiv:1805.10382 [hep-ph]} \BibitemShut
  {NoStop}%
\bibitem [{\citenamefont {Chang}\ and\ \citenamefont
  {Ng}(2018{\natexlab{b}})}]{Chang:2018wsw}%
  \BibitemOpen
  \bibfield  {author} {\bibinfo {author} {\bibfnamefont {W.-F.}\ \bibnamefont
  {Chang}}\ and\ \bibinfo {author} {\bibfnamefont {J.~N.}\ \bibnamefont {Ng}},\
  }\bibfield  {title} {\bibinfo {title} {{Neutrino masses and gauged
  $U(1)_\ell$ lepton number}},\ }\href
  {https://doi.org/10.1007/JHEP10(2018)015} {\bibfield  {journal} {\bibinfo
  {journal} {JHEP}\ }\textbf {\bibinfo {volume} {10}},\ \bibinfo {pages}
  {015}},\ \Eprint {https://arxiv.org/abs/1807.09439} {arXiv:1807.09439
  [hep-ph]} \BibitemShut {NoStop}%
\bibitem [{\citenamefont {Chang}\ and\ \citenamefont
  {Ng}(2019)}]{Chang:2018nid}%
  \BibitemOpen
  \bibfield  {author} {\bibinfo {author} {\bibfnamefont {W.-F.}\ \bibnamefont
  {Chang}}\ and\ \bibinfo {author} {\bibfnamefont {J.~N.}\ \bibnamefont {Ng}},\
  }\bibfield  {title} {\bibinfo {title} {{Alternative Perspective on Gauged
  Lepton Number and Implications for Collider Physics}},\ }\href
  {https://doi.org/10.1103/PhysRevD.99.075025} {\bibfield  {journal} {\bibinfo
  {journal} {Phys. Rev. D}\ }\textbf {\bibinfo {volume} {99}},\ \bibinfo
  {pages} {075025} (\bibinfo {year} {2019})},\ \Eprint
  {https://arxiv.org/abs/1808.08188} {arXiv:1808.08188 [hep-ph]} \BibitemShut
  {NoStop}%
\bibitem [{\citenamefont {Chang}\ and\ \citenamefont
  {Ng}(2014)}]{Chang:2014lxa}%
  \BibitemOpen
  \bibfield  {author} {\bibinfo {author} {\bibfnamefont {W.-F.}\ \bibnamefont
  {Chang}}\ and\ \bibinfo {author} {\bibfnamefont {J.~N.}\ \bibnamefont {Ng}},\
  }\bibfield  {title} {\bibinfo {title} {{Minimal model of Majoronic dark
  radiation and dark matter}},\ }\href
  {https://doi.org/10.1103/PhysRevD.90.065034} {\bibfield  {journal} {\bibinfo
  {journal} {Phys. Rev. D}\ }\textbf {\bibinfo {volume} {90}},\ \bibinfo
  {pages} {065034} (\bibinfo {year} {2014})},\ \Eprint
  {https://arxiv.org/abs/1406.4601} {arXiv:1406.4601 [hep-ph]} \BibitemShut
  {NoStop}%
\bibitem [{\citenamefont {Chang}\ and\ \citenamefont
  {Ng}(2016)}]{Chang:2016pya}%
  \BibitemOpen
  \bibfield  {author} {\bibinfo {author} {\bibfnamefont {W.-F.}\ \bibnamefont
  {Chang}}\ and\ \bibinfo {author} {\bibfnamefont {J.~N.}\ \bibnamefont {Ng}},\
  }\bibfield  {title} {\bibinfo {title} {{Renormalization Group Study of the
  Minimal Majoronic Dark Radiation and Dark Matter Model}},\ }\href
  {https://doi.org/10.1088/1475-7516/2016/07/027} {\bibfield  {journal}
  {\bibinfo  {journal} {JCAP}\ }\textbf {\bibinfo {volume} {07}},\ \bibinfo
  {pages} {027}},\ \Eprint {https://arxiv.org/abs/1604.02017} {arXiv:1604.02017
  [hep-ph]} \BibitemShut {NoStop}%
\bibitem [{CMS(2020)}]{CMS:2020gsy}%
  \BibitemOpen
  \bibfield  {title} {\bibinfo {title} {{Combined Higgs boson production and
  decay measurements with up to 137 fb$^{-1}$ of proton-proton collision data
  at $\sqrt s$ = 13 TeV, CMS-PAS-HIG-19-005}},\ }\href@noop {} {\  (\bibinfo
  {year} {2020})}\BibitemShut {NoStop}%
\bibitem [{ATL(2021)}]{ATLAS:2021vrm}%
  \BibitemOpen
  \bibfield  {title} {\bibinfo {title} {{Combined measurements of Higgs boson
  production and decay using up to $139$ fb$^{-1}$ of proton-proton collision
  data at $\sqrt{s}= 13$ TeV collected with the ATLAS experiment,
  ATLAS-CONF-2021-053}},\ }\href@noop {} {\  (\bibinfo {year}
  {2021})}\BibitemShut {NoStop}%
\bibitem [{\citenamefont {Barlow}(2004)}]{Barlow:2004wg}%
  \BibitemOpen
  \bibfield  {author} {\bibinfo {author} {\bibfnamefont {R.}~\bibnamefont
  {Barlow}},\ }\bibfield  {title} {\bibinfo {title} {{Asymmetric statistical
  errors}},\ }in\ \href@noop {} {\emph {\bibinfo {booktitle} {{Statistical
  Problems in Particle Physics, Astrophysics and Cosmology}}}}\ (\bibinfo
  {year} {2004})\ pp.\ \bibinfo {pages} {56--59},\ \Eprint
  {https://arxiv.org/abs/physics/0406120} {arXiv:physics/0406120} \BibitemShut
  {NoStop}%
\bibitem [{\citenamefont {Chang}\ and\ \citenamefont
  {Ng}(2005)}]{Chang:2005ag}%
  \BibitemOpen
  \bibfield  {author} {\bibinfo {author} {\bibfnamefont {W.-F.}\ \bibnamefont
  {Chang}}\ and\ \bibinfo {author} {\bibfnamefont {J.~N.}\ \bibnamefont {Ng}},\
  }\bibfield  {title} {\bibinfo {title} {{Lepton flavor violation in extra
  dimension models}},\ }\href {https://doi.org/10.1103/PhysRevD.71.053003}
  {\bibfield  {journal} {\bibinfo  {journal} {Phys. Rev. D}\ }\textbf {\bibinfo
  {volume} {71}},\ \bibinfo {pages} {053003} (\bibinfo {year} {2005})},\
  \Eprint {https://arxiv.org/abs/hep-ph/0501161} {arXiv:hep-ph/0501161}
  \BibitemShut {NoStop}%
\bibitem [{\citenamefont {Baldini}\ \emph {et~al.}(2016)\citenamefont {Baldini}
  \emph {et~al.}}]{TheMEG:2016wtm}%
  \BibitemOpen
  \bibfield  {author} {\bibinfo {author} {\bibfnamefont {A.~M.}\ \bibnamefont
  {Baldini}} \emph {et~al.} (\bibinfo {collaboration} {MEG}),\ }\bibfield
  {title} {\bibinfo {title} {{Search for the lepton flavour violating decay
  $\mu ^+ \rightarrow \mathrm {e}^+ \gamma $ with the full dataset of the MEG
  experiment}},\ }\href {https://doi.org/10.1140/epjc/s10052-016-4271-x}
  {\bibfield  {journal} {\bibinfo  {journal} {Eur. Phys. J. C}\ }\textbf
  {\bibinfo {volume} {76}},\ \bibinfo {pages} {434} (\bibinfo {year} {2016})},\
  \Eprint {https://arxiv.org/abs/1605.05081} {arXiv:1605.05081 [hep-ex]}
  \BibitemShut {NoStop}%
\bibitem [{\citenamefont {Aubert}\ \emph {et~al.}(2010)\citenamefont {Aubert}
  \emph {et~al.}}]{Aubert:2009ag}%
  \BibitemOpen
  \bibfield  {author} {\bibinfo {author} {\bibfnamefont {B.}~\bibnamefont
  {Aubert}} \emph {et~al.} (\bibinfo {collaboration} {BaBar}),\ }\bibfield
  {title} {\bibinfo {title} {{Searches for Lepton Flavor Violation in the
  Decays tau+- ---\ensuremath{>} e+- gamma and tau+- ---\ensuremath{>} mu+-
  gamma}},\ }\href {https://doi.org/10.1103/PhysRevLett.104.021802} {\bibfield
  {journal} {\bibinfo  {journal} {Phys. Rev. Lett.}\ }\textbf {\bibinfo
  {volume} {104}},\ \bibinfo {pages} {021802} (\bibinfo {year} {2010})},\
  \Eprint {https://arxiv.org/abs/0908.2381} {arXiv:0908.2381 [hep-ex]}
  \BibitemShut {NoStop}%
\bibitem [{\citenamefont {Baldini}\ \emph {et~al.}(2018)\citenamefont {Baldini}
  \emph {et~al.}}]{MEGII:2018kmf}%
  \BibitemOpen
  \bibfield  {author} {\bibinfo {author} {\bibfnamefont {A.~M.}\ \bibnamefont
  {Baldini}} \emph {et~al.} (\bibinfo {collaboration} {MEG II}),\ }\bibfield
  {title} {\bibinfo {title} {{The design of the MEG II experiment}},\ }\href
  {https://doi.org/10.1140/epjc/s10052-018-5845-6} {\bibfield  {journal}
  {\bibinfo  {journal} {Eur. Phys. J. C}\ }\textbf {\bibinfo {volume} {78}},\
  \bibinfo {pages} {380} (\bibinfo {year} {2018})},\ \Eprint
  {https://arxiv.org/abs/1801.04688} {arXiv:1801.04688 [physics.ins-det]}
  \BibitemShut {NoStop}%
\bibitem [{\citenamefont {Altmannshofer}\ \emph {et~al.}(2019)\citenamefont
  {Altmannshofer} \emph {et~al.}}]{Belle-II:2018jsg}%
  \BibitemOpen
  \bibfield  {author} {\bibinfo {author} {\bibfnamefont {W.}~\bibnamefont
  {Altmannshofer}} \emph {et~al.} (\bibinfo {collaboration} {Belle-II}),\
  }\bibfield  {title} {\bibinfo {title} {{The Belle II Physics Book}},\ }\href
  {https://doi.org/10.1093/ptep/ptz106} {\bibfield  {journal} {\bibinfo
  {journal} {PTEP}\ }\textbf {\bibinfo {volume} {2019}},\ \bibinfo {pages}
  {123C01} (\bibinfo {year} {2019})},\ \bibinfo {note} {[Erratum: PTEP 2020,
  029201 (2020)]},\ \Eprint {https://arxiv.org/abs/1808.10567}
  {arXiv:1808.10567 [hep-ex]} \BibitemShut {NoStop}%
\bibitem [{\citenamefont {de~Blas}\ \emph {et~al.}(2020)\citenamefont {de~Blas}
  \emph {et~al.}}]{deBlas:2019rxi}%
  \BibitemOpen
  \bibfield  {author} {\bibinfo {author} {\bibfnamefont {J.}~\bibnamefont
  {de~Blas}} \emph {et~al.},\ }\bibfield  {title} {\bibinfo {title} {{Higgs
  Boson Studies at Future Particle Colliders}},\ }\href
  {https://doi.org/10.1007/JHEP01(2020)139} {\bibfield  {journal} {\bibinfo
  {journal} {JHEP}\ }\textbf {\bibinfo {volume} {01}},\ \bibinfo {pages}
  {139}},\ \Eprint {https://arxiv.org/abs/1905.03764} {arXiv:1905.03764
  [hep-ph]} \BibitemShut {NoStop}%
\bibitem [{\citenamefont {Abada}\ \emph {et~al.}(2019)\citenamefont {Abada}
  \emph {et~al.}}]{FCC:2018byv}%
  \BibitemOpen
  \bibfield  {author} {\bibinfo {author} {\bibfnamefont {A.}~\bibnamefont
  {Abada}} \emph {et~al.} (\bibinfo {collaboration} {FCC}),\ }\bibfield
  {title} {\bibinfo {title} {{FCC Physics Opportunities}: {Future Circular
  Collider Conceptual Design Report Volume 1}},\ }\href
  {https://doi.org/10.1140/epjc/s10052-019-6904-3} {\bibfield  {journal}
  {\bibinfo  {journal} {Eur. Phys. J. C}\ }\textbf {\bibinfo {volume} {79}},\
  \bibinfo {pages} {474} (\bibinfo {year} {2019})}\BibitemShut {NoStop}%
\bibitem [{\citenamefont {Dong}\ \emph {et~al.}(2018)\citenamefont {Dong} \emph
  {et~al.}}]{CEPCStudyGroup:2018ghi}%
  \BibitemOpen
  \bibfield  {author} {\bibinfo {author} {\bibfnamefont {M.}~\bibnamefont
  {Dong}} \emph {et~al.} (\bibinfo {collaboration} {CEPC Study Group}),\
  }\bibfield  {title} {\bibinfo {title} {{CEPC Conceptual Design Report: Volume
  2 - Physics \& Detector}},\ }\href@noop {} {\  (\bibinfo {year} {2018})},\
  \Eprint {https://arxiv.org/abs/1811.10545} {arXiv:1811.10545 [hep-ex]}
  \BibitemShut {NoStop}%
\bibitem [{\citenamefont {Bambade}\ \emph {et~al.}(2019)\citenamefont {Bambade}
  \emph {et~al.}}]{Bambade:2019fyw}%
  \BibitemOpen
  \bibfield  {author} {\bibinfo {author} {\bibfnamefont {P.}~\bibnamefont
  {Bambade}} \emph {et~al.},\ }\bibfield  {title} {\bibinfo {title} {{The
  International Linear Collider: A Global Project}},\ }\href@noop {} {\
  (\bibinfo {year} {2019})},\ \Eprint {https://arxiv.org/abs/1903.01629}
  {arXiv:1903.01629 [hep-ex]} \BibitemShut {NoStop}%
\bibitem [{\citenamefont {Charles}\ \emph {et~al.}(2018)\citenamefont {Charles}
  \emph {et~al.}}]{CLICdp:2018cto}%
  \BibitemOpen
  \bibfield  {author} {\bibinfo {author} {\bibfnamefont {T.~K.}\ \bibnamefont
  {Charles}} \emph {et~al.} (\bibinfo {collaboration} {CLICdp, CLIC}),\
  }\bibfield  {title} {\bibinfo {title} {{The Compact Linear Collider (CLIC) -
  2018 Summary Report}}\ }\textbf {\bibinfo {volume} {2/2018}},\ \href
  {https://doi.org/10.23731/CYRM-2018-002} {10.23731/CYRM-2018-002} (\bibinfo
  {year} {2018}),\ \Eprint {https://arxiv.org/abs/1812.06018} {arXiv:1812.06018
  [physics.acc-ph]} \BibitemShut {NoStop}%
\bibitem [{\citenamefont {d'Enterria}\ \emph {et~al.}(2022)\citenamefont
  {d'Enterria}, \citenamefont {Poldaru},\ and\ \citenamefont
  {Wojcik}}]{dEnterria:2021xij}%
  \BibitemOpen
  \bibfield  {author} {\bibinfo {author} {\bibfnamefont {D.}~\bibnamefont
  {d'Enterria}}, \bibinfo {author} {\bibfnamefont {A.}~\bibnamefont
  {Poldaru}},\ and\ \bibinfo {author} {\bibfnamefont {G.}~\bibnamefont
  {Wojcik}},\ }\bibfield  {title} {\bibinfo {title} {{Measuring the electron
  Yukawa coupling via resonant s-channel Higgs production at FCC-ee}},\ }\href
  {https://doi.org/10.1140/epjp/s13360-021-02204-2} {\bibfield  {journal}
  {\bibinfo  {journal} {Eur. Phys. J. Plus}\ }\textbf {\bibinfo {volume}
  {137}},\ \bibinfo {pages} {201} (\bibinfo {year} {2022})},\ \Eprint
  {https://arxiv.org/abs/2107.02686} {arXiv:2107.02686 [hep-ex]} \BibitemShut
  {NoStop}%
\bibitem [{\citenamefont {Shifman}\ \emph {et~al.}(1979)\citenamefont
  {Shifman}, \citenamefont {Vainshtein}, \citenamefont {Voloshin},\ and\
  \citenamefont {Zakharov}}]{Shifman:1979eb}%
  \BibitemOpen
  \bibfield  {author} {\bibinfo {author} {\bibfnamefont {M.~A.}\ \bibnamefont
  {Shifman}}, \bibinfo {author} {\bibfnamefont {A.~I.}\ \bibnamefont
  {Vainshtein}}, \bibinfo {author} {\bibfnamefont {M.~B.}\ \bibnamefont
  {Voloshin}},\ and\ \bibinfo {author} {\bibfnamefont {V.~I.}\ \bibnamefont
  {Zakharov}},\ }\bibfield  {title} {\bibinfo {title} {{Low-Energy Theorems for
  Higgs Boson Couplings to Photons}},\ }\href@noop {} {\bibfield  {journal}
  {\bibinfo  {journal} {Sov. J. Nucl. Phys.}\ }\textbf {\bibinfo {volume}
  {30}},\ \bibinfo {pages} {711} (\bibinfo {year} {1979})}\BibitemShut
  {NoStop}%
\bibitem [{\citenamefont {Chang}\ \emph {et~al.}(2012)\citenamefont {Chang},
  \citenamefont {Ng},\ and\ \citenamefont {Wu}}]{Chang:2012ta}%
  \BibitemOpen
  \bibfield  {author} {\bibinfo {author} {\bibfnamefont {W.-F.}\ \bibnamefont
  {Chang}}, \bibinfo {author} {\bibfnamefont {J.~N.}\ \bibnamefont {Ng}},\ and\
  \bibinfo {author} {\bibfnamefont {J.~M.~S.}\ \bibnamefont {Wu}},\ }\bibfield
  {title} {\bibinfo {title} {{Constraints on New Scalars from the LHC 125 GeV
  Higgs Signal}},\ }\href {https://doi.org/10.1103/PhysRevD.86.033003}
  {\bibfield  {journal} {\bibinfo  {journal} {Phys. Rev. D}\ }\textbf {\bibinfo
  {volume} {86}},\ \bibinfo {pages} {033003} (\bibinfo {year} {2012})},\
  \Eprint {https://arxiv.org/abs/1206.5047} {arXiv:1206.5047 [hep-ph]}
  \BibitemShut {NoStop}%
\bibitem [{ATL(2020)}]{ATLAS:2020qdt}%
  \BibitemOpen
  \bibfield  {title} {\bibinfo {title} {{A combination of measurements of Higgs
  boson production and decay using up to $139$ fb$^{-1}$ of proton--proton
  collision data at $\sqrt{s}=$ 13 TeV collected with the ATLAS experiment,
  ATLAS-CONF-2020-027}},\ }\href@noop {} {\  (\bibinfo {year}
  {2020})}\BibitemShut {NoStop}%
\bibitem [{\citenamefont {Kuno}\ and\ \citenamefont
  {Okada}(2001)}]{Kuno:1999jp}%
  \BibitemOpen
  \bibfield  {author} {\bibinfo {author} {\bibfnamefont {Y.}~\bibnamefont
  {Kuno}}\ and\ \bibinfo {author} {\bibfnamefont {Y.}~\bibnamefont {Okada}},\
  }\bibfield  {title} {\bibinfo {title} {{Muon decay and physics beyond the
  standard model}},\ }\href {https://doi.org/10.1103/RevModPhys.73.151}
  {\bibfield  {journal} {\bibinfo  {journal} {Rev. Mod. Phys.}\ }\textbf
  {\bibinfo {volume} {73}},\ \bibinfo {pages} {151} (\bibinfo {year} {2001})},\
  \Eprint {https://arxiv.org/abs/hep-ph/9909265} {arXiv:hep-ph/9909265}
  \BibitemShut {NoStop}%
\bibitem [{\citenamefont {Bertl}\ \emph {et~al.}(2006)\citenamefont {Bertl}
  \emph {et~al.}}]{SINDRUMII:2006dvw}%
  \BibitemOpen
  \bibfield  {author} {\bibinfo {author} {\bibfnamefont {W.~H.}\ \bibnamefont
  {Bertl}} \emph {et~al.} (\bibinfo {collaboration} {SINDRUM II}),\ }\bibfield
  {title} {\bibinfo {title} {{A Search for muon to electron conversion in
  muonic gold}},\ }\href {https://doi.org/10.1140/epjc/s2006-02582-x}
  {\bibfield  {journal} {\bibinfo  {journal} {Eur. Phys. J. C}\ }\textbf
  {\bibinfo {volume} {47}},\ \bibinfo {pages} {337} (\bibinfo {year}
  {2006})}\BibitemShut {NoStop}%
\bibitem [{\citenamefont {Kitano}\ \emph {et~al.}(2002)\citenamefont {Kitano},
  \citenamefont {Koike},\ and\ \citenamefont {Okada}}]{Kitano:2002mt}%
  \BibitemOpen
  \bibfield  {author} {\bibinfo {author} {\bibfnamefont {R.}~\bibnamefont
  {Kitano}}, \bibinfo {author} {\bibfnamefont {M.}~\bibnamefont {Koike}},\ and\
  \bibinfo {author} {\bibfnamefont {Y.}~\bibnamefont {Okada}},\ }\bibfield
  {title} {\bibinfo {title} {{Detailed calculation of lepton flavor violating
  muon electron conversion rate for various nuclei}},\ }\href
  {https://doi.org/10.1103/PhysRevD.76.059902} {\bibfield  {journal} {\bibinfo
  {journal} {Phys. Rev. D}\ }\textbf {\bibinfo {volume} {66}},\ \bibinfo
  {pages} {096002} (\bibinfo {year} {2002})},\ \bibinfo {note} {[Erratum:
  Phys.Rev.D 76, 059902 (2007)]},\ \Eprint
  {https://arxiv.org/abs/hep-ph/0203110} {arXiv:hep-ph/0203110} \BibitemShut
  {NoStop}%
\bibitem [{\citenamefont {Suzuki}\ \emph {et~al.}(1987)\citenamefont {Suzuki},
  \citenamefont {Measday},\ and\ \citenamefont {Roalsvig}}]{Suzuki:1987jf}%
  \BibitemOpen
  \bibfield  {author} {\bibinfo {author} {\bibfnamefont {T.}~\bibnamefont
  {Suzuki}}, \bibinfo {author} {\bibfnamefont {D.~F.}\ \bibnamefont
  {Measday}},\ and\ \bibinfo {author} {\bibfnamefont {J.~P.}\ \bibnamefont
  {Roalsvig}},\ }\bibfield  {title} {\bibinfo {title} {{Total Nuclear Capture
  Rates for Negative Muons}},\ }\href
  {https://doi.org/10.1103/PhysRevC.35.2212} {\bibfield  {journal} {\bibinfo
  {journal} {Phys. Rev. C}\ }\textbf {\bibinfo {volume} {35}},\ \bibinfo
  {pages} {2212} (\bibinfo {year} {1987})}\BibitemShut {NoStop}%
\bibitem [{\citenamefont {Teshima}(2018)}]{Teshima:2018ise}%
  \BibitemOpen
  \bibfield  {author} {\bibinfo {author} {\bibfnamefont {N.}~\bibnamefont
  {Teshima}} (\bibinfo {collaboration} {DeeMe}),\ }\bibfield  {title} {\bibinfo
  {title} {{DeeMe experiment to search for muon to electron conversion at
  J-PARC MLF}},\ }\href {https://doi.org/10.22323/1.295.0109} {\bibfield
  {journal} {\bibinfo  {journal} {PoS}\ }\textbf {\bibinfo {volume}
  {NuFact2017}},\ \bibinfo {pages} {109} (\bibinfo {year} {2018})}\BibitemShut
  {NoStop}%
\bibitem [{\citenamefont {Abramishvili}\ \emph {et~al.}(2020)\citenamefont
  {Abramishvili} \emph {et~al.}}]{COMET:2018auw}%
  \BibitemOpen
  \bibfield  {author} {\bibinfo {author} {\bibfnamefont {R.}~\bibnamefont
  {Abramishvili}} \emph {et~al.} (\bibinfo {collaboration} {COMET}),\
  }\bibfield  {title} {\bibinfo {title} {{COMET Phase-I Technical Design
  Report}},\ }\href {https://doi.org/10.1093/ptep/ptz125} {\bibfield  {journal}
  {\bibinfo  {journal} {PTEP}\ }\textbf {\bibinfo {volume} {2020}},\ \bibinfo
  {pages} {033C01} (\bibinfo {year} {2020})},\ \Eprint
  {https://arxiv.org/abs/1812.09018} {arXiv:1812.09018 [physics.ins-det]}
  \BibitemShut {NoStop}%
\bibitem [{\citenamefont {Bartoszek}\ \emph {et~al.}(2014)\citenamefont
  {Bartoszek} \emph {et~al.}}]{Mu2e:2014fns}%
  \BibitemOpen
  \bibfield  {author} {\bibinfo {author} {\bibfnamefont {L.}~\bibnamefont
  {Bartoszek}} \emph {et~al.} (\bibinfo {collaboration} {Mu2e}),\ }\bibfield
  {title} {\bibinfo {title} {{Mu2e Technical Design Report}}\ }\href
  {https://doi.org/10.2172/1172555} {10.2172/1172555} (\bibinfo {year}
  {2014}),\ \Eprint {https://arxiv.org/abs/1501.05241} {arXiv:1501.05241
  [physics.ins-det]} \BibitemShut {NoStop}%
\bibitem [{\citenamefont {Kuno}(2005)}]{Kuno:2005mm}%
  \BibitemOpen
  \bibfield  {author} {\bibinfo {author} {\bibfnamefont {Y.}~\bibnamefont
  {Kuno}},\ }\bibfield  {title} {\bibinfo {title} {{PRISM/PRIME}},\ }\href
  {https://doi.org/10.1016/j.nuclphysbps.2005.05.073} {\bibfield  {journal}
  {\bibinfo  {journal} {Nucl. Phys. B Proc. Suppl.}\ }\textbf {\bibinfo
  {volume} {149}},\ \bibinfo {pages} {376} (\bibinfo {year}
  {2005})}\BibitemShut {NoStop}%
\bibitem [{\citenamefont {Petcov}(1977)}]{Petcov:1977ab}%
  \BibitemOpen
  \bibfield  {author} {\bibinfo {author} {\bibfnamefont {S.~T.}\ \bibnamefont
  {Petcov}},\ }\bibfield  {title} {\bibinfo {title} {{Heavy Neutral Lepton
  Mixing and mu --\ensuremath{>} 3 e Decay}},\ }\href
  {https://doi.org/10.1016/0370-2693(77)90495-6} {\bibfield  {journal}
  {\bibinfo  {journal} {Phys. Lett. B}\ }\textbf {\bibinfo {volume} {68}},\
  \bibinfo {pages} {365} (\bibinfo {year} {1977})}\BibitemShut {NoStop}%
\bibitem [{\citenamefont {Bellgardt}\ \emph {et~al.}(1988)\citenamefont
  {Bellgardt} \emph {et~al.}}]{SINDRUM:1987nra}%
  \BibitemOpen
  \bibfield  {author} {\bibinfo {author} {\bibfnamefont {U.}~\bibnamefont
  {Bellgardt}} \emph {et~al.} (\bibinfo {collaboration} {SINDRUM}),\ }\bibfield
   {title} {\bibinfo {title} {{Search for the Decay mu+ ---\ensuremath{>} e+ e+
  e-}},\ }\href {https://doi.org/10.1016/0550-3213(88)90462-2} {\bibfield
  {journal} {\bibinfo  {journal} {Nucl. Phys. B}\ }\textbf {\bibinfo {volume}
  {299}},\ \bibinfo {pages} {1} (\bibinfo {year} {1988})}\BibitemShut {NoStop}%
\bibitem [{\citenamefont {Blondel}\ \emph {et~al.}(2013)\citenamefont {Blondel}
  \emph {et~al.}}]{Blondel:2013ia}%
  \BibitemOpen
  \bibfield  {author} {\bibinfo {author} {\bibfnamefont {A.}~\bibnamefont
  {Blondel}} \emph {et~al.},\ }\bibfield  {title} {\bibinfo {title} {{Research
  Proposal for an Experiment to Search for the Decay $\mu \to eee$}},\
  }\href@noop {} {\  (\bibinfo {year} {2013})},\ \Eprint
  {https://arxiv.org/abs/1301.6113} {arXiv:1301.6113 [physics.ins-det]}
  \BibitemShut {NoStop}%
\bibitem [{\citenamefont {Hayasaka}\ \emph {et~al.}(2010)\citenamefont
  {Hayasaka} \emph {et~al.}}]{Hayasaka:2010np}%
  \BibitemOpen
  \bibfield  {author} {\bibinfo {author} {\bibfnamefont {K.}~\bibnamefont
  {Hayasaka}} \emph {et~al.},\ }\bibfield  {title} {\bibinfo {title} {{Search
  for Lepton Flavor Violating Tau Decays into Three Leptons with 719 Million
  Produced Tau+Tau- Pairs}},\ }\href
  {https://doi.org/10.1016/j.physletb.2010.03.037} {\bibfield  {journal}
  {\bibinfo  {journal} {Phys. Lett. B}\ }\textbf {\bibinfo {volume} {687}},\
  \bibinfo {pages} {139} (\bibinfo {year} {2010})},\ \Eprint
  {https://arxiv.org/abs/1001.3221} {arXiv:1001.3221 [hep-ex]} \BibitemShut
  {NoStop}%
\bibitem [{\citenamefont {Aaij}\ \emph {et~al.}(2018)\citenamefont {Aaij} \emph
  {et~al.}}]{LHCb:2018roe}%
  \BibitemOpen
  \bibfield  {author} {\bibinfo {author} {\bibfnamefont {R.}~\bibnamefont
  {Aaij}} \emph {et~al.} (\bibinfo {collaboration} {LHCb}),\ }\bibfield
  {title} {\bibinfo {title} {{Physics case for an LHCb Upgrade II -
  Opportunities in flavour physics, and beyond, in the HL-LHC era}},\
  }\href@noop {} {\  (\bibinfo {year} {2018})},\ \Eprint
  {https://arxiv.org/abs/1808.08865} {arXiv:1808.08865 [hep-ex]} \BibitemShut
  {NoStop}%
\bibitem [{\citenamefont {Beacham}\ \emph {et~al.}(2020)\citenamefont {Beacham}
  \emph {et~al.}}]{Beacham:2019nyx}%
  \BibitemOpen
  \bibfield  {author} {\bibinfo {author} {\bibfnamefont {J.}~\bibnamefont
  {Beacham}} \emph {et~al.},\ }\bibfield  {title} {\bibinfo {title} {{Physics
  Beyond Colliders at CERN: Beyond the Standard Model Working Group Report}},\
  }\href {https://doi.org/10.1088/1361-6471/ab4cd2} {\bibfield  {journal}
  {\bibinfo  {journal} {J. Phys. G}\ }\textbf {\bibinfo {volume} {47}},\
  \bibinfo {pages} {010501} (\bibinfo {year} {2020})},\ \Eprint
  {https://arxiv.org/abs/1901.09966} {arXiv:1901.09966 [hep-ex]} \BibitemShut
  {NoStop}%
\bibitem [{\citenamefont {Holdom}(1986)}]{Holdom:1985ag}%
  \BibitemOpen
  \bibfield  {author} {\bibinfo {author} {\bibfnamefont {B.}~\bibnamefont
  {Holdom}},\ }\bibfield  {title} {\bibinfo {title} {{Two U(1)'s and Epsilon
  Charge Shifts}},\ }\href {https://doi.org/10.1016/0370-2693(86)91377-8}
  {\bibfield  {journal} {\bibinfo  {journal} {Phys. Lett. B}\ }\textbf
  {\bibinfo {volume} {166}},\ \bibinfo {pages} {196} (\bibinfo {year}
  {1986})}\BibitemShut {NoStop}%
\bibitem [{\citenamefont {Chang}\ \emph {et~al.}(2006)\citenamefont {Chang},
  \citenamefont {Ng},\ and\ \citenamefont {Wu}}]{Chang:2006fp}%
  \BibitemOpen
  \bibfield  {author} {\bibinfo {author} {\bibfnamefont {W.-F.}\ \bibnamefont
  {Chang}}, \bibinfo {author} {\bibfnamefont {J.~N.}\ \bibnamefont {Ng}},\ and\
  \bibinfo {author} {\bibfnamefont {J.~M.~S.}\ \bibnamefont {Wu}},\ }\bibfield
  {title} {\bibinfo {title} {{A Very Narrow Shadow Extra Z-boson at
  Colliders}},\ }\href {https://doi.org/10.1103/PhysRevD.74.095005} {\bibfield
  {journal} {\bibinfo  {journal} {Phys. Rev. D}\ }\textbf {\bibinfo {volume}
  {74}},\ \bibinfo {pages} {095005} (\bibinfo {year} {2006})},\ \bibinfo {note}
  {[Erratum: Phys.Rev.D 79, 039902 (2009)]},\ \Eprint
  {https://arxiv.org/abs/hep-ph/0608068} {arXiv:hep-ph/0608068} \BibitemShut
  {NoStop}%
\bibitem [{\citenamefont {Chang}\ \emph {et~al.}(2007)\citenamefont {Chang},
  \citenamefont {Ng},\ and\ \citenamefont {Wu}}]{Chang:2007ki}%
  \BibitemOpen
  \bibfield  {author} {\bibinfo {author} {\bibfnamefont {W.-F.}\ \bibnamefont
  {Chang}}, \bibinfo {author} {\bibfnamefont {J.~N.}\ \bibnamefont {Ng}},\ and\
  \bibinfo {author} {\bibfnamefont {J.~M.~S.}\ \bibnamefont {Wu}},\ }\bibfield
  {title} {\bibinfo {title} {{Shadow Higgs from a scale-invariant hidden
  U(1)(s) model}},\ }\href {https://doi.org/10.1103/PhysRevD.75.115016}
  {\bibfield  {journal} {\bibinfo  {journal} {Phys. Rev. D}\ }\textbf {\bibinfo
  {volume} {75}},\ \bibinfo {pages} {115016} (\bibinfo {year} {2007})},\
  \Eprint {https://arxiv.org/abs/hep-ph/0701254} {arXiv:hep-ph/0701254}
  \BibitemShut {NoStop}%
\bibitem [{\citenamefont {Curtin}\ \emph {et~al.}(2015)\citenamefont {Curtin},
  \citenamefont {Essig}, \citenamefont {Gori},\ and\ \citenamefont
  {Shelton}}]{Curtin:2014cca}%
  \BibitemOpen
  \bibfield  {author} {\bibinfo {author} {\bibfnamefont {D.}~\bibnamefont
  {Curtin}}, \bibinfo {author} {\bibfnamefont {R.}~\bibnamefont {Essig}},
  \bibinfo {author} {\bibfnamefont {S.}~\bibnamefont {Gori}},\ and\ \bibinfo
  {author} {\bibfnamefont {J.}~\bibnamefont {Shelton}},\ }\bibfield  {title}
  {\bibinfo {title} {{Illuminating Dark Photons with High-Energy Colliders}},\
  }\href {https://doi.org/10.1007/JHEP02(2015)157} {\bibfield  {journal}
  {\bibinfo  {journal} {JHEP}\ }\textbf {\bibinfo {volume} {02}},\ \bibinfo
  {pages} {157}},\ \Eprint {https://arxiv.org/abs/1412.0018} {arXiv:1412.0018
  [hep-ph]} \BibitemShut {NoStop}%
\bibitem [{\citenamefont {Akerib}\ \emph {et~al.}(2017)\citenamefont {Akerib}
  \emph {et~al.}}]{LUX:2016ggv}%
  \BibitemOpen
  \bibfield  {author} {\bibinfo {author} {\bibfnamefont {D.~S.}\ \bibnamefont
  {Akerib}} \emph {et~al.} (\bibinfo {collaboration} {LUX}),\ }\bibfield
  {title} {\bibinfo {title} {{Results from a search for dark matter in the
  complete LUX exposure}},\ }\href
  {https://doi.org/10.1103/PhysRevLett.118.021303} {\bibfield  {journal}
  {\bibinfo  {journal} {Phys. Rev. Lett.}\ }\textbf {\bibinfo {volume} {118}},\
  \bibinfo {pages} {021303} (\bibinfo {year} {2017})},\ \Eprint
  {https://arxiv.org/abs/1608.07648} {arXiv:1608.07648 [astro-ph.CO]}
  \BibitemShut {NoStop}%
\bibitem [{\citenamefont {Aprile}\ \emph {et~al.}(2018)\citenamefont {Aprile}
  \emph {et~al.}}]{XENON:2018voc}%
  \BibitemOpen
  \bibfield  {author} {\bibinfo {author} {\bibfnamefont {E.}~\bibnamefont
  {Aprile}} \emph {et~al.} (\bibinfo {collaboration} {XENON}),\ }\bibfield
  {title} {\bibinfo {title} {{Dark Matter Search Results from a One Ton-Year
  Exposure of XENON1T}},\ }\href
  {https://doi.org/10.1103/PhysRevLett.121.111302} {\bibfield  {journal}
  {\bibinfo  {journal} {Phys. Rev. Lett.}\ }\textbf {\bibinfo {volume} {121}},\
  \bibinfo {pages} {111302} (\bibinfo {year} {2018})},\ \Eprint
  {https://arxiv.org/abs/1805.12562} {arXiv:1805.12562 [astro-ph.CO]}
  \BibitemShut {NoStop}%
\bibitem [{\citenamefont {Wang}\ \emph {et~al.}(2020)\citenamefont {Wang} \emph
  {et~al.}}]{PandaX-II:2020oim}%
  \BibitemOpen
  \bibfield  {author} {\bibinfo {author} {\bibfnamefont {Q.}~\bibnamefont
  {Wang}} \emph {et~al.} (\bibinfo {collaboration} {PandaX-II}),\ }\bibfield
  {title} {\bibinfo {title} {{Results of dark matter search using the full
  PandaX-II exposure}},\ }\href {https://doi.org/10.1088/1674-1137/abb658}
  {\bibfield  {journal} {\bibinfo  {journal} {Chin. Phys. C}\ }\textbf
  {\bibinfo {volume} {44}},\ \bibinfo {pages} {125001} (\bibinfo {year}
  {2020})},\ \Eprint {https://arxiv.org/abs/2007.15469} {arXiv:2007.15469
  [astro-ph.CO]} \BibitemShut {NoStop}%
\bibitem [{\citenamefont {Aghanim}\ \emph {et~al.}(2020)\citenamefont {Aghanim}
  \emph {et~al.}}]{Planck:2018vyg}%
  \BibitemOpen
  \bibfield  {author} {\bibinfo {author} {\bibfnamefont {N.}~\bibnamefont
  {Aghanim}} \emph {et~al.} (\bibinfo {collaboration} {Planck}),\ }\bibfield
  {title} {\bibinfo {title} {{Planck 2018 results. VI. Cosmological
  parameters}},\ }\href {https://doi.org/10.1051/0004-6361/201833910}
  {\bibfield  {journal} {\bibinfo  {journal} {Astron. Astrophys.}\ }\textbf
  {\bibinfo {volume} {641}},\ \bibinfo {pages} {A6} (\bibinfo {year} {2020})},\
  \bibinfo {note} {[Erratum: Astron.Astrophys. 652, C4 (2021)]},\ \Eprint
  {https://arxiv.org/abs/1807.06209} {arXiv:1807.06209 [astro-ph.CO]}
  \BibitemShut {NoStop}%
\end{thebibliography}%
\end{document}